\documentclass[a4paper,11pt]{article}
\pdfoutput=1 

\usepackage{jheppub} 

\usepackage[T1]{fontenc} 

\usepackage{hyperref}
\hypersetup{
  colorlinks=true,
  citecolor=cyan!50!blue,
  linkcolor=cyan!50!blue,
  filecolor=cyan!50!blue,      
  urlcolor=cyan!50!blue,
}
\usepackage{tikz}
\usetikzlibrary{snakes}
\usetikzlibrary{positioning}
\usetikzlibrary{arrows}
\usetikzlibrary{positioning,decorations.pathmorphing,decorations.markings,arrows}
\usetikzlibrary{arrows.meta,arrows}
\newcommand{\midarrow}{\tikz \draw[-stealth] (-1pt,0) -- (1pt,0);}
\usetikzlibrary{math}
\definecolor{neworange}{RGB}{255,110,30}
\definecolor{newteal}{RGB}{0,182,176}
\definecolor{newgold}{RGB}{169,108,8}
\definecolor{newgreen}{RGB}{10,127,76}
\tikzset{
mystyle/.style={
  circle,
  inner sep=0pt,
  text width=7mm,
  align=center,
  draw=black,
  fill=white
  }
}

\usepackage{amssymb}
\usepackage{pifont}
\newcommand{\cmark}{\ding{51}}%
\newcommand{\xmark}{\ding{55}}%

\tikzset{square left brace/.style={ncbar=0.5cm}}
\tikzset{square right brace/.style={ncbar=-0.5cm}}

\tikzset{round left paren/.style={ncbar=0.5cm,out=120,in=-120}}
\tikzset{round right paren/.style={ncbar=0.5cm,out=60,in=-60}}

\usepackage{comment} 
\usepackage{empheq}
\usepackage{amsmath}
\usepackage{mathtools}
\usepackage{bbm}
\usepackage{nicematrix}
\usepackage[normalem]{ulem}

\usepackage{float}

\newcommand{\tr}{\text{Tr}}
\newcommand{\angb}[1]{\langle #1 \rangle}
\newcommand{\sqrb}[2]{[#1]}
\newcommand{\ca}{\mathcal{A}}
\newcommand{\p}{\partial}
\newcommand{\bs}{\bar{\sigma}}
\newcommand{\ve}{\varepsilon}
\newcommand{\vp}{\varphi}
\newcommand{\dd}{\mathrm{d}}
\newcommand{\ha}{\hat{a}}
\newcommand{\hb}{\hat{b}}
\newcommand{\hp}{\hat{p}}
\newcommand{\he}{\hat{\eta}}
\newcommand{\rr}{\rangle}
\newcommand{\llg}{\langle}
\newcommand{\lips}[1]{\dd^3\text{LIPS}(#1)}
\newcommand{\ow}{\overline{\mathcal{W}}}
\newcommand{\bz}{\bar{z}}
\newcommand{\hx}{\hat{x}}
\newcommand{\hy}{\hat{y}}
\newcommand{\hz}{\hat{z}}
\newcommand{\scri}{\mathcal{I}}
\newcommand{\sw}{\mathcal{W}}
\newcommand{\bsw}{\overline{\mathcal{W}}}
\newcommand{\ang}[1]{\langle #1 \rangle}
\newcommand{\msdisk}{\overline{\mathcal{M}}_{0,n}(\mathbb{R})}

\newcommand*\widefbox[1]{\fbox{\hspace{1.2em}#1\hspace{1.2em}}}
\newcommand{\om}{\overline{m}}
\newcommand{\ua}{\bar{A}}
\newcommand{\ub}{\bar{B}}
\newcommand{\uc}{\bar{C}}
\newcommand{\ud}{\bar{D}}
\newcommand{\whp}{\widehat{p}}
\newcommand{\why}{\widehat{y}}
\newcommand{\mxx}[1]{\text{max}\left(#1\right)}
\newcommand{\ot}{\otimes}
\newcommand{\zbar}{\overline{z}}

\usepackage{color,soul}

\tikzmath{\qq=0.3;} 
\tikzmath{\change=0.15;}
\tikzmath{\ror=\qq*0.41421/1.41421;}

\allowdisplaybreaks

\author{Amit Suthar}
\emailAdd{amit.suthar@theory.tifr.res.in}
\affiliation{Department of Theoretical Physics, Tata Institute of Fundamental Research, Homi Bhabha Rd, Mumbai 400005, India}

\title{Curve integral formula for the Möbius strip}

\begin{document}
	\count\footins = 1000 
        
        \abstract{
        The scattering amplitudes for colored scalars can be calculated using the so-called curve integral formula, relying on simple combinatorics. It introduces a set of global Schwinger parameters for all Feynman diagrams that contribute to an amplitude. We extend this construction to non-orientable surfaces by making use of the quasi-cluster algebras defined for non-orientable surfaces. We embed the non-orientable surface in a doubled orientable surface, and project the appropriate features onto the non-orientable surface. The curve integral formula can also be thought of as the high-tension limit of an appropriate string amplitude. As a check of our construction, we take a superstring amplitude with the Möbius strip topology and take its field theory limit to obtain the same Feynman diagrams as in the corresponding curve integral. Our construction can be generalized to arbitrary higher genus non-orientable surfaces. To illustrate this, we list the possible curves and their dual momenta for a two-loop non-orientable surface, and construct the surface Symanzik polynomials using the surface generalization of spanning trees. 
        }

	\maketitle
	
	\newpage

\section{Introduction}
\label{section 1 intro}
Scattering amplitudes can be expressed as integrals over various moduli spaces. Factorization of the amplitudes follows from the factorization of the moduli spaces. For instance, scattering amplitudes in string theory are integrals of correlation functions of appropriate vertex operators, over the moduli space of Riemann surfaces with locations of vertex operators as marked points \cite{Polchinski:1998rq, Polchinski:1998rr, Green:1987sp, Green:1987mn}. In the field theory limit of string theory scattering amplitudes, the worldsheet moduli space turns into the Schwinger parameters for the Feynman diagrams, i.e., the moduli space of graphs. In more unusual examples, planar $\mathcal{N}=4$ super Yang-Mills amplitudes can been expressed as integrals over Grassmannians $Gr(k,n):$ the moduli space of $k$-planes in $n$-dimensions \cite{Arkani-Hamed:2010zjl, Arkani-Hamed:2012zlh}; the CHY formula expresses various amplitudes as integrals over the moduli space of the Riemann sphere with marked points \cite{Cachazo:2013hca, Cachazo:2013iea}, etc.

The field theory limit of string amplitudes has been studied extensively since the beginning of string theory. For instance, the field theory limit of type-I superstrings gives the $\mathcal{N}=4$ SYM\footnote{We refer to the gauge theory with sixteen supercharges as $\mathcal{N}=4$, even though in ten dimensions, there are not four spinors' worth of supercharges.} \cite{Green:1982sw}. In the high string tension limit, one recovers the Feynman diagrams in various regions of the worldsheet moduli space. String theory-inspired methods played a key role in understanding gauge theory and gravity amplitudes. Refer to \cite{Bern:1991an, Bern:1992ad, Bern:1993sx, DiVecchia:1996kf, DiVecchia:1996uq} for some of the earlier works. The field theory limit highlights the rich mathematical structure of the degenerations of moduli spaces of Riemann surfaces (for instance, see \cite{Tourkine:2013rda}). However, in all these studies, the integral over the worldsheet moduli space breaks up into a sum of different Feynman diagrams, and there is no sensible single 'moduli space of graphs' for a QFT amplitude. The \emph{curve integral formula}, discovered in \cite{Arkani-Hamed:2023CurveIntegral, Arkani-Hamed:2023Multiplicity}, constructs such a single moduli space of graphs, the \emph{global Schwinger space,} and a QFT amplitude is an integral over it. 

Even though the curve integral formula can be thought of as the infinite string tension limit of an appropriate worldsheet string integral, it can be constructed purely combinatorially, without resorting to the worldsheet or vertex operators \cite{Arkani-Hamed:2023CurveIntegral, Arkani-Hamed:2023Multiplicity}. It is inspired by the fact that the cluster algebra of surface type, constructed by the triangulations of a surface, can be used to parameterize the Teichmüller space of the surface \cite{fomin2001clusteralgebrasifoundations, fock2006modulispaceslocalsystems}. One can quotient out the Teichmüller space of the surface by the Mapping Class Group (MCG) and obtain the moduli space of the surface. Hence, drawing all possible chords or curves on a surface allows one to parameterize the moduli space of the surface and eventually write down an amplitude associated with the surface. See \cite{williams2013clusteralgebrasintroduction, fomin2024introductionclusteralgebraschapters} etc. for a mathematical introduction to the subject.  

The curve integral formula fits well into the quest of finding ‘alternative methods for scattering amplitudes'. For instance, in the \emph{positive geometry} program, one can find the volumes of certain convex polygons to compute sensible scattering amplitudes. The tree-level amplitude for colored scalars is calculated as the volume of (the dual of) the associahedron constructed in the kinematic space \cite{Arkani-Hamed:2017ABHY}. The moduli space of the tree-level $n$ point worldsheet (disk with $n$ insertions) is also an associahedron, and the aforementioned associahedron in the kinematic space can be obtained from the worldsheet associahedron by the mapping dictated by the CHY scattering equations \cite{Cachazo:2013hca, Cachazo:2013iea}. These investigations acted as a prelude to the curve integral formula \cite{Arkani-Hamed:2019_stringy_canonical_form, Arkani-Hamed:2019_binary_geometry, Arkani-Hamed:2019vag-AHST}. The subject has been termed \emph{surfaceology,} and connections with NLSM, and YM \cite{Arkani-Hamed:2023-HiddenZeros, Arkani-Hamed:2023-ScalarScaffolding, Arkani-Hamed:2024nhp, Laddha:2024qtn} amplitudes have been discovered. 

\paragraph{What is a curve integral formula?} Consider a QFT with matrix-valued field $\phi$, transforming as the adjoint of SU(N), with a $\tr \phi^3$ interaction. Feynman diagrams for this theory can be drawn with a double-line notation, and are called ribbon/fat graphs. The perturbation theory can be recast into a sum over topologies of the ribbon graphs. Let $S$ be a two-dimensional surface with at least one boundary component, and $n$ marked points on the boundaries, and $A_S$ be the contribution to the $n$-point amplitude with $S$ topology of ribbon graphs. The curve integral formula constructs the amplitude $A_S$ algorithmically using the curves and chords on the surface $S$, as a single integral. 

A tree-level trivalent Feynman diagram with $n$ legs is dual to a triangulation of an $n$-gon, or a disk with $n$ marked points. Each chord/curve connecting two marked points is dual to a propagator of the Feynman diagram, and has a well-defined momentum assigned to it. For an arbitrary order in perturbation theory, a trivalent Feynman diagram is dual to the triangulation of the surface $S$, and every curve $C$ has a definite momentum assignment $P_C$. Two curves that do not intersect each other are called mutually compatible and can appear together in a Feynman diagram. 

\begin{figure}[!t]
    \centering
\begin{tikzpicture}[scale=1.3,baseline={([yshift=-.5ex]current bounding box.center)}] 
\fill[teal!10!white] (0,0) -- (2.5,0) arc (0:90:2.5)  -- (0,0);
\fill[cyan!10!white] (0,0) -- (0,2.5) arc (90:135:2.5)  -- (0,0);
\fill[blue!10!white] (0,0) -- ({-2.5*sin(45)},{2.5*sin(45)}) arc (135:180:2.5)  -- (0,0);
\fill[purple!10!white] (0,0) -- (-2.5,0) arc (180:270:2.5)  -- (0,0);
\fill[orange!10!white] (0,0) -- (0,-2.5) arc (270:360:2.5)  -- (0,0);
\draw[gray, {Stealth[length=2mm, width=1.5mm]}-{Stealth[length=2mm, width=1.5mm]}] (-2.5,0) -- (2.5,0);
\draw[gray, {Stealth[length=2mm, width=1.5mm]}-{Stealth[length=2mm, width=1.5mm]}] (0,-2.5) -- (0,2.5);
\draw[purple!50!black,very thick,-{Stealth[length=3mm, width=2mm]}] (0,0) -- (0,2);
\draw[purple!50!black,very thick,-{Stealth[length=3mm, width=2mm]}] (0,0) -- (2,0);
\draw[purple!50!black,very thick,-{Stealth[length=3mm, width=2mm]}] (0,0) -- (0,-2);
\draw[purple!50!black,very thick,-{Stealth[length=3mm, width=2mm]}] (0,0) -- (-2,0);
\draw[purple!50!black,very thick,-{Stealth[length=3mm, width=2mm]}] (0,0) -- (-2,2);
\node[purple!50!black] at (0.3,2.1) {$g_{14}$};
\node[purple!50!black] at (2.1,-0.3) {$g_{13}$};
\node[purple!50!black] at (-2.1,-0.3) {$g_{25}$};
\node[purple!50!black] at (-0.3,-2.1) {$g_{35}$};
\node[purple!50!black] at (-2.2,2.2) {$g_{24}$};
\node[gray] at (2.7,0) {$t_1$};
\node[gray] at (0.3,2.5) {$t_2$};
\node at (1.2,1.2) {{\small $\begin{matrix}
    \alpha_{13}=t_1 \\
    \alpha_{14}=t_2 \\ 
    \qquad \alpha_{24},~ \alpha_{25},~\alpha_{35} = 0 
\end{matrix}$}};
\node at (1.3,-1.2) {{\small $\begin{matrix}
    \alpha_{13}=t_1 \\ 
    \alpha_{35} = -t_2\\
    \qquad \alpha_{14} = \alpha_{24} = \alpha_{25} = 0 
\end{matrix}$}};
\node at (-1.5,-1.2) {{\small $\begin{matrix}
    \alpha_{25} = -t_1 \\ 
    \alpha_{35} = -t_2 \\
    \alpha_{13} = \alpha_{14} = \alpha_{24} = 0
\end{matrix}$}};
\node at (-2.2,0.8) {{\small $\begin{matrix}
    \alpha_{24} =t_2 \\
    \alpha_{25}=-t_1-t_2 \\
    \alpha_{13} = \alpha_{14} = \alpha_{35} = 0
\end{matrix}$}};
\node at (-1,2) {{\small $\begin{matrix}
    \alpha_{13} = \alpha_{35} = 0\qquad\quad \\
    \alpha_{14}=t_1+t_2 \quad\\
    \alpha_{24}=-t_1 \\
    \qquad \alpha_{25} = 0 
\end{matrix}$}};
\node at (5,0) {$\begin{matrix}
    \vec{g}_{13} = (1,0) \\ \\
    \vec{g}_{14} = (0,1) \\ \\ 
    \vec{g}_{24} = (-1,1) \\ \\ 
    \vec{g}_{25} = (-1,0) \\ \\
    \vec{g}_{35} = (0,-1) 
\end{matrix}$};
\end{tikzpicture} 
    \caption{$g$-vectors for $S$ being disk with $5$ marked points. There are five vectors corresponding to five curves, and the resultant five cones are colored. The headlight functions evaluated in the cones are written.}
    \label{fig:headlight functions for tree level 5 points}
\end{figure}
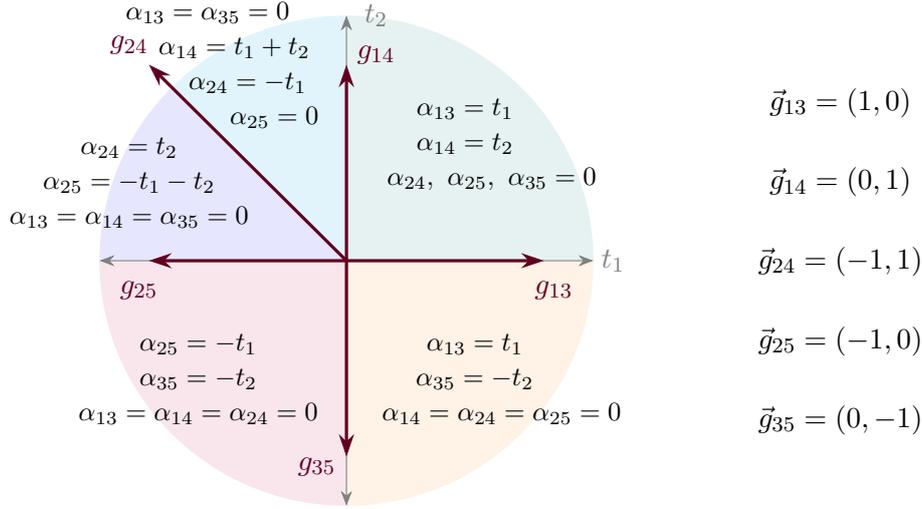

Let $E$ be the number of internal edges in the trivalent diagrams dual to triangulations of a surface $S$. The Schwinger parameters for a particular Feynman diagram lie in $\mathbb{R}^E_+$. We can have all the Feynman diagrams obtained from the triangulations of $S$ together in the $\mathbb{R}^E$ space as follows. Let us associate a particular vector $\vec{g}_C \in \mathbb{R}^E$ to every curve $C$ on the surface $S$ such that the $g$-vectors corresponding to a maximal set of mutually compatible curves make a cone\footnote{Given a set of vectors $\{\vec{g}_{i=1,\hdots,k}\}$, a (positive) cone is defined as the following region $\{\vec{x} = \sum_{i=1}^kc_i\vec{g}_i\ | \ c_i\geq 0\}$. } in $\mathbb{R}^E.$ No two cones have an $E$-dimensional intersection, and the union of all cones is the Euclidean space $\mathbb{R}^E$. Since a cone is made up of $g$-vectors of a set of $E$ mutually compatible curves, it is dual to a complete triangulation $T$ of the surface $S$, and corresponds to the Schwinger parameter space of the trivalent graph dual to $T$. For example, let $S$ be a disk with five marked points on its boundary. There are five curves on the $S$: $\{C_{13},~C_{14},~C_{24},~C_{25},~C_{35}\}$, and the dual trivalent graphs are five-point tree-level graphs with two propagators. The $g$-vectors in $\mathbb{R}^2$ are drawn in Figure \ref{fig:headlight functions for tree level 5 points}. $(t_1,t_2)$ are \emph{global} Schwinger parameters for the five-point tree diagrams. 

For every curve $C$, the \emph{headlight function} $\alpha_C$ is defined as the piecewise linear functions dual to the $g$-vectors: $\alpha_C(\vec{g}_{C'}) = \delta_{CC'}~.$ For instance, for the case of a disk with five insertions, the $g$-vectors given in Figure \ref{fig:headlight functions for tree level 5 points}, and $\alpha_C$ are as follows:
\begin{equation}
    \begin{aligned}
    \alpha_{13} &= \mxx{0,t_1} ~, \\
    \alpha_{14} &= -\mxx{0,t_1} + \mxx{0,t_1,t_1+t_2}~,\\
    \alpha_{24} &= -\mxx{0,t_1,t_1+t_2} + \mxx{0,t_1} + \mxx{0,t_2} ~,\\
    \alpha_{25} &= -t_1 - \mxx{0,t_2} + \mxx{0,t_1,t_1+t_2}~, \\
    \alpha_{35} &= -t_2 + \mxx{0,t_2}~.
    \end{aligned}
\end{equation}
Note that in a cone, only the headlight functions dual to the curves making up the cone are non-zero. Following this, we can write the following global Schwinger parameter integral, which evaluates to the sum of all diagrams:
\begin{align}
    A_S = \int_{\mathbb{R}^E} \dd^Et \int \dd^dl e^{-\sum_C \alpha_C P_C^2} ~.
\end{align}
For a generic surface $S$ representing loop processes, we have integration over loop momenta and a non-trivial Mapping Class Group, which needs to be quotiented out. All the ingredients for this integral, $P_C~, \alpha_C ~, \vec{g}_C$, can be determined algorithmically using combinatorics of curves on the surface $S$, and is called the \emph{curve integral formula.} We review the algorithms to determine $P_C~, \alpha_C ~, \vec{g}_C$ very briefly in Appendix \ref{appendix A review}. We can integrate out the loop momenta to get a nice form, which we revisit in Section \ref{section 4 surface symanzik}.

\paragraph{Color and orientation:}
The curve integral formula writes down scattering amplitudes for $\tr \phi^3$ scalars, which are equivalent to the adjoint valued $SU(N)$ scalars. The matrix-valued field $\phi$ carries a fundamental and an anti-fundamental index. We can draw the Feynman diagrams as ribbon graphs to illustrate the color structure. The $SU(N)$ fundamental indices dictate that the resulting ribbon graphs be orientable surfaces. Equivalently, only orientable string worldsheets contribute to amplitudes of $SU(N)$ Chan-Paton factors. However, in the case of $SO(N)$ Chan Paton factors, the state transforming under the adjoint representation carries two anti-symmetric indices, and in the case of $Sp(N)$, two symmetric indices. Since the two indices are on equal footing, one can contract them freely, and we also have non-orientable surfaces. From a purely QFT perspective, if we have $\tr \phi^3$ interactions with $\phi$ as a symmetric matrix, we have non-orientable surfaces contributing. For an antisymmetric $\phi$, the trivalent term $\phi_{ij}\phi_{jk}\phi_{ki}$ vanish, so we can focus on the symmetric matrix $\phi$. 

The amplitude for particles that allow for non-orientable surfaces is a sum over orientable and non-orientable ones. Inclusion of non-orientable surfaces results in single-trace contributions at subleading orders in $1/N$. \footnote{From the gauge theory amplitude perspective, consider the one-loop box diagram with the color structure $f^{ea_1b}f^{ba_2c}f^{ca_3d}f^{da_4e}~.$ For $SU(N)$, one can decompose it as follows:
\begin{align}
    f^{ea_1b}f^{ba_2c}f^{ca_3d}f^{da_4e} &= 2N\,\tr(T^{a_1}T^{a_2}T^{a_3}T^{a_4}) + 2\tr(T^{a_1}T^{a_2})\tr(T^{a_3}T^{a_4}) \nonumber \\
    &\qquad + 2\tr(T^{a_1}T^{a_3})\tr(T^{a_2}T^{a_4})+ 2\tr(T^{a_1}T^{a_4})\tr(T^{a_2}T^{a_3})
\end{align}
For $SO(N)$, we have:
\begin{align}
    f^{ea_1b}f^{ba_2c}f^{ca_3d}f^{da_4e} &= (N-4)\tr(T^{a_1}T^{a_2}T^{a_3}T^{a_4}) - \tr(T^{a_1}T^{a_3}T^{a_4}T^{a_2}) \nonumber \\
    & \qquad - \tr(T^{a_1}T^{a_4}T^{a_2}T^{a_3})+ \tr(T^{a_1}T^{a_2})\tr(T^{a_3}T^{a_4}) \nonumber \\
    &\qquad + \tr(T^{a_1}T^{a_3})\tr(T^{a_2}T^{a_4})+ \tr(T^{a_1}T^{a_4})\tr(T^{a_2}T^{a_3})
\end{align}
For $SO(N)$, we have single trace contributions at subleading order in $1/N$, which are absent for $SU(N).$ Look at \cite{Naculich:2024fiy, Bourjaily:2024jbt, Bourjaily:2025hvq} for details of color decomposition.} For instance, with a coupling constant $g$, the four-point one-loop amplitude can be expanded as follows:
\begin{align}
    A_4 &= \!\!\sum_{\sigma \in \substack{\text{color} \\ \text{orderings}}}\!\!\!\!\left(g^2\,A^{\text{Tree}}_4[\sigma] + Ng^4\,A_4^{\text{Annulus}}[\sigma] + g^4\,A_4^{\text{Möbius}}[\sigma] + \mathcal{O}(g^6)\right)\tr\big[T_{\sigma(1)}T_{\sigma(2)}T_{\sigma(3)}T_{\sigma(4)}\big] \nonumber \\
    &\hspace{3cm} + g^4\left(\text{multi-trace annulus} \right) + \mathcal{O}(g^6)~.
\end{align}
The Möbius strip has a single boundary component and does not contribute to the amplitudes with multi-trace color structures. 

\paragraph{Curve integral for non-orientable surfaces:} The usual cluster algebras and the curve integral formula are defined for orientable surfaces. The cluster algebras were generalized to non-orientable surfaces in \cite{dupont2015quasi}. Triangulations of Möbius strips have been studied in \cite{bazier2023number-of-triangulations}. We denote the combinatorial polytope, obtained from the triangulation of Mobius strip with $n$ marked points on its boundary as $M_n$. It is the analogue of the associahedron $A_n$ for disk with $n$ marked points. One is tempted to ask if these quasi-cluster algebras and $M_n$ polytope can be used to construct sensible amplitudes. From a stringy perspective, it is expected that the Feynman diagrams appearing in the field theory limit form the same combinatorial structure as the quasi-cluster algebras. We construct the curve integral formula corresponding to the non-orientable surfaces in this article. As a check, we take the field theory limit of the Möbius strip string amplitude and see that we obtain the same diagrams as in the curve integral formula. 

A Möbius strip can be represented either by a rectangular strip with two opposite ends identified with a twist, or equivalently as a disk with a cross-cap in the middle. We depict a Möbius strip with two marked points on its boundary in the two representations below:
\begin{align}
    \begin{tikzpicture}[scale=0.45,baseline={([yshift=-.5ex]current bounding box.center)}]
    \fill[cyan!6!white] (0,0) circle (2);
    \draw (0,0) circle (2);
    \node at (0,0) {$\ot$};
    \node at (0,2) {$\times$};
    \node at (0,-2) {$\times$};
    \node at (0.4,2.6) {1};
    \node at (-0.4,-2.6) {2};
\end{tikzpicture} \quad \cong \quad  \begin{tikzpicture}[scale=0.6,baseline={([yshift=-1ex]current bounding box.center)}]
    \fill[cyan!6!white] (0,0) -- (0,2) -- (3,2) -- (3,0) -- (0,0);
    \draw[-{Stealth[length=2mm, width=1.5mm]}] (0,0) -- (0,1.2);
    \draw (0,0.95) -- (0,2) -- (3,2) -- (3,0.95);
    \draw [-{Stealth[length=2mm, width=1.5mm]}] (3,2) -- (3,0.8);
    \draw (3,0) -- (3,1.05);
    \draw (0,0) -- (3,0);
    \node at (1,2) {$\times$};
    \node at (2,2) {$\times$};
    \node at (1,2.6) {$1$};
    \node at (2,2.6) {$2$};
\end{tikzpicture} ~.
\end{align}
In the rectangular strip case, the single boundary component appears to be split into an upper and a lower part. 

Orientation of the surface is crucial in order to define the necessary ingredients for the curve integral formula. We deal with the non-orientable surfaces by doubling them up into an orientable surface and projecting out the key data from there. For instance, as depicted in \eqref{abstract doubling of Möbius strip}, the Mobius strip with $n$ marked points can be doubled into an annulus with $n$ marked points on each of the two boundaries.

The curves/chords on the Mobius strip are given by $C_{ij}~, C^{\ot}_{ij}~, C^{\ot}~,$ see Figure \ref{fig:curves on Mobius2}. The curves with superscript ${}^{\ot}$ denote the fact that they passed through the cross-cap. We classify curves on the surface only upto homotopy. So a curve on the Möbius strip passing through the cross-cap twice is equivalent to a curve not passing through the cross-cap. The curve $C^{\ot}$ is a closed curve, and appears only in the tadpoles. The momentum assignment to the curves carrying loop momentum $C_{ij}^{\ot}$ is as follows:
\begin{align}
    P_{ij}^{\ot} = l+ (p_1+ \hdots + p_{i-1}) + (p_1+ \hdots + p_{j-1})~. 
\end{align}
The momentum associated to $C_{ij}$ is $P_{ij}=(p_i+\hdots+p_{j-1})~.$ The kinematic variables are $X_{ik}$ and $X_{ik}^{\ot}$, where $X_I = P_I^2~.$ 

The $g$-vectors and the headlight functions can be calculated using the projection from the doubled orientable surface. Using this, we demonstrate that appropriate projections of the curve integral for orientable surfaces contains the one for non-orientable surfaces as well. The construction can be followed in the same manner for higher genus surfaces as well. We work out the case of a two-loop non-orientable surface. 

The field theory limit of appropriate string amplitudes is expected to yield the curve integral formula. A generic loop curve integral formula, being a QFT amplitude, will have bubbles and tadpoles; however, sensible superstring loop amplitudes for massless external states do not have such bubbles and tadpoles. One can start with a bosonic loop open string amplitudes, though they will have some unwanted features. We analyze the well-understood type-I open superstring amplitude with massless external states, and consider its field theory limit. Our objective is to see the emergence of the box diagrams constructed from the quasi-cluster algebras. For a four-particle amplitude, there are eight box diagrams, and we check that different regions of the moduli space integral give the expected eight box diagrams from the quasi-cluster algebras and the curve integral. We identify the regions in the moduli space that would have led to triangles and bubbles, though the integral is zero in those regions. 

\paragraph{Outline of the article:}
In Section \ref{section 2 review of mobius}, we review the triangulations of the Mobius strip. In Section \ref{section 3 curve integral for mobius}, we find the associated momenta to the curves on the Mobius strip, construct the $g$-vectors and headlight functions, and write the curve integral formula. In Section \ref{section 4 surface symanzik}, we construct the surface Symanzik polynomials for the Mobius strip. In Section \ref{section 5 two loop}, we analyze the two-loop non-orientable surface, write down all its curves, their associated momenta, and the surface Symanzik polynomials for the surface. In Section \ref{section 6 string amplitudes}, we study the field theory limit of the Mobius strip superstring amplitude, and check that we obtain the same box diagrams expected from the curve integral. 

\section{Review of triangulations of the Möbius strip}
\label{section 2 review of mobius}
Traditionally, cluster algebras of surface type have been defined only for orientable surfaces. The natural reason is that we need a handedness on the surface to construct a quiver for a triangulation of the surface. Keeping the curve integral in mind, the turning direction (left/right) of a curve is not defined on a non-orientable surface. The cluster algebras of the surface types were generalized to quasi-cluster algebras for non-orientable surfaces in \cite{dupont2015quasi}. We describe the chords that are used to triangulate a Möbius strip below. 

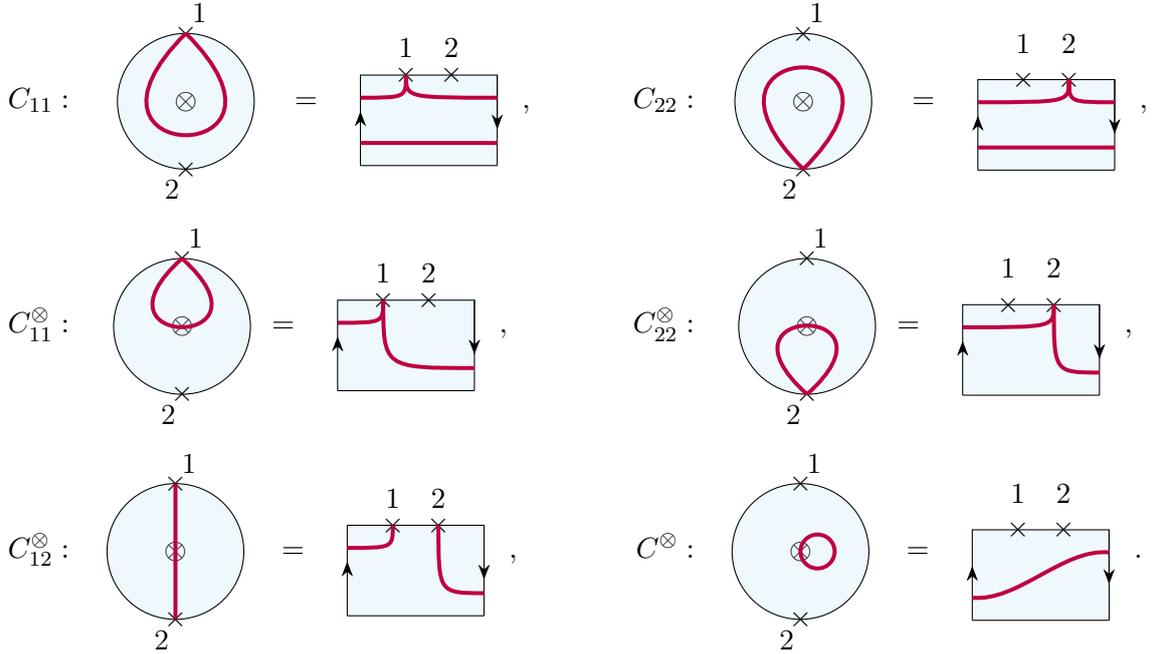
\begin{figure}[H]
    \centering
\begin{align*}
    C_{11} &: \hspace{-0.4cm}\begin{tikzpicture}[scale=0.45,baseline={([yshift=-.5ex]current bounding box.center)}]
    \fill[cyan!6!white] (0,0) circle (2);
    \draw (0,0) circle (2);
    \node at (0,0) {$\ot$};
    \node at (0,2) {$\times$};
    \node at (0,-2) {$\times$};
    \node at (0.4,2.6) {1};
    \node at (-0.4,-2.6) {2};
    \draw[line width=1.5, purple] (0,2) .. controls (-4,-2) and (4,-2) .. (0,2);
\end{tikzpicture} \hspace{-0.5cm} =\quad  \begin{tikzpicture}[scale=0.6,baseline={([yshift=-.8ex]current bounding box.center)}]
    \fill[cyan!6!white] (0,0) -- (0,2) -- (3,2) -- (3,0) -- (0,0);
    \draw[-{Stealth[length=2mm, width=1.5mm]}] (0,0) -- (0,1.2);
    \draw (0,0.95) -- (0,2) -- (3,2) -- (3,0.95);
    \draw [-{Stealth[length=2mm, width=1.5mm]}] (3,2) -- (3,0.8);
    \draw (3,0) -- (3,1.05);
    \draw (0,0) -- (3,0);
    \node at (1,2) {$\times$};
    \node at (2,2) {$\times$};
    \draw[line width=1.5, purple] (0,1.5) .. controls (1,1.5) .. (1,2) .. controls (1,1.5) .. (3,1.5);
    \draw[line width=1.5, purple] (0,0.5) -- (3,0.5);
    \node at (1,2.6) {$1$};
    \node at (2,2.6) {$2$};
\end{tikzpicture} ~~, & \hspace{1cm} 
    C_{22} &: \hspace{-0.5cm} \begin{tikzpicture}[scale=0.45,baseline={([yshift=-.5ex]current bounding box.center)}]
    \fill[cyan!6!white] (0,0) circle (2);
    \draw (0,0) circle (2);
    \node at (0,0) {$\ot$};
    \node at (0,2) {$\times$};
    \node at (0,-2) {$\times$};
    \node at (0.4,2.6) {1};
    \node at (-0.4,-2.6) {2};
    \draw[line width=1.5, purple] (0,-2) .. controls (-4,2) and (4,2) .. (0,-2);
\end{tikzpicture} \hspace{-0.5cm}=\quad  \begin{tikzpicture}[scale=0.6,baseline={([yshift=-.8ex]current bounding box.center)}]
    \fill[cyan!6!white] (0,0) -- (0,2) -- (3,2) -- (3,0) -- (0,0);
    \draw[-{Stealth[length=2mm, width=1.5mm]}] (0,0) -- (0,1.2);
    \draw (0,0.95) -- (0,2) -- (3,2) -- (3,0.95);
    \draw [-{Stealth[length=2mm, width=1.5mm]}] (3,2) -- (3,0.8);
    \draw (3,0) -- (3,1.05);
    \draw (0,0) -- (3,0);
    \node at (1,2) {$\times$};
    \node at (2,2) {$\times$};
    \draw[line width=1.5, purple] (0,1.5) .. controls (2,1.5) .. (2,2) .. controls (2,1.5) .. (3,1.5);
    \draw[line width=1.5, purple] (0,0.5) -- (3,0.5);
    \node at (1,2.8) {$1$};
    \node at (2,2.8) {$2$};
\end{tikzpicture} ~~, \\
    C_{11}^{\ot} &: \begin{tikzpicture}[scale=0.45,baseline={([yshift=-.5ex]current bounding box.center)}]
    \fill[cyan!6!white] (0,0) circle (2);
    \draw (0,0) circle (2);
    \node at (0,0) {$\ot$};
    \node at (0,2) {$\times$};
    \node at (0,-2) {$\times$};
    \node at (0.4,2.6) {1};
    \node at (-0.4,-2.6) {2};
    \draw[line width=1.5, purple] (0,2) .. controls (-3,-0.7) and (3,-0.7) .. (0,2);
\end{tikzpicture} \hspace{-0.3cm}=\quad  \begin{tikzpicture}[scale=0.6,baseline={([yshift=-.8ex]current bounding box.center)}]
    \fill[cyan!6!white] (0,0) -- (0,2) -- (3,2) -- (3,0) -- (0,0);
    \draw[-{Stealth[length=2mm, width=1.5mm]}] (0,0) -- (0,1.2);
    \draw (0,0.95) -- (0,2) -- (3,2) -- (3,0.95);
    \draw [-{Stealth[length=2mm, width=1.5mm]}] (3,2) -- (3,0.8);
    \draw (3,0) -- (3,1.05);
    \draw (0,0) -- (3,0);
    \node at (1,2) {$\times$};
    \node at (2,2) {$\times$};
    \draw[line width=1.5, purple] (1,2) .. controls (1,1.5) .. (0,1.5);
    \draw[line width=1.5, purple] (1,2) .. controls (1,0.5) .. (3,0.5);
    \node at (1,2.6) {$1$};
    \node at (2,2.6) {$2$};
\end{tikzpicture} ~~, &\hspace{1cm} 
    C_{22}^{\ot} &:  \begin{tikzpicture}[scale=0.45,baseline={([yshift=-.5ex]current bounding box.center)}]
    \fill[cyan!6!white] (0,0) circle (2);
    \draw (0,0) circle (2);
    \node at (0,0) {$\ot$};
    \node at (0,2) {$\times$};
    \node at (0,-2) {$\times$};
    \node at (0.4,2.6) {1};
    \node at (-0.4,-2.6) {2};
    \draw[line width=1.5, purple] (0,-2) .. controls (-3,0.7) and (3,0.7) .. (0,-2);
\end{tikzpicture} \hspace{-0.3cm}=\quad  \begin{tikzpicture}[scale=0.6,baseline={([yshift=-.8ex]current bounding box.center)}]
    \fill[cyan!6!white] (0,0) -- (0,2) -- (3,2) -- (3,0) -- (0,0);
    \draw[-{Stealth[length=2mm, width=1.5mm]}] (0,0) -- (0,1.2);
    \draw (0,0.95) -- (0,2) -- (3,2) -- (3,0.95);
    \draw [-{Stealth[length=2mm, width=1.5mm]}] (3,2) -- (3,0.8);
    \draw (3,0) -- (3,1.05);
    \draw (0,0) -- (3,0);
    \node at (1,2) {$\times$};
    \node at (2,2) {$\times$};
    \draw[line width=1.5, purple] (2,2) .. controls (2,1.5) .. (0,1.5);
    \draw[line width=1.5, purple] (2,2) .. controls (2,0.5) .. (3,0.5);
    \node at (1,2.8) {$1$};
    \node at (2,2.8) {$2$};
\end{tikzpicture} ~~, \\
C_{12}^{\ot} &: \quad \begin{tikzpicture}[scale=0.45,baseline={([yshift=-.5ex]current bounding box.center)}]
    \fill[cyan!6!white] (0,0) circle (2);
    \draw (0,0) circle (2);
    \node at (0,0) {$\ot$};
    \node at (0,2) {$\times$};
    \node at (0,-2) {$\times$};
    \node at (0.4,2.6) {1};
    \node at (-0.4,-2.6) {2};
    \draw[line width=1.5, purple] (0,2) -- (0,-2);
\end{tikzpicture} \quad =\quad  \begin{tikzpicture}[scale=0.6,baseline={([yshift=-.8ex]current bounding box.center)}]
    \fill[cyan!6!white] (0,0) -- (0,2) -- (3,2) -- (3,0) -- (0,0);
    \draw[-{Stealth[length=2mm, width=1.5mm]}] (0,0) -- (0,1.2);
    \draw (0,0.95) -- (0,2) -- (3,2) -- (3,0.95);
    \draw [-{Stealth[length=2mm, width=1.5mm]}] (3,2) -- (3,0.8);
    \draw (3,0) -- (3,1.05);
    \draw (0,0) -- (3,0);
    \node at (1,2) {$\times$};
    \node at (2,2) {$\times$};
    \draw[line width=1.5, purple] (0,1.5) .. controls (1,1.5) .. (1,2);
    \draw[line width=1.5, purple] (2,2) .. controls (2,0.5) .. (3,0.5);
    \node at (1,2.6) {$1$};
    \node at (2,2.6) {$2$};
\end{tikzpicture} ~~, &\hspace{1cm} 
    C^{\ot} &:  \quad \begin{tikzpicture}[scale=0.45,baseline={([yshift=-.5ex]current bounding box.center)}]
    \fill[cyan!6!white] (0,0) circle (2);
    \draw (0,0) circle (2);
    \node at (0,0) {$\ot$};
    \node at (0,2) {$\times$};
    \node at (0,-2) {$\times$};
    \node at (0.4,2.6) {1};
    \node at (-0.4,-2.6) {2};
    \draw[line width=1.5, purple] (0.5,0) circle (0.5);
\end{tikzpicture} \quad =\quad  \begin{tikzpicture}[scale=0.6,baseline={([yshift=-.8ex]current bounding box.center)}]
    \fill[cyan!6!white] (0,0) -- (0,2) -- (3,2) -- (3,0) -- (0,0);
    \draw[-{Stealth[length=2mm, width=1.5mm]}] (0,0) -- (0,1.2);
    \draw (0,0.95) -- (0,2) -- (3,2) -- (3,0.95);
    \draw [-{Stealth[length=2mm, width=1.5mm]}] (3,2) -- (3,0.8);
    \draw (3,0) -- (3,1.05);
    \draw (0,0) -- (3,0);
    \node at (1,2) {$\times$};
    \node at (2,2) {$\times$};
    \draw[line width=1.5, purple] (0,0.5) .. controls (1,0.4) and (2,1.6) .. (3,1.5);
    \node at (1,2.8) {$1$};
    \node at (2,2.8) {$2$};
\end{tikzpicture} ~~.
\end{align*}
    \caption{List of all the curves on the Möbius strip with two marked points.}
    \label{fig:curves on Mobius2}
\end{figure}

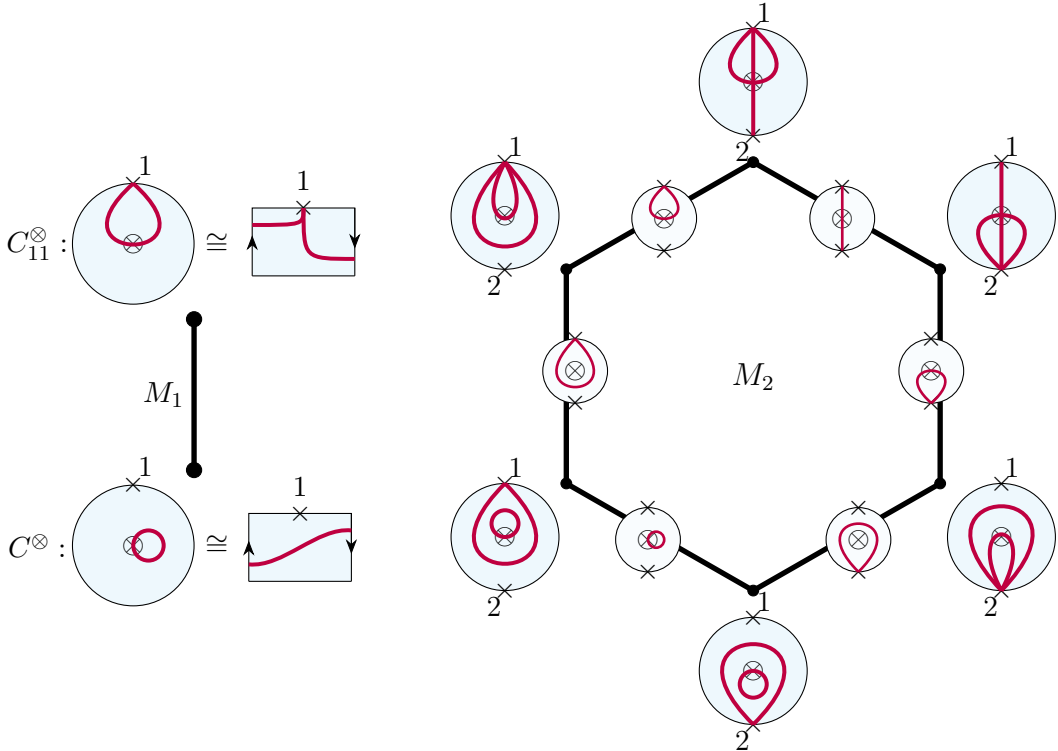
\begin{figure}[H]
    \centering
\begin{tikzpicture}[scale=1,baseline={([yshift=-.5ex]current bounding box.center)}]
	\begin{scope}[scale=0.4,xshift=-2cm, yshift=5cm]
		\node at (-3,0) {$C^{\ot}_{11}:~$};
        \fill[cyan!6!white] (0,0) circle (2);
	    \draw (0,0) circle (2);
   	 \node at (0,0) {$\ot$};
    	\node at (0,2) {$\times$};
    	\node at (0.4,2.6) {1};
    \draw[line width=1.5, purple] (0,2) .. controls (-3,-0.7) and (3,-0.7) .. (0,2);
    \node at (2.75,0) {$\cong$};
    	\end{scope}
	\begin{scope}[scale=0.45,xshift=1.7cm, yshift=3.5cm]
	\fill[cyan!6!white] (0,0) -- (0,2) -- (3,2) -- (3,0) -- (0,0);
    \draw[-{Stealth[length=2mm, width=1.5mm]}] (0,0) -- (0,1.2);
    \draw (0,0.95) -- (0,2) -- (3,2) -- (3,0.95);
    \draw [-{Stealth[length=2mm, width=1.5mm]}] (3,2) -- (3,0.8);
    \draw (3,0) -- (3,1.05);
    \draw (0,0) -- (3,0);
    \node at (1.5,2) {$\times$};
    \draw[line width=1.5, purple] (1.5,2) .. controls (1.5,1.5) .. (0,1.5);
    \draw[line width=1.5, purple] (1.5,2) .. controls (1.5,0.5) .. (3,0.5);
    \node at (1.5,2.8) {$1$};
    \end{scope}
    \draw[line width=2,black] (0,-1) -- (0,1);
        \draw[fill=black] (0,-1) circle (0.1);
        \draw[fill=black] (0,1) circle (0.1);
        \node at (-0.4,0) {$M_1$};
    \begin{scope}[scale=0.4,xshift=-2cm, yshift=-5cm]
        \node at (-3,0) {$C^{\ot}:~$};
		\fill[cyan!6!white] (0,0) circle (2);
	    \draw (0,0) circle (2);
   	 \node at (0,0) {$\ot$};
    	\node at (0,2) {$\times$};
    	\node at (0.4,2.6) {1};
	\node at (2.75,0) {$\cong$};
    	\draw[line width=1.5, purple] (0.5,0) circle (0.5);
	\end{scope}
	\begin{scope}[scale=0.45,xshift=1.6cm, yshift=-5.5cm]
	\fill[cyan!6!white] (0,0) -- (0,2) -- (3,2) -- (3,0) -- (0,0);
    \draw[-{Stealth[length=2mm, width=1.5mm]}] (0,0) -- (0,1.2);
    \draw (0,0.95) -- (0,2) -- (3,2) -- (3,0.95);
    \draw [-{Stealth[length=2mm, width=1.5mm]}] (3,2) -- (3,0.8);
    \draw (3,0) -- (3,1.05);
    \draw (0,0) -- (3,0);
    \node at (1.5,2) {$\times$};
    \draw[line width=1.5, purple] (0,0.5) .. controls (1,0.4) and (2,1.6) .. (3,1.5);
    \node at (1.5,2.8) {$1$};
    \end{scope}
\end{tikzpicture}
\hspace{0.3cm}
\begin{tikzpicture}[scale=0.71,baseline={([yshift=-.5ex]current bounding box.center)}]
        \draw[line width=2pt, black] (-3.464,0) -- (0,-2) -- (3.464,0) -- (3.464,4) -- (0,6) -- (-3.464,4) -- (-3.464,0);
        \draw[fill=black] (-3.464,0) circle (0.1);
        \draw[fill=black] (0,-2) circle (0.1);
        \draw[fill=black]  (3.464,0) circle (0.1);
        \draw[fill=black] (3.464,4) circle (0.1);
        \draw[fill=black] (0,6) circle (0.1);
        \draw[fill=black] (-3.464,4) circle (0.1);
        \node at (0,2) {\large $M_2$};
        \begin{scope}[scale=0.5,xshift=-9.2cm, yshift=-2cm]
	        \fill[cyan!6!white] (0,0) circle (2);
    		\draw (0,0) circle (2);
    		\node at (0,0) {$\ot$};
    		\node at (0,2) {$\times$};
    		\node at (0.4,2.6) {1};
    		\node at (0,-2) {$\times$};
    		\node at (-0.4,-2.6) {2};
		\draw[line width=1.5, purple] (0,2) .. controls (-4,-2) and (4,-2) .. (0,2);
    		\draw[line width=1.5, purple] (0,0.5) circle (0.5);
    	\end{scope} 
        \begin{scope}[scale=0.3,xshift=-11cm, yshift=7cm]
	        \fill[cyan!2!white] (0,0) circle (2);
    		\draw (0,0) circle (2);
    		\node at (0,0) {$\ot$};
    		\node at (0,2) {$\times$};
    		\node at (0,-2) {$\times$};
		\draw[line width=1, purple] (0,2) .. controls (-4,-2) and (4,-2) .. (0,2);
        \end{scope}
        \begin{scope}[scale=0.5,xshift=-9.2cm, yshift=10cm]
	        \fill[cyan!6!white] (0,0) circle (2);
    		\draw (0,0) circle (2);
    		\node at (0,0) {$\ot$};
    		\node at (0,2) {$\times$};
    		\node at (0.4,2.6) {1};
    		\node at (0,-2) {$\times$};
    		\node at (-0.4,-2.6) {2};	
    		\draw[line width=1.5, purple] (0,2) .. controls (-1.5,-0.8) and (1.5,-0.8) .. (0,2);
    		\draw[line width=1.5, purple] (0,2) .. controls (-4,-2.2) and (4,-2.2) .. (0,2);
        \end{scope}
        \begin{scope}[scale=0.3,xshift=-5.5cm, yshift=16.5cm]
	        \fill[cyan!2!white] (0,0) circle (2);
    		\draw (0,0) circle (2);
    		\node at (0,0) {$\ot$};
    		\node at (0,2) {$\times$};
    		\node at (0,-2) {$\times$};
        		\draw[line width=1, purple] (0,2) .. controls (-3,-0.7) and (3,-0.7) .. (0,2);
        \end{scope}
        \begin{scope}[scale=0.5,xshift=0cm, yshift=15cm]
	        \fill[cyan!6!white] (0,0) circle (2);
    		\draw (0,0) circle (2);
    		\node at (0,0) {$\ot$};
    		\node at (0,2) {$\times$};
    		\node at (0.4,2.6) {1};
    		\node at (0,-2) {$\times$};
    		\node at (-0.4,-2.6) {2};
        		\draw[line width=1.5, purple] (0,2) .. controls (-3,-0.7) and (3,-0.7) .. (0,2);
		\draw[line width=1.5, purple] (0,2) -- (0,-2);
        \end{scope}
        \begin{scope}[scale=0.3,xshift=5.5cm, yshift=16.5cm]
	        \fill[cyan!2!white] (0,0) circle (2);
    		\draw (0,0) circle (2);
    		\node at (0,0) {$\ot$};
    		\node at (0,2) {$\times$};
    		\node at (0,-2) {$\times$};
		\draw[line width=1, purple] (0,2) -- (0,-2);
        \end{scope}
        \begin{scope}[scale=0.5,xshift=9.2cm, yshift=10cm]
	        \fill[cyan!6!white] (0,0) circle (2);
    		\draw (0,0) circle (2);
    		\node at (0,0) {$\ot$};
    		\node at (0,2) {$\times$};
    		\node at (0.4,2.6) {1};
    		\node at (0,-2) {$\times$};
    		\node at (-0.4,-2.6) {2};
		\draw[line width=1.5, purple] (0,2) -- (0,-2);
    		\draw[line width=1.5, purple] (0,-2) .. controls (-3,0.7) and (3,0.7) .. (0,-2);
        \end{scope}
        
        \begin{scope}[scale=0.3,xshift=11cm, yshift=7cm]
	        \fill[cyan!2!white] (0,0) circle (2);
    		\draw (0,0) circle (2);
    		\node at (0,0) {$\ot$};
    		\node at (0,2) {$\times$};
    		\node at (0,-2) {$\times$};
		\draw[line width=1, purple] (0,-2) .. controls (-3,0.7) and (3,0.7) .. (0,-2);
        \end{scope}
        \begin{scope}[scale=0.5,xshift=9.2cm, yshift=-2cm]
	        \fill[cyan!6!white] (0,0) circle (2);
    		\draw (0,0) circle (2);
    		\node at (0,0) {$\ot$};
    		\node at (0,2) {$\times$};
    		\node at (0.4,2.6) {1};
    		\node at (0,-2) {$\times$};
    		\node at (-0.4,-2.6) {2};
    		\draw[line width=1.5, purple] (0,-2) .. controls (-1.5,0.8) and (1.5,0.8) .. (0,-2);
    		\draw[line width=1.5, purple] (0,-2) .. controls (-4,2.2) and (4,2.2) .. (0,-2);
        \end{scope}
        \begin{scope}[scale=0.3,xshift=6.5cm, yshift=-3.5cm]
	        \fill[cyan!2!white] (0,0) circle (2);
    		\draw (0,0) circle (2);
    		\node at (0,0) {$\ot$};
    		\node at (0,2) {$\times$};
    		\node at (0,-2) {$\times$};
		\draw[line width=1, purple] (0,-2) .. controls (-4,2) and (4,2) .. (0,-2);
        \end{scope}
        \begin{scope}[scale=0.5,xshift=0cm, yshift=-7cm]
	        \fill[cyan!6!white] (0,0) circle (2);
    		\draw (0,0) circle (2);
    		\node at (0,0) {$\ot$};
    		\node at (0,2) {$\times$};
    		\node at (0.4,2.6) {1};
    		\node at (0,-2) {$\times$};
    		\node at (-0.4,-2.6) {2};
		\draw[line width=1.5, purple] (0,-0.5) circle (0.5);
		\draw[line width=1.5, purple] (0,-2) .. controls (-4,2) and (4,2) .. (0,-2);
        \end{scope}
        \begin{scope}[scale=0.3,xshift=-6.5cm, yshift=-3.5cm]
	        \fill[cyan!2!white] (0,0) circle (2);
    		\draw (0,0) circle (2);
    		\node at (0,0) {$\ot$};
    		\node at (0,2) {$\times$};
    		\node at (0,-2) {$\times$};
		\draw[line width=1, purple] (0.5,0) circle (0.5);
        \end{scope}
    \end{tikzpicture}
        \caption{Combinatorial polytopes $M_{1,2}$ obtained from (quasi-)mutations of the Möbius strip with one and two marked points.}
    \label{fig: Mn polytope for 1 and 2}
\end{figure}

\paragraph{What are the possible chords?} Note that the Möbius strip has one boundary component, and the triangulation and moduli space are well defined for a Möbius strip with at least one marked point on its boundary component. As explained in \cite{dupont2015quasi, bazier2023number-of-triangulations, bazierquasi-overview}, there are three kinds of chords/curves used for triangulations of a Möbius strip with marked points:
\begin{itemize}
    \item $C_{ik}^{\ot}~:~~$ The chord $C_{ik}^{\ot}$ joins the marked points $i$ and $k$ passing through the cross-cap. (Hence the superscript ${}^{\ot}$ in the notation.) While drawing the Möbius strip as a rectangular strip, such curves pass once through the identified vertical lines. Intuitively, $C_{ij}^{\ot}=C_{ji}^{\ot}~.$ With $n$ marked points, there are $n(n+1)/2$ such chords. 
    \begin{align}
        \text{For} ~ i,j\in[1,n] ~; \qquad  C_{ij}^{\ot} ~:\quad     \begin{tikzpicture}[scale=0.45,baseline={([yshift=-.5ex]current bounding box.center)}]
    \fill[cyan!6!white] (0,0) circle (2);
    \draw (0,0) circle (2);
    \node at (0,0) {$\ot$};
    \node at (0,2) {$\times$};
    \node at (0,2.9) {1};
    \node at ({2* cos(45)},{2* sin(45)}) {$\times$};
    \node at ({2* cos(45)+0.5},{2* sin(45)+0.5}) {2};
    \node at ({-2* cos(45)},{2* sin(45)}) {$\times$};
    \node at ({-2* cos(45)-0.5},{2* sin(45)+0.5}) {n};
    \node at ({2* cos(45)},{-2* sin(45)}) {$\times$};
    \node at ({2* cos(45)+0.5},{-2* sin(45)-0.5}) {i};
    \node at ({-2* cos(45)},{-2* sin(45)}) {$\times$};
    \node at ({-2* cos(45)-0.5},{-2* sin(45)-0.5}) {j};
    \node at (0,-2.4) {$\hdots$};
    \node at (2.4,0) {$\vdots$};
    \node at (-2.4,0) {$\vdots$};
    \draw[line width=1.5, purple] ({2* cos(45)},{-2* sin(45)}) .. controls (1,0.5) and (-1,0.5) .. ({-2* cos(45)},{-2* sin(45)});
\end{tikzpicture} \quad \cong \quad \begin{tikzpicture}[scale=0.65,baseline={([yshift=-.8ex]current bounding box.center)}]
    \fill[cyan!6!white] (0,0) -- (0,2) -- (3,2) -- (3,0) -- (0,0);
    \draw[-{Stealth[length=2mm, width=1.5mm]}] (0,0) -- (0,1.2);
    \draw (0,0.95) -- (0,2) -- (3,2) -- (3,0.95);
    \draw [-{Stealth[length=2mm, width=1.5mm]}] (3,2) -- (3,0.8);
    \draw (3,0) -- (3,1.05);
    \draw (0,0) -- (3,0);
    \node at (1,2) {$\times$};
    \node at (2,2) {$\times$};
    \node at (0.8,2.5) {$i$};
    \node at (2.2,2.5) {$j$};
    \node at (1.5,2.4) {$\hdots$};
    \draw[line width=1.5, purple] (1,2) .. controls (1,1.5) .. (0,1.5);
    \draw[line width=1.5, purple] (2,2) .. controls (2,0.5) .. (3,0.5);
    \end{tikzpicture}~.
    \end{align}
    \item $C_{ik} ~:~~$ $C_{ik}$ are the same chords we use in the case of triangulations of an annulus. A chord $C_{ik}$ joins the marked points $i$ and $k$ passing around the cross-cap\footnote{In the case of an annulus, cross-cap gets replaced by the hole (second boundary component) of the annulus.}. Since there are two ways around the cross-cap, $C_{ik}$ and $C_{ki}$ are distinct. While drawing the Möbius strip as a rectangular strip, such curves must either pass through the identified vertical lines twice or not at all. There are $n(n-1)$ such curves, following the same counting as in the case of an annulus. 
    \begin{align}
        \begin{matrix}
            i\in[1,n]~, \ \  j\in[i+2,i+n]~; ~~~  \textcolor{teal}{C_{ij}} ~: \\ \quad \\ j\in[1,n]~, \ \  i\in[j+2,j+n]~; ~~~  \textcolor{purple}{C_{ji}} ~:
        \end{matrix} \quad \begin{tikzpicture}[scale=0.45,baseline={([yshift=-.5ex]current bounding box.center)}]
    \fill[cyan!6!white] (0,0) circle (2);
    \draw (0,0) circle (2);
    \node at (0,0) {$\ot$};
    \node at (0,2) {$\times$};
    \node at (0,2.9) {1};
    \node at ({2* cos(45)},{2* sin(45)}) {$\times$};
    \node at ({2* cos(45)+0.5},{2* sin(45)+0.5}) {2};
    \node at ({-2* cos(45)},{2* sin(45)}) {$\times$};
    \node at ({-2* cos(45)-0.5},{2* sin(45)+0.5}) {n};
    \node at ({2* cos(45)},{-2* sin(45)}) {$\times$};
    \node at ({2* cos(45)+0.5},{-2* sin(45)-0.5}) {i};
    \node at ({-2* cos(45)},{-2* sin(45)}) {$\times$};
    \node at ({-2* cos(45)-0.5},{-2* sin(45)-0.5}) {j};
    \node at (0,-2.4) {$\hdots$};
    \node at (2.4,0) {$\vdots$};
    \node at (-2.4,0) {$\vdots$};
    \draw[line width=1.5, purple] ({2* cos(45)},{-2* sin(45)}) .. controls (1.2,2) and (-1.2,2) .. ({-2* cos(45)},{-2* sin(45)});
    \draw[line width=1.5, teal] ({2* cos(45)},{-2* sin(45)}) .. controls (1,-0.5) and (-1,-0.5) .. ({-2* cos(45)},{-2* sin(45)});
\end{tikzpicture} \ \ \cong \quad \begin{tikzpicture}[scale=0.65,baseline={([yshift=-.8ex]current bounding box.center)}]
    \fill[cyan!6!white] (0,0) -- (0,2) -- (3,2) -- (3,0) -- (0,0);
    \draw[-{Stealth[length=2mm, width=1.5mm]}] (0,0) -- (0,1.2);
    \draw (0,0.95) -- (0,2) -- (3,2) -- (3,0.95);
    \draw [-{Stealth[length=2mm, width=1.5mm]}] (3,2) -- (3,0.8);
    \draw (3,0) -- (3,1.05);
    \draw (0,0) -- (3,0);
    \node at (1,2) {$\times$};
    \node at (2,2) {$\times$};
    \node at (0.8,2.5) {$i$};
    \node at (2.2,2.5) {$j$};
    \node at (1.5,2.4) {$\hdots$};
    \draw[line width=1.5, purple] (1,2) .. controls (1,1.5) .. (0,1.5);
    \draw[line width=1.5, purple] (2,2) .. controls (2,1.5) .. (3,1.5);
    \draw[line width=1.5, purple] (0,0.5) -- (3,0.5);
    \draw[line width=1.5, teal] (1,2) .. controls (1,1.5) and (2,1.5) .. (2,2);
    \end{tikzpicture}~.
    \end{align}
    \item $C^{\ot}~:~~$ There is a single chord $C^{\ot}$ that cuts the Möbius strip \emph{in the middle~.} \footnote{Upon cutting a Möbius strip along its length, instead of obtaining two disconnected pieces, we obtain a single orientable surface of double the length of the original Möbius strip with two twists. The surface obtained after the cut is topologically an annulus. We should revisit it afterwards.} Inclusion of this curve is necessary to obtain a consistent combinatorial structure.
    \begin{align}
        C^{\ot} ~:~~ \begin{tikzpicture}[scale=0.45,baseline={([yshift=-.5ex]current bounding box.center)}]
    \fill[cyan!6!white] (0,0) circle (2);
    \draw (0,0) circle (2);
    \node at (0,0) {$\ot$};
    \node at (0,2) {$\times$};
    \node at (0,2.9) {1};
    \node at ({2* cos(45)},{2* sin(45)}) {$\times$};
    \node at ({2* cos(45)+0.5},{2* sin(45)+0.5}) {2};
    \node at ({-2* cos(45)},{2* sin(45)}) {$\times$};
    \node at ({-2* cos(45)-0.5},{2* sin(45)+0.5}) {n};
    \node at ({2* cos(45)},{-2* sin(45)}) {$\times$};
    \node at ({2* cos(45)+0.5},{-2* sin(45)-0.5}) {i};
    \node at ({-2* cos(45)},{-2* sin(45)}) {$\times$};
    \node at ({-2* cos(45)-0.5},{-2* sin(45)-0.5}) {j};
    \node at (0,-2.4) {$\hdots$};
    \node at (2.4,0) {$\vdots$};
    \node at (-2.4,0) {$\vdots$};
    \draw[line width=1.5, purple] (0.5,0) circle (0.5);
\end{tikzpicture} \ \ \cong \quad \begin{tikzpicture}[scale=0.65,baseline={([yshift=-.8ex]current bounding box.center)}]
    \fill[cyan!6!white] (0,0) -- (0,2) -- (3,2) -- (3,0) -- (0,0);
    \draw[-{Stealth[length=2mm, width=1.5mm]}] (0,0) -- (0,1.2);
    \draw (0,0.95) -- (0,2) -- (3,2) -- (3,0.95);
    \draw [-{Stealth[length=2mm, width=1.5mm]}] (3,2) -- (3,0.8);
    \draw (3,0) -- (3,1.05);
    \draw (0,0) -- (3,0);
    \node at (1,2) {$\times$};
    \node at (2,2) {$\times$};
    \node at (0.8,2.5) {$i$};
    \node at (2.2,2.5) {$j$};
    \node at (1.5,2.4) {$\hdots$};
    \draw[line width=1.5, purple] (0,0.5) .. controls (1,0.4) and (2,1.6) .. (3,1.5);
    \end{tikzpicture}~~.
    \end{align}
\end{itemize}
To illustrate the idea, Figure \ref{fig:curves on Mobius2} list all the possible chords for a Möbius strip with two marked points. 
\noindent For a Möbius strip with a single marked point, there are two chords possible: $C^{\ot}$ and $C^{\ot}_{11}~,$ as depicted in Figure \ref{fig: Mn polytope for 1 and 2}.

\paragraph{Which chords are compatible?} Two curves are compatible with each other if they can be drawn on the surface without intersections. The cross-cap region makes the compatibility of curves opaque. The rectangular strip picture makes it comparatively clearer. Note that a cross-cap is defined by identifying the anti-podal points of a region, so \textbf{any curve entering a cross-cap must exit from the anti-podal point.} We can end up with incompatible curves intersecting nowhere away from the cross-cap, as in \eqref{cx is not compatible with cx12}, if we don't draw curves passing through the cross-caps consistently. Note that multiple compatible curves can pass through the cross-cap as follows:  
\begin{center}
 {\huge \xmark} \ \ \begin{tikzpicture}[scale=0.45,baseline={([yshift=-.5ex]current bounding box.center)}]
    \node at (0,0) {$\ot$};
    \draw[line width=1.5, purple] (-1,1) arc (90:-90:1);
    \draw[line width=1.5, teal] (1,1) arc (90:270:1);
\end{tikzpicture} \ $\equiv$  \ 
\begin{tikzpicture}[scale=0.45,baseline={([yshift=-.5ex]current bounding box.center)}]
    \node at (0,0) {$\ot$};
    \draw[line width=1.5, purple] (-1,1) -- (0.5,-0.5);
    \draw[line width=1.5, purple] (0.5,-0.5) .. controls (0.8,-0.8) .. (-1,-1.3);
    \draw[line width=1.5, teal] (1,1) -- (-0.5,-0.5);
    \draw[line width=1.5, teal] (-0.5,-0.5) .. controls (-0.8,-0.8) .. (1,-1.3);
\end{tikzpicture}  \hspace{2cm}  
{\huge \cmark} \ \ \begin{tikzpicture}[scale=0.45,baseline={([yshift=-.5ex]current bounding box.center)}]
    \node at (0,0) {$\ot$};
    \draw[line width=1.5, purple] (-1,1) -- (1,-1);
    \draw[line width=1.5, teal] (1,1) -- (-1,-1);
\end{tikzpicture}       \hspace{2cm}
{\huge \cmark} \ \ \begin{tikzpicture}[scale=0.45,baseline={([yshift=-.5ex]current bounding box.center)}]
    \node at (0,0) {$\ot$};
    \draw[line width=1.5, purple] (-1,1) -- (1,-1);
    \draw[line width=1.5, teal] (1,1) -- (-1,-1);
    \draw[line width=1.5, purple!50!teal] (0,1.4) -- (0,-1.4);
\end{tikzpicture}      
\end{center}    
We depict the compatibility of curves using two examples below for Mobius strip with $n=2$:
\begin{itemize}
    \item $C^{\ot}_{12}$ is compatible with $C_{11}^{\ot}$ and $C_{22}^{\ot}$, but not with $C_{11}$ and $C_{22}$:
    \begin{align}
    \begin{tikzpicture}[scale=0.45,baseline={([yshift=-.5ex]current bounding box.center)}]
    \draw (0,0) circle (2);
    \node at (0,0) {$\ot$};
    \node at (0,2) {$\times$};
    \node at (0,-2) {$\times$};
    \node at (0.4,2.6) {1};
    \node at (-0.4,-2.6) {2};
    \draw[line width=1.5, purple] (0,2) -- (0,-2);
    \draw[line width=1.5, teal] (0,2) .. controls (-3,-0.7) and (3,-0.7) .. (0,2);
\end{tikzpicture}  \hspace{-0.2cm}\cong \ \  \begin{tikzpicture}[scale=0.8,baseline={([yshift=-1.9ex]current bounding box.center)}]
    \fill[yellow!20] (0,0.5) .. controls (0.9,0.5) .. (1,2) .. controls (0.8,1.5) .. (0,1.5);
    \fill[yellow!20] (1,2) .. controls (1.1,0.5) .. (3,0.5) -- (3,1.5) .. controls (2,1.5) .. (2,2) -- (1,2);
    \fill[red!10] (3,1.5) .. controls (2,1.5) .. (2,2) -- (3,2) -- (3,1.5);
    \fill[red!10] (0,1.5) -- (0,2) -- (1,2) .. controls (0.8,1.5) .. (0,1.5);
    \fill[red!10] (1,2) .. controls (1.1,0.5) .. (3,0.5) -- (3,0) -- (0,0) -- (0,0.5) .. controls (0.9,0.5) .. (1,2);
    \draw[-{Stealth[length=2mm, width=1.5mm]}] (0,0) -- (0,1.2);
    \draw (0,0.95) -- (0,2) -- (3,2) -- (3,0.95);
    \draw [-{Stealth[length=2mm, width=1.5mm]}] (3,2) -- (3,0.8);
    \draw (3,0) -- (3,1.05);
    \draw (0,0) -- (3,0);
    \node at (1,2) {$\times$};
    \node at (2,2) {$\times$};
    \draw[line width=1.5, purple] (0,0.5) .. controls (0.9,0.5) .. (1,2);
    \draw[line width=1.5, purple] (2,2) .. controls (2,1.5) .. (3,1.5);
    \draw[line width=1.5, teal] (0,1.5) .. controls (0.8,1.5) .. (1,2);
    \draw[line width=1.5, teal] (1,2) .. controls (1.1,0.5) .. (3,0.5);
    \node at (1,2.5) {$1$};
    \node at (2,2.5) {$2$};
\end{tikzpicture} ~~, \ \ 
    \begin{tikzpicture}[scale=0.45,baseline={([yshift=-.5ex]current bounding box.center)}]
    \draw (0,0) circle (2);
    \node at (0,0) {$\ot$};
    \node at (0,2) {$\times$};
    \node at (0,-2) {$\times$};
    \node at (0.4,2.6) {1};
    \node at (-0.4,-2.6) {2};
    \draw[line width=1.5, purple] (0,2) -- (0,-2);
    \draw[line width=1.5, teal] (0,2) .. controls (-4,-2) and (4,-2) .. (0,2);
\end{tikzpicture}  \hspace{-0.7cm}\cong \ \  \begin{tikzpicture}[scale=0.8,baseline={([yshift=-1.9ex]current bounding box.center)}]
    \draw[-{Stealth[length=2mm, width=1.5mm]}] (0,0) -- (0,1.2);
    \draw (0,0.95) -- (0,2) -- (3,2) -- (3,0.95);
    \draw [-{Stealth[length=2mm, width=1.5mm]}] (3,2) -- (3,0.8);
    \draw (3,0) -- (3,1.05);
    \draw (0,0) -- (3,0);
    \node at (1,2) {$\times$};
    \node at (2,2) {$\times$};
    \draw[line width=1.5, purple] (0,0.5) .. controls (0.9,0.5) .. (1,2);
    \draw[line width=1.5, purple] (2,2) .. controls (2,1.5) .. (3,1.5);
    \draw[line width=1.5, teal] (0,1.3) .. controls (0.8,1.3) .. (1,2);
    \draw[line width=1.5, teal] (1,2) .. controls (1.1,1.3) .. (3,1.3);
    \draw[line width=1.5, teal] (0,0.7) -- (3,0.7);
    \node at (1,2.5) {$1$};
    \node at (2,2.5) {$2$};
\end{tikzpicture} 
\end{align}
We have colored the two triangles obtained by triangulation using $C_{12}^{\ot}$ and $C_{11}^{\ot}$ in the left picture. For the right picture, one can convince oneself that there is no way to draw $C_{11}$ and $C_{12}^{\ot}$ on a rectangular strip without them crossing. 
\item Even though one can draw $C_{12}^{\ot}$ and $C^{\ot}$ such that they do not intersect anywhere apart from the cross-cap, they are not compatible:
\begin{align}
    \begin{tikzpicture}[scale=0.45,baseline={([yshift=-.5ex]current bounding box.center)}]
    \draw (0,0) circle (2);
    \node at (0,0) {$\ot$};
    \node at (0,2) {$\times$};
    \node at (0,-2) {$\times$};
    \node at (0.4,2.6) {1};
    \node at (-0.4,-2.6) {2};
    \draw[line width=1.5, purple] (0,2) -- (0,-2);
    \draw[line width=1.5, teal] (0.5,0) circle (0.5);
    \end{tikzpicture}  \quad \cong \quad  \begin{tikzpicture}[scale=0.8,baseline={([yshift=-1.9ex]current bounding box.center)}]
    \draw[-{Stealth[length=2mm, width=1.5mm]}] (0,0) -- (0,1.2);
    \draw (0,0.95) -- (0,2) -- (3,2) -- (3,0.95);
    \draw [-{Stealth[length=2mm, width=1.5mm]}] (3,2) -- (3,0.8);
    \draw (3,0) -- (3,1.05);
    \draw (0,0) -- (3,0);
    \node at (1,2) {$\times$};
    \node at (2,2) {$\times$};
    \draw[line width=1.5, purple] (0,0.5) .. controls (0.9,0.5) .. (1,2);
    \draw[line width=1.5, purple] (2,2) .. controls (2,1.5) .. (3,1.5);
    \draw[line width=1.5, teal] (0,0.7) .. controls (1,0.7) and (2,1.3) .. (3,1.3);
    \node at (1,2.5) {$1$};
    \node at (2,2.5) {$2$};
    \end{tikzpicture} 
    \label{cx is not compatible with cx12}
\end{align}
\end{itemize}

\paragraph{How to define quasi-mutations and the combinatorial polytope?} To define the quasi-cluster algebra, we need to generalize the idea of mutations to the chords on unorientable surfaces. 

Instead of describing the detailed rules of such quasi-mutations, we write down the examples of the combinatorial polytopes $M_1$ and $M_2~,$ obtained from quasi-triangulations of Mobius strip with $1$ and $2$ marked points respectively. $M_{1,2}$ are shown in Figure \ref{fig: Mn polytope for 1 and 2}. Apart from the usual mutations we encounter in the orientable surface triangulations, the new addition is the \emph{quasi-mutations} of $C_{11}^{\ot} \longleftrightarrow C^{\ot}~.$ In the next subsection, we should see that these quasi-mutations can be naturally understood as obtained from the projection of triangulations of an orientable surface. 

Let us add up the total number of chords possible $C_{ij}^{\ot}~, ~ C_{ij} ~,~ C^{\ot}$ for Möbius strip with $n$ insertions.
\begin{align}
    \text{Number of chords on Möbius}_n = \frac{n(n+1)}{2} + n(n-1) + 1 = \frac{3n^2-n+2}{2}~.
\end{align}
This is the number of codimension-one faces of the combinatorial polytope $M_n$. The total number of vertices of this $M_n$ polytope has been counted in \cite{bazier2023number-of-triangulations} to be the following:
\begin{align}
    \text{Number of total triangulations of Möbius}_n = 4^{n-1} + {}^{2n-2}C_{n-1}~.
\end{align}
Let us compare it with the planar one-loop ($n$ marked points on one boundary of the annulus, and none on the other) case of $\hat{D}_n:$ \cite{Laddha:2024qtn}
\begin{align}
    \text{Number of total triangulations of Annulus}_n = 2(2n-1)\,C_{n-1} = \frac{2(2n-1)!}{n!\,(n-1)!} ~.
\end{align}
Note that $C_n$ is the Catalan number. The total number of triangulations for $n=1$ and $n=2$ is the same for the annulus and the Möbius strip. For $n=3$, there are a total of 22 triangulations for the Möbius strip and 20 for the annulus. There are two more triangulations in the Möbius strip dual to the triangle graphs. 

\section{Curve integral for Möbius strip}
\label{section 3 curve integral for mobius}
In this section, we use the mathematical structures built on the triangulations of the Möbius strip to study the scattering amplitudes defined as integrals over the moduli space of non-orientable surfaces with marked points. We wish to mimic the curve integral formula studied in \cite{Arkani-Hamed:2023CurveIntegral, Arkani-Hamed:2023Multiplicity}. We need to associate a momentum, a $g$-vector, and eventually a headlight function to every chord/curve on the Möbius strip. Since these structures are well understood for orientable surfaces, we embed our Möbius strip in an orientable surface, an annulus, and project out the required information from the orientable surface. 

\paragraph{Doubling the Möbius strip into an annulus:} We wish to embed the unorientable surface into a bigger orientable surface. We proceed as depicted in \eqref{abstract doubling of Möbius strip}. We begin with the Möbius strip, take its mirror image, and glue the mirror image to the original surface. As also discussed in \cite{bazierquasi-overview}, in terms of the disk with a cross-cap, the doubling amounts to blowing up the cross-cap into a boundary and putting identical marked points on the new boundary. The ordering of the marked points on the inside boundary is also clockwise, analogous to the original ones.  
\begin{align}
\label{abstract doubling of Möbius strip}
    \begin{tikzpicture}[scale=0.55,baseline={([yshift=-1.9ex]current bounding box.center)}]
        \shade[left color=white,right color=cyan!35!white] (0,0) rectangle (3,2);
        \draw[-{Stealth[length=2mm, width=1.5mm]}] (0,0) -- (0,1.2);
        \draw (0,0.95) -- (0,2) -- (3,2) -- (3,0.95);
        \draw [-{Stealth[length=2mm, width=1.5mm]}] (3,2) -- (3,0.8);
        \draw (3,0) -- (3,1.05);
        \draw (0,0) -- (3,0);
        \node at (1,2) {$\times$};
        \node at (2,2) {$\times$};
        \node at (1,2.5) {$1$};
        \node at (2,2.5) {$2$};
    \end{tikzpicture} 
    \ \ \to \ \
    \begin{tikzpicture}[scale=0.55,baseline={([yshift=-1.9ex]current bounding box.center)}]
        \shade[left color=white,right color=cyan!35!white] (0,0) rectangle (3,2);
        \draw[-{Stealth[length=2mm, width=1.5mm]}] (0,0) -- (0,1.2);
        \draw (0,0.95) -- (0,2) -- (3,2) -- (3,0.95);
        \draw [-{Stealth[length=2mm, width=1.5mm]}] (3,2) -- (3,0.8);
        \draw (3,0) -- (3,1.05);
        \draw (0,0) -- (3,0);
        \node at (1,2) {$\times$};
        \node at (2,2) {$\times$};
        \node at (1,2.5) {$1$};
        \node at (2,2.5) {$2$};
        \draw (-0.5,-0.75) -- (3.5,-0.75);
        \foreach \x in {0,...,20}
            \draw ({-0.5+\x*0.2},-0.75) -- ({-0.5+\x*0.2+0.15},-0.9);
            \shade[left color=white,right color=cyan!35!white] (0,-1.5) rectangle (3,-3.5);
        \draw[-{Stealth[length=2mm, width=1.5mm]}] (0,-1.5) -- (0,-2.7);
        \draw (0,-1.5-0.95) -- (0,-3.5) -- (3,-3.5) -- (3,-1.5-0.95);
        \draw [-{Stealth[length=2mm, width=1.5mm]}] (3,-3.5) -- (3,-1.5-0.8);
        \draw (3,-1.5) -- (3,-1.5-1.05);
        \draw (0,-1.5) -- (3,-1.5);
        \node at (1,-3.5) {$\times$};
        \node at (2,-3.5) {$\times$};
        \node at (1,-4) {$1'$};
        \node at (2,-4) {$2'$};
    \end{tikzpicture}  \ \ \to \ \ 
    \begin{tikzpicture}[scale=0.55,baseline={([yshift=-1.9ex]current bounding box.center)}]
        \shade[left color=white,right color=cyan!35!white] (0,0) rectangle (3,2);
        \shade[left color=white,right color=cyan!35!white] (3,0) rectangle (6,2);
        \draw[-{Stealth[length=2mm, width=1.5mm]}] (0,0) -- (0,1.2);
        \draw (0,0.95) -- (0,2) -- (3,2) -- (3,0.95);
        \draw [-{Stealth[length=2mm, width=1.5mm]}] (3,2) -- (3,0.8);
        \draw (3,0) -- (3,1.05);
        \draw (0,0) -- (3,0);
        \node at (1,2) {$\times$};
        \node at (2,2) {$\times$};
        \node at (1,2.5) {$1$};
        \node at (2,2.5) {$2$};
        \draw (3,0.95) -- (3,2) -- (6,2) -- (6,0.95);
        \draw [-{Stealth[length=2mm, width=1.5mm]}] (6,0) -- (6,1.2);
        \draw (6,0) -- (6,1.05);
        \draw (3,0) -- (6,0);
        \node at (4,0) {$\times$};
        \node at (5,0) {$\times$};
        \node at (4,-0.5) {$1'$};
        \node at (5,-0.5) {$2'$};
    \end{tikzpicture} \ \ \cong \ \ 
    \begin{tikzpicture}[scale=0.9,baseline={([yshift=-1.9ex]current bounding box.center)}]
        \shade[left color=white,right color=cyan!35!white] (0.8,0) -- (1.2,0) arc (0:180:1.2) -- (-0.8,0) arc (180:0:0.8);
        \shade[left color=cyan!35!white,right color=white] (0.8,0) -- (1.2,0) arc (0:-180:1.2) -- (-0.8,0) arc (-180:0:0.8);
        \draw [-{Stealth[length=2mm, width=1.5mm]}] (0.8,0) -- (1.15,0);
        \draw (1,0) -- (1.2,0);
        \draw (-1,0) -- (-0.8,0);
        \draw[-{Stealth[length=2mm, width=1.5mm]}] (-1.2,0) -- (-0.85,0);
        \draw (0,0) circle (1.2);
        \draw (0,0) circle (0.8);
        \node at ({1.2*cos(60)},{1.2*sin(60)}) {$\times$};
        \node at ({1.2*cos(120)},{1.2*sin(120)}) {$\times$};
        \node at ({1.2*cos(60)},{1.2*sin(60)+0.45}) {$2$};
        \node at ({1.2*cos(120)},{1.2*sin(120)+0.45}) {$1$};
        \node at ({0.8*cos(240)},{0.8*sin(240)}) {$\times$};
        \node at ({0.8*cos(300)},{0.8*sin(300)}) {$\times$};
        \node at ({0.8*cos(240)},{0.8*sin(240)+0.4}) {$2'$};
        \node at ({0.8*cos(300)},{0.8*sin(300)+0.4}) {$1'$};
    \end{tikzpicture}
\end{align}

\paragraph{Calculating the $g$-vectors:} Without the doubling of the Möbius strip and the embedding into an orientable surface, it is not clear how to define $g$-vectors for the curves on the Möbius strip. In the absence of an orientation, it is meaningless to say that the curve took a left or a right turn. 

Along with the doubling of the Möbius strip, a curve on the Möbius strip gets doubled into a set of two curves. We depict this using the example of $C_{12}^{\ot}$ for the case of Möbius${}_{2}$ in \eqref{doubling of curve Cx12 for Möbius2}. The curve has been mapped to $C_{12'}$ and $C_{21'}$ on the annulus obtained from the doubling. It is a general feature: apart from $C^{\ot}$, every curve on the Möbius strip maps to two distinct curves on the doubled annulus. 
\begin{align}
\label{doubling of curve Cx12 for Möbius2}
    \begin{tikzpicture}[scale=0.55,baseline={([yshift=-1.9ex]current bounding box.center)}]
        \shade[left color=white,right color=cyan!35!white] (0,0) rectangle (3,2);
        \draw[-{Stealth[length=2mm, width=1.5mm]}] (0,0) -- (0,1.2);
        \draw (0,0.95) -- (0,2) -- (3,2) -- (3,0.95);
        \draw [-{Stealth[length=2mm, width=1.5mm]}] (3,2) -- (3,0.8);
        \draw (3,0) -- (3,1.05);
        \draw (0,0) -- (3,0);
        \node at (1,2) {$\times$};
        \node at (2,2) {$\times$};
        \node at (1,2.5) {$1$};
        \node at (2,2.5) {$2$};
        \draw[line width=1.6, purple] (0,1.5) .. controls (1,1.5) .. (1,2);
        \draw[line width=1.6, purple] (2,2) .. controls (2,0.5) .. (3,0.5);
    \end{tikzpicture} 
    \ \ \to \ \
    \begin{tikzpicture}[scale=0.55,baseline={([yshift=-1.9ex]current bounding box.center)}]
        \shade[left color=white,right color=cyan!35!white] (0,0) rectangle (3,2);
        \draw[-{Stealth[length=2mm, width=1.5mm]}] (0,0) -- (0,1.2);
        \draw (0,0.95) -- (0,2) -- (3,2) -- (3,0.95);
        \draw [-{Stealth[length=2mm, width=1.5mm]}] (3,2) -- (3,0.8);
        \draw (3,0) -- (3,1.05);
        \draw (0,0) -- (3,0);
        \node at (1,2) {$\times$};
        \node at (2,2) {$\times$};
        \node at (1,2.5) {$1$};
        \node at (2,2.5) {$2$};
        \draw[line width=1.6, purple] (0,1.5) .. controls (1,1.5) .. (1,2);
        \draw[line width=1.6, purple] (2,2) .. controls (2,0.5) .. (3,0.5);
        \draw (-0.5,-0.75) -- (3.5,-0.75);
        \foreach \x in {0,...,20}
            \draw ({-0.5+\x*0.2},-0.75) -- ({-0.5+\x*0.2+0.15},-0.9);
            \shade[left color=white,right color=cyan!35!white] (0,-1.5) rectangle (3,-3.5);
        \draw[-{Stealth[length=2mm, width=1.5mm]}] (0,-1.5) -- (0,-2.7);
        \draw (0,-1.5-0.95) -- (0,-3.5) -- (3,-3.5) -- (3,-1.5-0.95);
        \draw [-{Stealth[length=2mm, width=1.5mm]}] (3,-3.5) -- (3,-1.5-0.8);
        \draw (3,-1.5) -- (3,-1.5-1.05);
        \draw (0,-1.5) -- (3,-1.5);
        \node at (1,-3.5) {$\times$};
        \node at (2,-3.5) {$\times$};
        \node at (1,-4) {$1'$};
        \node at (2,-4) {$2'$};
        \draw[line width=1.6, purple] (0,-3) .. controls (1,-3) .. (1,-3.5);
        \draw[line width=1.6, purple] (3,-2) .. controls (2,-2) .. (2,-3.5);
    \end{tikzpicture}  \ \ \to \ \ 
    \begin{tikzpicture}[scale=0.55,baseline={([yshift=-1.9ex]current bounding box.center)}]
        \shade[left color=white,right color=cyan!35!white] (0,0) rectangle (3,2);
        \shade[left color=white,right color=cyan!35!white] (3,0) rectangle (6,2);
        \draw[-{Stealth[length=2mm, width=1.5mm]}] (0,0) -- (0,1.2);
        \draw (0,0.95) -- (0,2) -- (3,2) -- (3,0.95);
        \draw [-{Stealth[length=2mm, width=1.5mm]}] (3,2) -- (3,0.8);
        \draw (3,0) -- (3,1.05);
        \draw (0,0) -- (3,0);
        \node at (1,2) {$\times$};
        \node at (2,2) {$\times$};
        \node at (1,2.5) {$1$};
        \node at (2,2.5) {$2$};
        \draw (3,0.95) -- (3,2) -- (6,2) -- (6,0.95);
        \draw [-{Stealth[length=2mm, width=1.5mm]}] (6,0) -- (6,1.2);
        \draw (6,0) -- (6,1.05);
        \draw (3,0) -- (6,0);
        \node at (4,0) {$\times$};
        \node at (5,0) {$\times$};
        \node at (4,-0.5) {$1'$};
        \node at (5,-0.5) {$2'$};
        \draw[line width=1.6, purple] (0,1.5) .. controls (1,1.5) .. (1,2);
        \draw[line width=1.6, purple] (2,2) .. controls (2,0.5) .. (3,0.5);
        \draw[line width=1.6, purple] (3,0.5) .. controls (4,0.5) .. (4,0);
        \draw[line width=1.6, purple] (6,1.5) .. controls (5,1.5) .. (5,0);
    \end{tikzpicture} \ \ \cong \ \ 
    \begin{tikzpicture}[scale=0.9,baseline={([yshift=-1.9ex]current bounding box.center)}]
        \shade[left color=white,right color=cyan!35!white] (0.8,0) -- (1.2,0) arc (0:180:1.2) -- (-0.8,0) arc (180:0:0.8);
        \shade[left color=cyan!35!white,right color=white] (0.8,0) -- (1.2,0) arc (0:-180:1.2) -- (-0.8,0) arc (-180:0:0.8);
        \draw [-{Stealth[length=2mm, width=1.5mm]}] (0.8,0) -- (1.15,0);
        \draw (1,0) -- (1.2,0);
        \draw (-1,0) -- (-0.8,0);
        \draw[-{Stealth[length=2mm, width=1.5mm]}] (-1.2,0) -- (-0.85,0);
        \draw (0,0) circle (1.2);
        \draw (0,0) circle (0.8);
        \node at ({1.2*cos(60)},{1.2*sin(60)}) {$\times$};
        \node at ({1.2*cos(120)},{1.2*sin(120)}) {$\times$};
        \node at ({1.2*cos(60)},{1.2*sin(60)+0.45}) {$2$};
        \node at ({1.2*cos(120)},{1.2*sin(120)+0.45}) {$1$};
        \node at ({0.8*cos(240)},{0.8*sin(240)}) {$\times$};
        \node at ({0.8*cos(300)},{0.8*sin(300)}) {$\times$};
        \node at ({0.8*cos(240)},{0.8*sin(240)+0.4}) {$2'$};
        \node at ({0.8*cos(300)},{0.8*sin(300)+0.4}) {$1'$};
        \draw[line width=1.6, purple] ({1.2*cos(60)},{1.2*sin(60)}) -- ({cos(60)},{sin(60)}) arc (60:-60:1) -- ({0.8*cos(-60)},{0.8*sin(-60)});
        \draw[line width=1.6, purple] ({1.2*cos(120)},{1.2*sin(120)}) -- ({cos(120)},{sin(120)}) arc (120:240:1) -- ({0.8*cos(240)},{0.8*sin(240)});
    \end{tikzpicture}
\end{align}

The $g$-vector fan of the Möbius strip should be obtained from an appropriate projection of the $g$-vector fan of the doubled annulus with double the marked points. The projection is determined as follows: \emph{the $g$-vectors of the two curves on the annulus, mapping to the same curve on the Möbius strip, should become equal under the projection.} We quantify this projection by working out some examples. By inspection, we deduce the appropriate projection to be the following:
\begin{align}
    t_i+t_i' = 0 ~, \qquad \forall\quad  t_i~.
\end{align}
It projects out an $n$-dimensional space in the $2n$-dimensional space in which the $g$-vectors for the doubled annulus reside. {  We illustrate the idea pictorially for $n=1$ in Figure \ref{fig:projection picture for mobius1}. The $g$-vector fan for annulus with one marked point on each boundary is constructed in Appendix \ref{appendix A review}, \eqref{fan for annulus with 2 marked points}, and we project the fan onto $t+t'=0$:
\begin{figure}[h]
    \centering
     \begin{tikzpicture}[scale=1.2,baseline={([yshift=-.5ex]current bounding box.center)}]
        \draw[<->] (-1.2,0) -- (1.2,0);
        \draw[<->] (0,-1.2) -- (0,1.2);
        \draw[thick,teal,-{Stealth[length=2mm, width=1.5mm]}] (0,0) -- (1,0);
        \draw[thick,teal,-{Stealth[length=2mm, width=1.5mm]}] (0,0) -- (0,1);
        \draw[thick,teal,-{Stealth[length=2mm, width=1.5mm]}] (0,0) -- (-1,2);
        \draw[thick,teal,-{Stealth[length=2mm, width=1.5mm]}] (0,0) -- (-2,3);
        \draw[thick,purple,-{Stealth[length=2mm, width=1.5mm]}] (0,0) -- (0,-1);
        \draw[thick,purple,-{Stealth[length=2mm, width=1.5mm]}] (0,0) -- (-1,0);
        \draw[thick,purple,-{Stealth[length=2mm, width=1.5mm]}] (0,0) -- (-2,1);
        \draw[thick,purple,-{Stealth[length=2mm, width=1.5mm]}] (0,0) -- (-3,2);        
        \node[purple, rotate=120] at (-2.5,2.1) {$\hdots$};
        \node[teal, rotate=-20] at (-2.1,2.5) {$\hdots$};
        \draw[thick,-{Stealth[length=2mm, width=1.5mm]}] (0,0) -- (-1,1);
        \node at (1.3,0.2) {$t$};
        \node at (0.2,1.3) {$t'$};
        \node[teal] at (0.9,-0.3) {$C_0'$};
        \node[teal] at (0.3,0.9) {$C_1'$};
        \node[teal] at (-0.6,1.9) {$C_2'$};
        \node[teal] at (-1.5,2.9) {$C_3'$};
        \node[purple] at (0.3,-0.9) {$C_0$};
        \node[purple] at (-0.9,-0.3) {$C_1$};
        \node[purple] at (-1.9,0.6) {$C_2$};
        \node[purple] at (-2.9,1.6) {$C_3$};
        \node at (-1.2,1.2) {$\Delta$};
        \draw[dotted, thick] (1.2,-1.2) -- (-2.7,2.7);
        \node at (-3.5,2.7) {$t+t'=0$};
    \end{tikzpicture}    $\ \ \xrightarrow[t+t'=0]{\text{project to}} \ \ $
    \begin{tikzpicture}[scale=1.2,baseline={([yshift=-.5ex]current bounding box.center)}]
        \draw[<->] (-1.2,1.2) -- (1.2,-1.2);
        \draw[very thick,{Stealth[length=2mm, width=1.5mm]}-{Stealth[length=2mm, width=1.5mm]},purple!50!black] (1,-1) -- (-1,1);
        \draw (0.1,0.1) -- (-0.1,-0.1);
        \node[purple!50!black] at (1,-0.5) {$C_{11}^{\ot}$};
        \node[purple!50!black] at (-0.5,1) {$C^{\ot}$};
        \node at (1.35,-1.35) {$t$};
        \node at (-0.25,-0.25) {0};
    \end{tikzpicture} 
    \caption{The $g$-vector fan for an annulus (left) with one marked point on either boundary projected along the line $t+t'=0$ yields the $g$-vector fan for a Mobius strip (right) with a single marked point on its boundary.}
    \label{fig:projection picture for mobius1}
\end{figure}

The pairs of curves, $C_i$ and $C_i'$, for $n>0$, maps to the same
curve $C^{\ot}$ on the Mobius strip, and hence is counted only once. 

The projection condition $t_i+t_i'=0$ is independent of the reference triangulation.} For a generic reference triangulation of the doubled annulus with coordinates $\{t_i,t_i'\}~,$ where $t_i'$ and $t_i$ are parameters for the two edges related to each other by a flip of two boundaries. For detailed worked examples, refer to Figures \ref{fig:g-vectors for Möbius1} and \ref{fig:g-vectors for Möbius2} for $n=1$ and $n=2$, respectively. Finally, Figure \ref{fig:g-vector fan for Möbius1,2} depicts the resultant $g$-vector fan for the two cases. 

The $g$-vectors for the Möbius strip are defined for a reference triangulation specified by the annulus reference triangulation. Here, in Figures \ref{fig:g-vectors for Möbius1}, \ref{fig:g-vectors for Möbius2} and \ref{fig:g-vector fan for Möbius1,2}, the reference triangulation is the analogue of the 'wheel-graph': the $n$-gon.
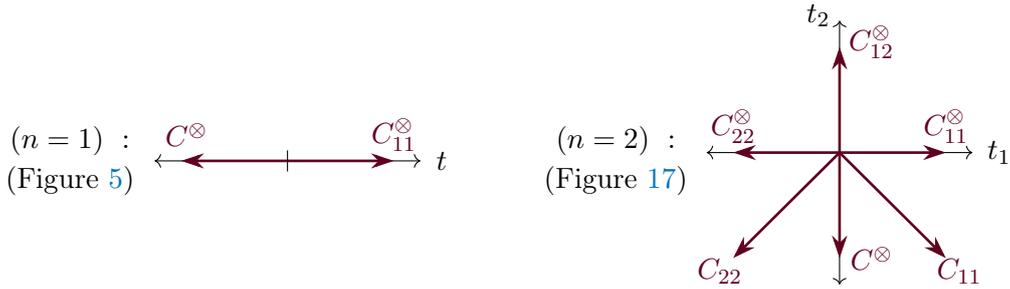
\begin{figure}[h]
    \centering
\begin{align*}
    \begin{matrix}
    (n=1) ~: \\ (\text{Figure }\ref{fig:g-vectors for Möbius1})
\end{matrix} \ \ \begin{tikzpicture}[scale=0.7,baseline={([yshift=-1.9ex]current bounding box.center)}]
    \draw[<->] (-2.5,0) -- (2.5,0);
    \draw[line width=1, purple!50!black, -{Stealth[length=3mm, width=2mm]}] (0,0) -- (-2,0);
    \node[purple!50!black] at (-1.9,0.5) {$C^{\ot}$};
    \draw[line width=1, purple!50!black, -{Stealth[length=3mm, width=2mm]}] (0,0) -- (2,0);
    \node[purple!50!black] at (2,0.5) {$C^{\ot}_{11}$};
    \draw (0,-0.2) -- (0,0.2);
    \node at (2.9,0) {$t$};
\end{tikzpicture} \qquad \quad \begin{matrix}
    (n=2) ~: \\ (\text{Figure }\ref{fig:g-vectors for Möbius2})
\end{matrix} \begin{tikzpicture}[scale=0.7,baseline={([yshift=-1.9ex]current bounding box.center)}]
    \draw[<->] (-2.5,0) -- (2.5,0);
    \draw[<->] (0,-2.5) -- (0,2.5);
    \draw[line width=1, purple!50!black, -{Stealth[length=3mm, width=2mm]}] (0,0) -- (0,-2);
    \node[purple!50!black] at (0.6,-2) {$C^{\ot}$};
    \draw[line width=1, purple!50!black, -{Stealth[length=3mm, width=2mm]}] (0,0) -- (-2,0);
    \node[purple!50!black] at (-2,0.5) {$C^{\ot}_{22}$};
    \draw[line width=1, purple!50!black, -{Stealth[length=3mm, width=2mm]}] (0,0) -- (2,0);
    \node[purple!50!black] at (2,0.5) {$C^{\ot}_{11}$};
    \draw[line width=1, purple!50!black, -{Stealth[length=3mm, width=2mm]}] (0,0) -- (0,2);
    \node[purple!50!black] at (0.6,2.1) {$C^{\ot}_{12}$};
    \draw[line width=1, purple!50!black, -{Stealth[length=3mm, width=2mm]}] (0,0) -- (2,-2);
    \node[purple!50!black] at (2.25,-2.25) {$C_{11}$};
    \draw[line width=1, purple!50!black, -{Stealth[length=3mm, width=2mm]}] (0,0) -- (-2,-2);
    \node[purple!50!black] at (-2.25,-2.25) {$C_{22}$};
    \node at (3,0) {$t_1$};
    \node at (-0.4,2.6) {$t_2$};
\end{tikzpicture}
\end{align*}
    \caption{$g$-vector fans for the Möbius strip with one or two marked points on its boundary, with the 'wheel' graph ($n$-gon) as the reference triangulation.}
    \label{fig:g-vector fan for Möbius1,2}
\end{figure}

\begin{figure}[!th]
    \centering
\begin{align*}
    C_{11}^{\ot}:\quad \begin{tikzpicture}[scale=0.6,baseline={([yshift=-.5ex]current bounding box.center)}]
    \fill[cyan!10!white] (0,0) -- (0,2) -- (3,2) -- (3,0) -- (0,0);
    \draw[-{Stealth[length=2mm, width=1.5mm]}] (0,0) -- (0,1.2);
    \draw (0,0.95) -- (0,2) -- (6,2) -- (6,0.95);
    \draw[-{Stealth[length=2mm, width=1.5mm]}] (6,0) -- (6,1.2); 
    \draw (0,0) -- (6,0);
    \draw [-{Stealth[length=2mm, width=1.5mm]}] (3,2) -- (3,0.8);
    \draw (3,0) -- (3,1.05);
    \node at (1.5,2) {$\times$};
    \node at (4.5,0) {$\times$};
    \draw[line width=1.6, purple] (1.5,2) .. controls (1.5,1.5) .. (0,1.5);
    \draw[line width=1.6, purple] (4.5,0) .. controls (4.5,1.5) .. (6,1.5);
    \draw[line width=1.6, teal] (1.5,2) .. controls (1.5,-0.8) and (4.5,1.2) .. (4.5,0);
    \node at (1.5,2.6) {$1$};
    \node at (4.5,-0.6) {$1'$};
\end{tikzpicture} &\qquad 
\begin{tikzpicture}[scale=0.51,baseline={([yshift=-.5ex]current bounding box.center)}]
    \def \rin {1.7};
    \def \rout {2.5};
    \def \rone {1.9667};
    \def \rtwo {2.2333};
    \def \r {2.1};
    \def \deltaout {9};
    \def \deltain {13};
    \def \lenout {0.6};
    \def \lenin {0.6};     
    \fill[cyan!10!white] ({\rout * cos(90+\deltaout)},{\rout * sin(90+\deltaout) + \lenout}) -- ({\rout * cos(90+\deltaout)},{\rout * sin(90+\deltaout)}) arc (90+\deltaout : 180 : \rout) -- ({-\rin},0 ) arc (180:0:\rin) -- ({\rout},0) arc (0:90-\deltaout : \rout) -- ({\rout * cos(90-\deltaout)},{\rout * sin(90-\deltaout)  + \lenout});
    \draw ({\rout * cos(90+\deltaout)},{\rout * sin(90+\deltaout) + \lenout}) -- ({\rout * cos(90+\deltaout) },{\rout * sin(90+\deltaout)}) arc (90+\deltaout : 360+90-\deltaout : \rout) -- ({\rout * cos(90-\deltaout)},{\rout * sin(90-\deltaout) + \lenout});
    \draw ({\rin * cos(270+\deltain)},{\rin * sin(270+\deltain) + \lenin}) -- ({\rin * cos(270+\deltain)},{\rin * sin(270+\deltain)}) arc (270+\deltain : 360+270-\deltain : \rin) -- ({\rin * cos(270-\deltain)},{\rin * sin(270-\deltain) + \lenin});
    \draw[-stealth] (\rin,0) -- ({0.85*\rout+0.15*\rin},0);     
    \draw ({0.8*\rout+0.2*\rin},0) -- (\rout,0);     
    \draw ({-0.2*\rout-0.8*\rin},0) -- (-\rin,0);     
    \draw[-stealth] (-\rout,0) -- ({-0.15*\rout-0.85*\rin},0) ;  
    \node at (0.7,\rout +0.6) {$1$};
    \node at (0.7,{-\rin+0.8}) {$1'$};
    \node at (\rout+0.5, 0) {$t$};
    \node at (-\rout-0.5, 0) {$t'$};
    \draw[line width=1.6, purple] ({\r * cos(90+0.33*\deltaout)},{\r * sin(90+0.33*\deltaout) + 2*\lenout}) -- ({\r * cos(90+0.33*\deltaout)},{\r * sin(90+0.33*\deltaout)}) arc (90+0.33*\deltaout : 270-0.33*\deltaout : \r) -- ({\r * cos(270-0.33*\deltaout)},{\r * sin(270-0.33*\deltaout) + 2*\lenout});
    \draw[line width=1.6, teal] ({\r * cos(90-0.33*\deltaout)},{\r * sin(90-0.33*\deltaout) + 2*\lenout}) -- ({\r * cos(90-0.33*\deltaout)},{\r * sin(90-0.33*\deltaout)}) arc (90-0.33*\deltaout : -90+0.33*\deltaout : \r) -- ({\r * cos(270+0.33*\deltaout)},{\r * sin(270+0.33*\deltaout) + 2*\lenout}); 
\end{tikzpicture} \qquad \begin{matrix}
    \textcolor{teal}{\vec{g}^{\text{aux}}_{11'}= \big(1,0\big)} \\ \textcolor{purple}{\vec{g}^{\text{aux}}_{1'1}= \big(0,-1\big)} \\ \quad \\ \xrightarrow{\text{  projection  }} \vec{g}^{\ot}_{11} = (1)~. 
\end{matrix}\\
        C^{\ot}:\quad \begin{tikzpicture}[scale=0.6,baseline={([yshift=-.5ex]current bounding box.center)}]
    \fill[cyan!10!white] (0,0) -- (0,2) -- (3,2) -- (3,0) -- (0,0);
    \draw[-{Stealth[length=2mm, width=1.5mm]}] (0,0) -- (0,1.2);
    \draw (0,0.95) -- (0,2) -- (6,2) -- (6,0.95);
    \draw[-{Stealth[length=2mm, width=1.5mm]}] (6,0) -- (6,1.2); 
    \draw (0,0) -- (6,0);
    \draw [-{Stealth[length=2mm, width=1.5mm]}] (3,2) -- (3,0.8);
    \draw (3,0) -- (3,1.05);
    \node at (1.5,2) {$\times$};
    \node at (4.5,0) {$\times$};
    \node at (1.5,2.6) {$1$};
    \node at (4.5,-0.6) {$1'$};
    \draw[line width=1.6, purple] (0,0.5) .. controls (1,0.4) and (2,1.6) .. (3,1.5);
    \draw[line width=1.6, purple] (3,1.5) .. controls (4,1.6) and (5,0.4) .. (6,0.5);
\end{tikzpicture} &\qquad 
\begin{tikzpicture}[scale=0.51,baseline={([yshift=-.5ex]current bounding box.center)}]
    \def \rin {1.7};
    \def \rout {2.5};
    \def \rone {1.9667};
    \def \rtwo {2.2333};
    \def \r {2.1};
    \def \deltaout {9};
    \def \deltain {13};
    \def \lenout {0.6};
    \def \lenin {0.6};     
    \fill[cyan!10!white] ({\rout * cos(90+\deltaout)},{\rout * sin(90+\deltaout) + \lenout}) -- ({\rout * cos(90+\deltaout)},{\rout * sin(90+\deltaout)}) arc (90+\deltaout : 180 : \rout) -- ({-\rin},0 ) arc (180:0:\rin) -- ({\rout},0) arc (0:90-\deltaout : \rout) -- ({\rout * cos(90-\deltaout)},{\rout * sin(90-\deltaout)  + \lenout});
    \draw ({\rout * cos(90+\deltaout)},{\rout * sin(90+\deltaout) + \lenout}) -- ({\rout * cos(90+\deltaout) },{\rout * sin(90+\deltaout)}) arc (90+\deltaout : 360+90-\deltaout : \rout) -- ({\rout * cos(90-\deltaout)},{\rout * sin(90-\deltaout) + \lenout});
    \draw ({\rin * cos(270+\deltain)},{\rin * sin(270+\deltain) + \lenin}) -- ({\rin * cos(270+\deltain)},{\rin * sin(270+\deltain)}) arc (270+\deltain : 360+270-\deltain : \rin) -- ({\rin * cos(270-\deltain)},{\rin * sin(270-\deltain) + \lenin});
    \draw[-stealth] (\rin,0) -- ({0.85*\rout+0.15*\rin},0);     
    \draw ({0.8*\rout+0.2*\rin},0) -- (\rout,0);     
    \draw ({-0.2*\rout-0.8*\rin},0) -- (-\rin,0);     
    \draw[-stealth] (-\rout,0) -- ({-0.15*\rout-0.85*\rin},0) ;  
    \node at (0.7,\rout +0.6) {$1$};
    \node at (0.7,{-\rin+0.8}) {$1'$};
    \node at (\rout+0.5, 0) {$t$};
    \node at (-\rout-0.5, 0) {$t'$};
    \draw[line width=1.6, purple] (0,0) circle ({\r});
\end{tikzpicture}  \qquad \begin{matrix}
   \textcolor{purple}{\vec{g}^{\text{aux}}_{\Delta}= \big(-1,1\big)} \\ \quad \\ \xrightarrow{\text{  projection  }} \vec{g}^{\ot} = (-1)~. 
\end{matrix}
\end{align*}
    \caption{Evaluation of $g$-vectors for Möbius${}_1$ by embedding it into an annulus. The $g$-vectors for the annulus have been calculated explicitly in Appendix \ref{appendix A review}.}
    \label{fig:g-vectors for Möbius1}
\end{figure}

Similarly, we write down the $g$-vectors for $n=3$ with the $n$-gon reference triangulation in the Appendix \ref{appendix B details}: \eqref{g-vectors for n=3 with wheel as reference}.

\begin{figure}[!ht]
    \centering
    \includegraphics[width=0.95\linewidth]{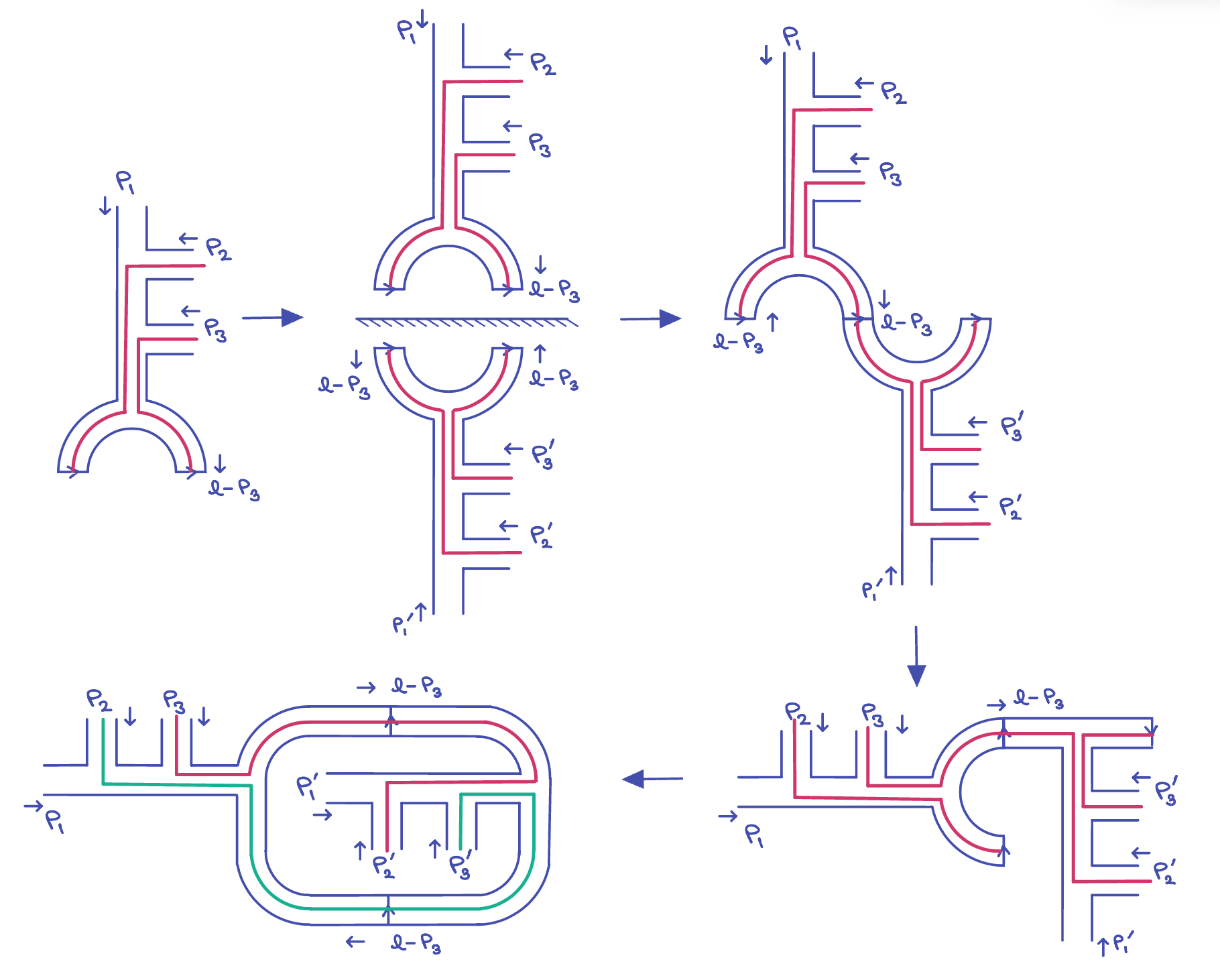}
    \caption{A particular Feynman diagram for Möbius${}_3$, a tadpole, doubled to obtain an annulus with 6 insertions. The curve \textcolor{purple}{$C^{\ot}_{23}$} gets doubled to \textcolor{purple}{$C_{2'3}$} and \textcolor{teal}{$C_{23'}$} on the annulus.}
    \label{fig:3-pt-tadpole-doubled-momenta}
\end{figure}

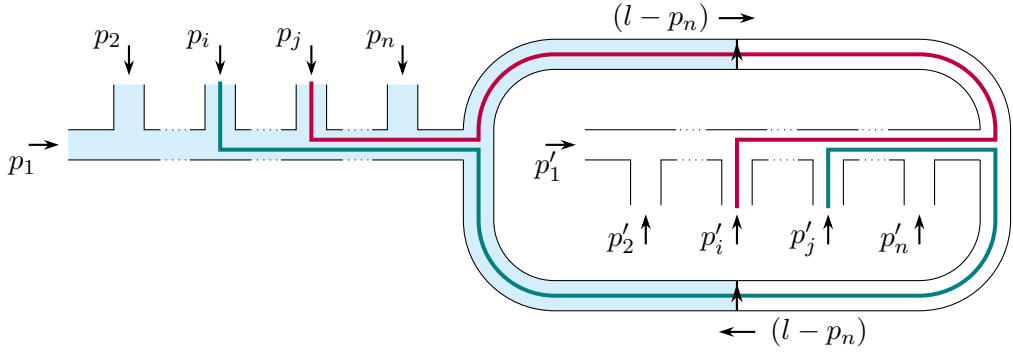
\begin{figure}[!ht]
    \centering
\begin{tikzpicture}[scale=0.4,baseline={([yshift=-1.9ex]current bounding box.center)}]
    \fill[cyan!15!white] (0,0) -- (1.5,0) -- (1.5,1.5) -- (2.5,1.5) -- (2.5,0) -- (4.5,0)  -- (4.5,1.5) -- (5.5,1.5) -- (5.5,0) -- (7.5,0) -- (7.5,1.5) -- (8.5,1.5) -- (8.5,0) -- (10.5,0) -- (10.5,1.5) -- (11.5,1.5) -- (11.5,0) -- (13,0) arc (180:90:3) -- (22,3) -- (22,2) -- (16,2) arc (90:180:2) -- (14,0) -- (14,-3) arc (180:270:2) -- (22,-5) -- (22,-6) -- (16,-6) arc (-90:-180:3) -- (13,-1) -- (0,-1) -- (0,0);
    \draw (0,0) -- (1.5,0) -- (1.5,1.5);
    \draw (2.5,1.5) -- (2.5,0) -- (3,0);
    \draw[dotted] (3,0) -- (4,0);
    \draw (4,0) -- (4.5,0) -- (4.5,1.5);
    \draw (5.5,1.5) -- (5.5,0) -- (6,0);
    \draw[dotted] (6,0) -- (7,0);
    \draw (7,0) -- (7.5,0) -- (7.5,1.5);
    \draw (8.5,1.5) -- (8.5,0) -- (9,0);
    \draw[dotted] (9,0) -- (10,0);
    \draw (10,0) -- (10.5,0) -- (10.5,1.5);
    \draw (11.5,1.5) -- (11.5,0) -- (13,0);
    \draw (0,-1) -- (3,-1);
    \draw[dotted] (3,-1) -- (4,-1);
    \draw (4,-1) -- (6,-1);
    \draw[dotted] (6,-1) -- (7,-1);
    \draw (7,-1) -- (9,-1);
    \draw[dotted] (9,-1) -- (10,-1);
    \draw (10,-1) -- (13,-1);
    \draw (13,0) arc (180:90:3) -- (28,3) arc (90:0:3) -- (31,0) -- (31,-3) arc (0:-90:3) -- (16,-6) arc (-90:-180:3) -- (13,-1);
    \draw (14,0) arc (180:90:2) -- (28,2) arc (90:0:2);
    \draw (30,-1) -- (30,-3) arc (0:-90:2) -- (27,-5) -- (16,-5) arc (-90:-180:2) -- (14,0);
    \draw (30,0) -- (27,0);
    \draw[dotted] (27,0) -- (26,0);
    \draw (26,0) -- (24,0);
    \draw[dotted] (24,0) -- (23,0);
    \draw (23,0) -- (21,0);
    \draw[dotted] (21,0) -- (20,0);
    \draw (20,0) -- (17,0);
    \draw (17,-1) -- (18.5,-1) -- (18.5,-2.5);
    \draw (19.5,-2.5) -- (19.5,-1) -- (20,-1);
    \draw[dotted] (20,-1) -- (21,-1);
    \draw (21,-1) -- (21.5,-1) -- (21.5,-2.5);
    \draw (22.5,-2.5) -- (22.5,-1) -- (23,-1);
    \draw[dotted] (23,-1) -- (24,-1);
    \draw (24,-1) -- (24.5,-1) -- (24.5,-2.5);
    \draw (25.5,-2.5) -- (25.5,-1) -- (26,-1);
    \draw[dotted] (26,-1) -- (27,-1);
    \draw (27,-1) -- (27.5,-1) -- (27.5,-2.5);
    \draw (28.5,-2.5) -- (28.5,-1) -- (30,-1);
    \draw[thick,-{Stealth[length=2mm, width=1.3mm]}] (-1.3,-0.5) -- (-0.3,-0.5);
    \draw[thick,-{Stealth[length=2mm, width=1.3mm]}] (2,2.8) -- (2,1.8);
    \draw[thick,-{Stealth[length=2mm, width=1.3mm]}] (5,2.8) -- (5,1.8);
    \draw[thick,-{Stealth[length=2mm, width=1.3mm]}] (8,2.8) -- (8,1.8);
    \draw[thick,-{Stealth[length=2mm, width=1.3mm]}] (11,2.8) -- (11,1.8);
    \draw[thick,-{Stealth[length=2mm, width=1.3mm]}] (19,-3.8) -- (19,-2.8);
    \draw[thick,-{Stealth[length=2mm, width=1.3mm]}] (22,-3.8) -- (22,-2.8);
    \draw[thick,-{Stealth[length=2mm, width=1.3mm]}] (25,-3.8) -- (25,-2.8);
    \draw[thick,-{Stealth[length=2mm, width=1.3mm]}] (28,-3.8) -- (28,-2.8);
    \draw[thick,-{Stealth[length=2mm, width=1.3mm]}] (15.7,-0.5) --(16.7,-0.5);
    \node at (-1.5,-1.2) {$p_1$};
    \node at (1.3,3) {$p_2$};
    \node at (4.3,3) {$p_i$};
    \node at (7.3,3) {$p_j$};
    \node at (10.3,3) {$p_n$};
    \node at (15.8,-1.2) {$p_1'$};
    \node at (18.2,-3.6) {$p_2'$};
    \node at (21.2,-3.6) {$p_i'$};
    \node at (24.2,-3.6) {$p_j'$};
    \node at (27.2,-3.6) {$p_n'$};
    \draw[line width=1.5, purple] (22,-2.6) -- (22,-0.33) -- (30.5,-0.33) -- (30.5,0) arc (0:90:2.5) -- (16,2.5) arc (90:180:2.5) -- (13.5,-0.33) -- (8,-0.33) -- (8,1.6);
    \draw[line width=1.5, teal] (25,-2.6) -- (25,-0.66) -- (30.5,-0.66) -- (30.5,-3) arc (0:-90:2.5) -- (16,-5.5) arc (-90:-180:2.5) -- (13.5,-0.66) -- (5,-0.66) -- (5,1.6);
    \draw[thick,-{Stealth[length=2mm, width=1.3mm]}] (21.4,3.7) -- (22.6,3.7); 
    \node at (19.5,3.7) {$(l-p_n)$};
    \draw[thick,-{Stealth[length=2mm, width=1.3mm]}] (22.6,-6.7) -- (21.4,-6.7); 
    \node at (24.7,-6.7) {$(l-p_n)$};
    \draw[thick,-{Stealth[length=2.5mm, width=1.5mm]}] (22,2) -- (22,2.9);
    \draw[thick] (22,2.5) -- (22,3);
    \draw[thick,-{Stealth[length=2.5mm, width=1.5mm]}] (22,-6) -- (22,-5.1);
    \draw[thick] (22,-5) -- (22,-6);
\end{tikzpicture}
    \caption{Finding the momenta of the curve $C_{ij}^{\ot}:$ The momenta of the doubled curves on the annulus are \textcolor{purple}{$P_{ji'}=(p_1'+\hdots+p_i')+l-p_n+(p_1+\hdots+p_{j-1})$} and \textcolor{teal}{$P_{ij'}=(p_1'+\hdots+p_j')+l-p_n+(p_1+\hdots+p_{i-1})$}. Under the identification $p_i'=p_{i-1}$, we obtain \textcolor{purple}{$P_{ji'}$}, \textcolor{teal}{$P_{ij'}$} $\to P_{ij}^{\ot} = l+(p_1+\hdots+p_{i-1})+(p_1+\hdots+p_{j-1})$~.}
    \label{fig:general-n-tadpole-doubled-momenta}
\end{figure}

\paragraph{Momentum assignment to the curves:} As of now, we have discussed mathematical objects without making much contact with scattering amplitudes. As expected, a complete triangulation of Möbius is also dual to a trivalent graph. For instance, below is a bubble for $n=2$:
\begin{align}
\label{bubble in n=2 from triangulation}
    \begin{tikzpicture}[scale=0.45,baseline={([yshift=-.5ex]current bounding box.center)}]
    \draw (0,0) circle (2);
    \node at (0,0) {$\ot$};
    \node at (0,2) {$\times$};
    \node at (0,-2) {$\times$};
    \node[gray] at (0.4,2.7) {1};
    \node[gray] at (-0.4,-2.7) {2};
    \draw[line width=1.5, purple] (0,2) -- (0,-2);
    \draw[line width=1.5, teal] (0,2) .. controls (-3,-0.7) and (3,-0.7) .. (0,2);
\end{tikzpicture}  = \qquad  \begin{tikzpicture}[scale=0.8,baseline={([yshift=-1.9ex]current bounding box.center)}]
    \fill[yellow!20] (0,0.5) .. controls (0.9,0.5) .. (1,2) .. controls (0.8,1.5) .. (0,1.5);
    \fill[yellow!20] (1,2) .. controls (1.1,0.5) .. (3,0.5) -- (3,1.5) .. controls (2,1.5) .. (2,2) -- (1,2);
    \fill[red!10] (3,1.5) .. controls (2,1.5) .. (2,2) -- (3,2) -- (3,1.5);
    \fill[red!10] (0,1.5) -- (0,2) -- (1,2) .. controls (0.8,1.5) .. (0,1.5);
    \fill[red!10] (1,2) .. controls (1.1,0.5) .. (3,0.5) -- (3,0) -- (0,0) -- (0,0.5) .. controls (0.9,0.5) .. (1,2);
    \draw[-{Stealth[length=2mm, width=1.5mm]}] (0,0) -- (0,1.2);
    \draw (0,0.95) -- (0,2) -- (3,2) -- (3,0.95);
    \draw [-{Stealth[length=2mm, width=1.5mm]}] (3,2) -- (3,0.8);
    \draw (3,0) -- (3,1.05);
    \draw (0,0) -- (3,0);
    \node at (1,2) {$\times$};
    \node at (2,2) {$\times$};
    \draw[line width=1.5, purple] (0,0.5) .. controls (0.9,0.5) .. (1,2);
    \draw[line width=1.5, purple] (2,2) .. controls (2,1.5) .. (3,1.5);
    \draw[line width=1.5, teal] (0,1.5) .. controls (0.8,1.5) .. (1,2);
    \draw[line width=1.5, teal] (1,2) .. controls (1.1,0.5) .. (3,0.5);
    \node[gray] at (1,2.5) {$1$};
    \node[gray] at (2,2.5) {$2$};
\end{tikzpicture} \quad \to \quad  \begin{tikzpicture}[scale=0.8,baseline={([yshift=-.8ex]current bounding box.center)}]
    \fill[yellow!20] (0,0.5) .. controls (0.9,0.5) .. (1,2) .. controls (0.8,1.5) .. (0,1.5);
    \fill[yellow!20] (1,2) .. controls (1.1,0.5) .. (3,0.5) -- (3,1.5) .. controls (2,1.5) .. (2,2) -- (1,2);
    \fill[red!10] (3,1.5) .. controls (2,1.5) .. (2,2) -- (3,2) -- (3,1.5);
    \fill[red!10] (0,1.5) -- (0,2) -- (1,2) .. controls (0.8,1.5) .. (0,1.5);
    \fill[red!10] (1,2) .. controls (1.1,0.5) .. (3,0.5) -- (3,0) -- (0,0) -- (0,0.5) .. controls (0.9,0.5) .. (1,2);
    \draw[-{Stealth[length=2mm, width=1.5mm]}] (0,0) -- (0,1.2);
    \draw (0,0.95) -- (0,2) -- (3,2) -- (3,0.95);
    \draw [-{Stealth[length=2mm, width=1.5mm]}] (3,2) -- (3,0.8);
    \draw (3,0) -- (3,1.05);
    \draw (0,0) -- (3,0);
    \node at (1,2) {$\times$};
    \node at (2,2) {$\times$};
    \draw[line width=1, purple] (0,0.5) .. controls (0.9,0.5) .. (1,2);
    \draw[line width=1, purple] (2,2) .. controls (2,1.5) .. (3,1.5);
    \draw[line width=1, teal] (0,1.5) .. controls (0.8,1.5) .. (1,2);
    \draw[line width=1, teal] (1,2) .. controls (1.1,0.5) .. (3,0.5);
    \node[gray] at (0.8,2.4) {$1$};
    \node[gray] at (2.2,2.4) {$2$};
    \draw[line width=1.6] (1.1,-0.6) -- (1.1,0.4) .. controls (0.2,0.5) and (1,1) .. (0,1.3);
    \draw[line width=1.6] (1.5,2.6) .. controls (1.6,1) .. (3,0.7);
    \draw[line width=1.6] (1.1,0.4) .. controls (1.6,1) .. (1.7,1.2);
    \node at (1.5,3) {$\widetilde{1}$};
    \node at (1.1,-0.9) {$\widetilde{2}$};
\end{tikzpicture}
\quad \cong \quad  
\begin{tikzpicture}[scale=0.55,baseline={([yshift=-.0ex]current bounding box.center)}]
\draw[line width=1.6] (0,0) circle (1);
\draw[line width=1.6] (0,1) -- (0,2.2);
\draw[line width=1.6] (0,-1) -- (0,-2.2);
    \node at (0.4,2.5) {$\tilde{1}$};
    \node at (-0.4,-2.5) {$\tilde{2}$};
\end{tikzpicture} ~.
\end{align}
We associate momenta with the curves on the Möbius strip. The usual rule to associate momentum with a curve on an orientable surface uses handedness and can not be used for a Möbius strip. We obtain the momentum by resorting to the doubled annulus. A curve on the Möbius strip gets mapped to two curves on the doubled annulus. Initially, we let the doubled insertions on the annulus have arbitrary momenta $\{p_i'\}$, different from the original momenta $\{p_i\}$ on the Möbius strip. We determine the momenta $\{p_i'\}$ in terms of $\{p_i\}$ by demanding that the momenta of the two doubled curves should match. To concretize the proposal, and keeping in mind the expected \emph{telescopic property} of the curve integral, we draw a tadpole graph with explicit momenta assignments in Figure \ref{fig:3-pt-tadpole-doubled-momenta}, and double it to obtain the annulus. Note that the momentum associated with a curve does not depend on any reference triangulation. We have made an odd choice for loop momenta, $l-p_3$, for later convenience. We can draw all the curves and their doubled ones on the annulus, calculate the momenta, and find an appropriate projection on the momenta as follows:
\begin{align}
    p_{i}'=p_{i-1}~.
\end{align}
It should be understood that $p_1'=p_n~.$ Analogous to the doubling of a tadpole graph for $n=3$ in Figure \ref{fig:3-pt-tadpole-doubled-momenta}, we depict the doubled annulus obtained from the doubling of a tadpole graph of a Möbius strip with an arbitrary number of marked points in Figure \ref{fig:general-n-tadpole-doubled-momenta}. Calculating the momenta for doubled curves, and using the identification $p_{i}'=p_{i-1}$, we obtain the following:
\begin{align}
    \label{momentum assignment to Pikot}
    P_{ij}^{\ot} = l+(p_1+\hdots+p_{i-1})+(p_1+\hdots+p_{j-1}) ~.
\end{align}
$P_{ij}^{\ot}$ is symmetric in $i$ and $j$, as it should be. Had we chosen the loop momentum on the tadpole leg as $l$ rather than $l-p_n$, we would have obtained $P_{ij}^{\ot} = l+p_n+(p_1+\hdots+p_{i-1})+(p_1+\hdots+p_{j-1}) ~.$ $C_{ij}^{\ot}$ (and $C^{\ot}$) are the only curves carrying loop momenta. While dealing with the one-loop kinematic space in the planar case, we usually double up the external momenta and impose the identification at the end, as done in \cite{mrunmay2304, mrunmay2206}, etc. The doubling of the kinematic variables is distinct from the doubling of non-orientable surfaces onto the orientable ones, discussed in this section. The kinematic space constructed here: $\{X_{ik},~X_{ik}^{\ot}\}$ needs to be doubled further, along the lines of \cite{mrunmay2304, mrunmay2206}, etc., to have a kinematic space suitable for tadpoles and bubbles on external legs. Though we do not pursue this in this manuscript. 

Let us perform a cross-check on our momenta assignments to the curves $C_{ik}^{\ot}~.$ Let us choose $n=3$, and draw a complete triangulation corresponding to a triangle graph, as shown in Figure \ref{fig:momenta-determination-Mobius3}. Let us choose a momenta $P_{11}^{\ot}\equiv l$ as the definition of our $l~.$ This fixes momentum assignments to other curves in the same triangulation. Mutating once gets us to another triangulation with a single different curve. Momentum conservation in the new triangulation gives us the momentum assignment for the new curve in the new triangulation. Following the pattern, we can find all the $P_{ik}^{\ot}:$
\begin{equation}
\begin{aligned}
    P_{11}^{\ot} &= l~, \quad & \quad P_{12}^{\ot} &= l+p_1~, \\
    P_{22}^{\ot} &= l+2p_1~, \quad & \quad P_{23}^{\ot} &= l+p_1-p_3~,  \\
    P_{33}^{\ot} &= l-2p_3~, \quad & \quad P_{31}^{\ot} &= l-p_3~.
\end{aligned}    
\end{equation}
It agrees with our earlier answer \eqref{momentum assignment to Pikot}.

The momenta associated with the curves $C_{ij}$ are the same as the planar one-loop case. Since these curves steer clear of the cross-cap, even if we replace the cross-cap with another boundary component, they remain unaffected, and their momenta also remain the same. So, we have:
\begin{align}
    P_{ij} = p_i + p_{i+1} + \hdots + p_{j-1}~.
\end{align}
The curve $C^{\ot}$ appears in the tadpole as the loop leg, hence the associated momentum $P^{\ot}$ should carry loop momentum. $P^{\ot}$ does not carry any particle label, and hence it must be independent of any external particle label. So, it should be $P^{\ot}\propto l~.$ It is not possible to determine this momentum via mutations, as done in Figure \ref{fig:momenta-determination-Mobius3}. Let us reiterate that these quasi-triangulations with $C^{\ot}$, all of them being tadpoles, behave in a non-standard way, and we will find a consistent way to avoid these. 

\paragraph{Headlight functions}
We have embedded the Möbius strip with $n$ marked points on its boundary into an annulus with $n$ marked points on either of its boundaries. With the 'wheel' graph as the reference triangulation, we have realized how the global Schwinger parameters for the Möbius strip are embedded into the global Schwinger parameters for the doubled annulus: $t_i'+t_i=0~.$ We can go ahead and write down the headlight functions $\alpha_C$ for the doubled annulus and project them onto the $t_i'+t_i=0$ space to obtain the headlight functions for the curves on the Möbius strip. For the $g$-vectors calculated with respect to the wheel graph as the reference triangulation in Figure \ref{fig:g-vector fan for Möbius1,2}, we calculate the headlight functions to obtain the following:
\begin{equation}
\begin{aligned}
    \text{For } n=1: \qquad \vec{g}^{\ot} &= -1~, &\quad \vec{g}^{\ot}_{11} &= +1~; \\ 
    \alpha^{\ot}(t) &= \mxx{0,t}-t ~, & \alpha^{\ot}_{11}(t) &= \mxx{0,t}~.
\end{aligned}    
\end{equation}
For $n=2$, we have:
\begin{equation}
    \begin{aligned}
        \vec{g}_{11} &= (1,-1) ~~, \hspace{2cm} \vec{g}_{22} = (-1,-1) ~~, \\
        \vec{g}_{12}^{\otimes} &= (0,1) ~~, \hspace{2.3cm} \vec{g}^{\otimes}_{11} = (1,0)~~, \\
\vec{g}_{22}^{\ot} &= (-1,0) ~~, \hspace{2cm} \vec{g}^{\ot} = (0,-1)~.
    \end{aligned}
    \label{g-vectors for n=2 with wheel as reference}
\end{equation}
\begin{equation}
\begin{aligned}
    \alpha_{11} &= \mxx{0,t_2} + 2\, \mxx{0,t_1} - \mxx{0,\,t_1,\,t_2+2\,\mxx{0,t_1}} ~, \\
    \alpha_{22} &= \mxx{0,t_2} + 2\, \mxx{0,t_1} - \mxx{t_2,\,t_1+\mxx{0,t_1}+\mxx{0,t_2}}~, \\
    \alpha^{\otimes}_{12} &= \mxx{0,t_2} ~, \\
    \alpha^{\otimes}_{11} &= \mxx{0,\,t_2,\,t_1+t_2} - \mxx{0,t_2}~, \\
    \alpha^{\otimes}_{22} &= \mxx{t_1,\,t_2,\,t_1+t_2} -t_1- \mxx{0,t_2}~, \\
    \alpha^{\otimes} &= \mxx{t_1,\,t_2+\mxx{0,t_2} + 2\,\mxx{0,t_1}} - t_2 - \mxx{0,t_2} - 2\,\mxx{0,t_1}~.
\end{aligned}    
\label{headlight functions for n=2 with wheel as reference}
\end{equation}
We write down the explicit headlight functions for $n=3$ obtained similarly, in the Appendix \eqref{headlights for n=3 with wheel as reference}.

We have calculated the momenta associated with the curves and their corresponding $g$-vectors and headlight functions. There is no non-trivial Mapping Class Group for the Möbius strip. With this much data, we can write down the curve integral formula:

{  
\begin{align}
    A_n^{\text{Möbius}}=\int \dd^n t_i\int \dd^Dl\,e^{-\sum_C \alpha_CP_C^2}~.
\end{align}
The sum in the exponent is over all the curves; $\alpha_C$ and $P_C$ are the headlight functions and momenta associated with the curve $C$.}\footnote{{   The momenta squared: $P_C^2$, i.e., the generalized Mandelstam variables, are also often represented by $X_C(\equiv P_C^2)$.}} We carry out the explicit loop integration in the Appendix \eqref{explicit loop integration 1} - \eqref{explicit loop integration 3}~, and express the curve integral in terms of the surface Symanzik polynomials. We discuss this further in Section \ref{section 4 surface symanzik}, where we construct these surface Symanzik polynomials from \emph{spanning subsurfaces}. 

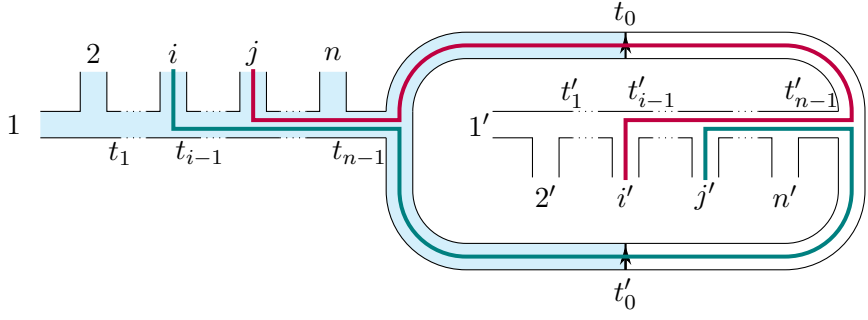
\begin{figure}[!t]
    \centering
\begin{tikzpicture}[scale=0.35,baseline={([yshift=-1.9ex]current bounding box.center)}]
    \fill[cyan!15!white] (0,0) -- (1.5,0) -- (1.5,1.5) -- (2.5,1.5) -- (2.5,0) -- (4.5,0)  -- (4.5,1.5) -- (5.5,1.5) -- (5.5,0) -- (7.5,0) -- (7.5,1.5) -- (8.5,1.5) -- (8.5,0) -- (10.5,0) -- (10.5,1.5) -- (11.5,1.5) -- (11.5,0) -- (13,0) arc (180:90:3) -- (22,3) -- (22,2) -- (16,2) arc (90:180:2) -- (14,0) -- (14,-3) arc (180:270:2) -- (22,-5) -- (22,-6) -- (16,-6) arc (-90:-180:3) -- (13,-1) -- (0,-1) -- (0,0);
    \draw (0,0) -- (1.5,0) -- (1.5,1.5);
    \draw (2.5,1.5) -- (2.5,0) -- (3,0);
    \draw[dotted] (3,0) -- (4,0);
    \draw (4,0) -- (4.5,0) -- (4.5,1.5);
    \draw (5.5,1.5) -- (5.5,0) -- (6,0);
    \draw[dotted] (6,0) -- (7,0);
    \draw (7,0) -- (7.5,0) -- (7.5,1.5);
    \draw (8.5,1.5) -- (8.5,0) -- (9,0);
    \draw[dotted] (9,0) -- (10,0);
    \draw (10,0) -- (10.5,0) -- (10.5,1.5);
    \draw (11.5,1.5) -- (11.5,0) -- (13,0);
    \draw (0,-1) -- (3,-1);
    \draw[dotted] (3,-1) -- (4,-1);
    \draw (4,-1) -- (6,-1);
    \draw[dotted] (6,-1) -- (7,-1);
    \draw (7,-1) -- (9,-1);
    \draw[dotted] (9,-1) -- (10,-1);
    \draw (10,-1) -- (13,-1);
    \draw (13,0) arc (180:90:3) -- (28,3) arc (90:0:3) -- (31,0) -- (31,-3) arc (0:-90:3) -- (16,-6) arc (-90:-180:3) -- (13,-1);
    \draw (14,0) arc (180:90:2) -- (28,2) arc (90:0:2);
    \draw (30,-1) -- (30,-3) arc (0:-90:2) -- (27,-5) -- (16,-5) arc (-90:-180:2) -- (14,0);
    \draw (30,0) -- (27,0);
    \draw[dotted] (27,0) -- (26,0);
    \draw (26,0) -- (24,0);
    \draw[dotted] (24,0) -- (23,0);
    \draw (23,0) -- (21,0);
    \draw[dotted] (21,0) -- (20,0);
    \draw (20,0) -- (17,0);
    \draw (17,-1) -- (18.5,-1) -- (18.5,-2.5);
    \draw (19.5,-2.5) -- (19.5,-1) -- (20,-1);
    \draw[dotted] (20,-1) -- (21,-1);
    \draw (21,-1) -- (21.5,-1) -- (21.5,-2.5);
    \draw (22.5,-2.5) -- (22.5,-1) -- (23,-1);
    \draw[dotted] (23,-1) -- (24,-1);
    \draw (24,-1) -- (24.5,-1) -- (24.5,-2.5);
    \draw (25.5,-2.5) -- (25.5,-1) -- (26,-1);
    \draw[dotted] (26,-1) -- (27,-1);
    \draw (27,-1) -- (27.5,-1) -- (27.5,-2.5);
    \draw (28.5,-2.5) -- (28.5,-1) -- (30,-1);
    \draw[thick,-{Stealth[length=2.5mm, width=1.5mm]}] (22,2) -- (22,2.9);
    \draw[thick] (22,2.5) -- (22,3);
    \draw[thick,-{Stealth[length=2.5mm, width=1.5mm]}] (22,-6) -- (22,-5.1);
    \draw[thick] (22,-5) -- (22,-6);
    \node at (-1,-0.5) {1};
    \node at (2,2.2) {2};
    \node at (5,2.2) {$i$};
    \node at (8,2.2) {$j$};
    \node at (11,2.2) {$n$};
    \node at (3,-1.6) {$t_1$};
    \node at (6,-1.6) {$t_{i-1}$};
    \node at (12,-1.6) {$t_{n-1}$};
    \node at (16.5,-0.5) {$1'$};
    \node at (19,-3.2) {$2'$};
    \node at (22,-3.2) {$i'$};
    \node at (25,-3.2) {$j'$};
    \node at (28,-3.2) {$n'$};
    \node at (20,0.7) {$t_1'$};
    \node at (23,0.7) {$t_{i-1}'$};
    \node at (29,0.7) {$t_{n-1}'$};
    \node at (22,3.7) {$t_0$};
    \node at (22,-6.9) {$t_0'$};
    \draw[line width=1.5, purple] (22,-2.6) -- (22,-0.33) -- (30.5,-0.33) -- (30.5,0) arc (0:90:2.5) -- (16,2.5) arc (90:180:2.5) -- (13.5,-0.33) -- (8,-0.33) -- (8,1.6);
    \draw[line width=1.5, teal] (25,-2.6) -- (25,-0.66) -- (30.5,-0.66) -- (30.5,-3) arc (0:-90:2.5) -- (16,-5.5) arc (-90:-180:2.5) -- (13.5,-0.66) -- (5,-0.66) -- (5,1.6);
\end{tikzpicture}
    \caption{The Schwinger parameters on the doubled annulus and the two curves obtained from doubling of $C_{ij}^{\ot}$ are shown.}
    \label{fig:Schwinger parameters for generic n with tadpole as reference}
\end{figure}

\paragraph{Telescopic property} Let us perform the analysis of calculating the $g$-vectors and the headlight functions with a different reference triangulation, namely a tadpole depicted in Figure \ref{fig:general-n-tadpole-doubled-momenta}. It will help us understand the properties of the curve integral that are common to all $n$. Figure \ref{fig:Schwinger parameters for generic n with tadpole as reference} depicts the Schwinger parameter labels on the doubled annulus graph. 

We employ the same rule that the $g$-vectors of the two curves on the doubled annulus should map to the same vector under the projection. The same projection $t_i'+t_i=0 ~, \ \ \forall \ i\in[0,(n-1)]$ works. To demonstrate it, following Figure \ref{fig:Schwinger parameters for generic n with tadpole as reference}, let us find the $g$-vector of $C_{ij}^{\ot}$ ($i,j\neq 1,n$) for generic $n$. 
\begin{align}
    \begin{matrix}
        \textcolor{purple}{ \vec{g}^{\text{aux}}_{i'j} =\big(\stackrel{t_0}{1},\hdots, \stackrel{t_{j-1}}{1},\hdots,\stackrel{t_{n-1}}{-1}~;~\hdots,\stackrel{t'_{i-1}}{-1},\hdots\big) } \\ \quad \\ 
        \textcolor{teal}{ \vec{g}^{\text{aux}}_{ij'} = \big(\hdots,\stackrel{t_{i-1}}{1},\hdots~;~\stackrel{t'_0}{-1},\hdots,\stackrel{t'_{j-1}}{-1},\hdots,\stackrel{t'_{n-1}}{1}\big)}
    \end{matrix} \ \Rightarrow \ \begin{matrix}\vec{g}_{ij}^{\ot} = \big(\stackrel{t_0}{1},\hdots,\stackrel{t_{i-1}}{1},\hdots,\stackrel{t_{j-1}}{1},\hdots,\stackrel{t_{n-1}}{-1}\big)  \\  {}_{(i,j \neq 1,n)} \end{matrix}~.
    \label{projection of random two curves}
\end{align}
{  Refer to Appendix \ref{appendix A review}, \eqref{curves from figure 9} for some more details on paths traversed by the doubled curves and their resultant $g$-vectors. } We have written only the non-zero components of the vectors, and all the omitted ones, denoted as $\hdots$, are zero. For $i,j$ taking the values $1$ or $n$, the $g$-vectors are changed to be the following:
\begin{equation}
    \begin{aligned}
    \vec{g}_{11}^{\ot} &= \big(\stackrel{t_0}{1},\hdots,\stackrel{t_{n-1}}{-1}\big) ~, \\
    \vec{g}_{1j}^{\ot} &= \big(\stackrel{t_0}{1},\hdots,\stackrel{t_{j-1}}{1},\hdots,\stackrel{t_{n-1}}{-1}\big) \qquad j\in[2,\hdots,n-1] ~, \\
    \vec{g}_{1n}^{\ot} &= \big(\stackrel{t_0}{1},\hdots\big) ~, \\
    \vec{g}_{in}^{\ot} &= \big(\stackrel{t_0}{1},\hdots,\stackrel{t_{i-1}}{1},\hdots\big) \quad \qquad i\in[2,\hdots,n]~.
\end{aligned}
\end{equation}

The curves $C_{ij}$ (and the doubled curves on the annulus) never take any turn at $t_0$ (and $t_0'$). They trace out the same path as the curves $C_{ij}$ in the case of the planar one-loop case, and thus have the same $g$-vectors:  
    \begin{align}
        \text{For } k\in[3,n+1]~:\qquad \vec{g}_{1k} &= \big(\hdots, \stackrel{t_{k-2}}{-1},\hdots \big) ~,\nonumber\\
        \text{For } i\in[2,n-1]~, ~~k\in[i+2,n+1]~:\qquad \vec{g}_{ik} &= \big(\hdots,\stackrel{t_{i-1}}{1},\hdots,\stackrel{t_{k-2}}{-1},\hdots\big) ~, \nonumber\\
        \text{For } i\in[2,n-1]~, ~~k\in[n+2,n+i]~:\qquad \vec{g}_{ik} &= \big(\hdots,\stackrel{t_{i-1}}{1},\hdots,\stackrel{t_{j-1}}{1},\hdots,\stackrel{t_{n-1}}{-1},\hdots\big) ~, \nonumber\\
        \text{For } k\in[2,n]~:\qquad \vec{g}_{nk} &= \big(\hdots, \stackrel{t_{k-1}}{1},\hdots \big) ~.
    \end{align}
Note that the indices $i,k$ are defined modulo $n$, so that $C_{1(n+1)}\equiv C_{11}~.$ Finally, inspecting the closed $\Delta$ curve in the doubled annulus, we find the $g$-vector for the $C^{\ot}$:
\begin{align}
    \vec{g}^{\ot} = \big(\stackrel{t_0}{-1},\hdots\big)~.
\end{align}
We draw the $g$-vector fan for $n=1$ and $n=2$ in Figure \ref{fig:g-vector fan for Möbius1,2 with tadpole reference} and write down the $g$-vectors for $n=3$ in the Appendix. 

\begin{figure}[h]
    \centering
\begin{align*}
    (n=1): \ \ \begin{tikzpicture}[scale=0.7,baseline={([yshift=-1.9ex]current bounding box.center)}]
    \draw[<->] (-2.5,0) -- (2.5,0);
    \draw[line width=1, purple!50!black, -{Stealth[length=3mm, width=2mm]}] (0,0) -- (-2,0);
    \node[purple!50!black] at (-1.9,0.5) {$C^{\ot}$};
    \draw[line width=1, purple!50!black, -{Stealth[length=3mm, width=2mm]}] (0,0) -- (2,0);
    \node[purple!50!black] at (2,0.5) {$C^{\ot}_{11}$};
    \draw (0,-0.2) -- (0,0.2);
    \node at (2.9,0) {$t$};
\end{tikzpicture} \qquad \quad (n=2): \ \ \begin{tikzpicture}[scale=0.7,baseline={([yshift=-1.9ex]current bounding box.center)}]
    \draw[<->] (-2.5,0) -- (2.5,0);
    \draw[<->] (0,-2.5) -- (0,2.5);
    \draw[line width=1, purple!50!black, -{Stealth[length=3mm, width=2mm]}] (0,0) -- (0,-2);
    \node[purple!50!black] at (-0.6,-2) {$C_{11}$};
    \draw[line width=1, purple!50!black, -{Stealth[length=3mm, width=2mm]}] (0,0) -- (-2,0);
    \node[purple!50!black] at (-2,0.5) {$C^{\ot}$};
    \draw[line width=1, purple!50!black, -{Stealth[length=3mm, width=2mm]}] (0,0) -- (2,0);
    \node[purple!50!black] at (2,0.5) {$C^{\ot}_{12}$};
    \draw[line width=1, purple!50!black, -{Stealth[length=3mm, width=2mm]}] (0,0) -- (0,2);
    \node[purple!50!black] at (-0.6,2.1) {$C_{22}$};
    \draw[line width=1, purple!50!black, -{Stealth[length=3mm, width=2mm]}] (0,0) -- (2,-2);
    \node[purple!50!black] at (2.25,-2.25) {$C^{\ot}_{11}$};
    \draw[line width=1, purple!50!black, -{Stealth[length=3mm, width=2mm]}] (0,0) -- (2,2);
    \node[purple!50!black] at (2.25,2.25) {$C^{\ot}_{22}$};
    \node at (3,0) {$t_0$};
    \node at (0.4,2.6) {$t_1$};
\end{tikzpicture}
\end{align*}
    \caption{$g$-vector fans for the Möbius strip with one or two marked points on its boundary, with a tadpole graph as the reference triangulation.}
    \label{fig:g-vector fan for Möbius1,2 with tadpole reference}
\end{figure}
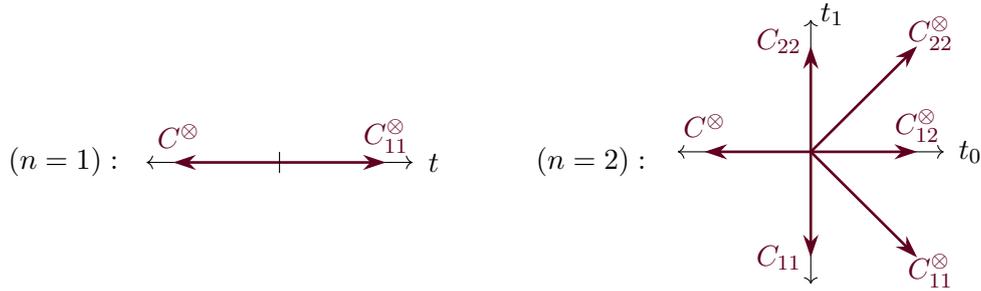

Since the curves on the Möbius strip and the doubled annulus are going to be 'tree-curves' attached to the tadpole, the curve matrices and hence the headlight functions can be evaluated using the information of tree-curves. \textbf{All the curves not carrying any loop momentum, namely the $C_{ij}$ curves, lie on a co-dimension one plane: $t_0=0~.$} It is a common feature in the annulus and Möbius strip. However, the $g$-vectors for $C_{ij}$ curves do not lie on a co-dimension one space for the reference triangulation 'wheel' graph \ref{fig:g-vector fan for Möbius1,2}. The tadpole reference triangulation simplifies the surface Symanzik polynomial: $\mathcal{U}=t_0$, as we shall see in the next section. Also, the only curve with $t_0<0$ is $C^{\ot}$ for all $n~.$ \textbf{We restrict ourselves to the $t_0>0$ region to avoid tadpoles arising from quasi-triangulations. }

\section{Constructing surface Symanzik polynomials}
\label{section 4 surface symanzik}
In this section, we briefly review the construction of surface Symanzik polynomials from simple topological considerations, and repeat the same for the Mobius strip. 

\paragraph{Definition of surface Symanzik polynomials:}
A generic curve integral for $n$ external particles,  associated with a surface $S$, prior to the loop momenta integration is as follows:
\begin{align}
    A_n^{S}=\int  \bigwedge\dd^{E} t_i\,\mathcal{K}(t_i)\int \prod_i\dd^Dl_i\,e^{-\sum_C \alpha_CX_C}~,
\end{align}
where $E$ is the dimension of the Schwinger space (number of internal edges), $D$ is the number of spacetime dimensions, and $\mathcal{K}(t_i)$ is the MCG fixing kernel. One can carry out the loop momenta to obtain the following form of the curve integral:
\begin{align}
    A_n^{S}=\int  \bigwedge\dd^{E} t_i\,\mathcal{K}(t_i)\,\frac{1}{\mathcal{U}_S^{D/2}}e^{-\mathcal{F}_S/\mathcal{U}_S - \mathcal{Z}_S}~.
\end{align}
$\mathcal{U}_S,~\mathcal{F}_S,~\mathcal{Z}_S$ are called the surface Symanzik polynomials. $\mathcal{U}_S$ and $\mathcal{F}_S$ are the global Schwinger generalizations of Symanzik polynomials for a particular graph. 

\paragraph{Symanzik polynomials for a graph:}
Symanzik polynomials $U_\gamma$ and $F_\gamma$ for a particular graph $\gamma$ can be constructed by simple topological considerations. Let $\{s_i\}$ be the set of Schwinger parameters for the internal edges of $\gamma$. Spanning-1 trees are defined to be the tree subgraphs $\gamma_1\subset\gamma$ obtained from the minimal deletion of edges. For instance, for a four-point box diagram, there are four spanning-1 trees obtained from deleting either one of the four loop legs. Similarly, a spanning-2 tree is defined to be a disjoint sum of two tree graphs $\gamma_1\sqcup \gamma_2 \subset \gamma$ obtained from minimal deletion of edges. We have the following:
\begin{equation}
\begin{aligned}
    U_\gamma &= \sum_{\gamma_1\in\left\{\substack{\text{spanning-} 1 \\ \text{trees}}\right\}}\prod_{s_i\in\gamma/\gamma_1} s_i~, \\
    F_{\gamma} &= \sum_{\gamma_1\sqcup \gamma_2\in\left\{\substack{\text{spanning-} 2 \\ \text{trees}}\right\}} p_{\gamma_1}.p_{\gamma_2} \prod_{s_i\in \gamma/(\gamma_1\sqcup \gamma_2)}s_i ~.
\end{aligned}
\end{equation}
$p_{\gamma_{1(2)}}$ is the sum of external momenta flowing in the subgraph $\gamma_{1(2)}~.$ These remarkable properties let us write down the Schwinger integrand for a given loop diagram without doing the actual loop integration. Refer to \cite{weinzierl, smirnovbook1, smirnovbook2} for more details. 

\paragraph{Surface Symanzik polynomials determined from spanning-surfaces}
Analogous to the Symanzik polynomials for a particular graph, the surface Symanzik polynomials can also be written in terms of surface generalizations of spanning trees \cite{Arkani-Hamed:2023CurveIntegral, Arkani-Hamed:2023Multiplicity}. We review it for the orientable surfaces with the examples of planar one and two-loop $n$-point amplitudes, and then use it for non-orientable surfaces. The examples also show the efficiency of writing down higher loop global Schwinger \emph{integrands} for the full amplitude.

Given a surface $S$, we have
\begin{itemize}
    \item A spanning-1 surface $s_1$ is defined to be a subsurface of $S$, with the inclusion of a maximal set of mutually compatible curves such that the curves do not separate $S$ into two disjoint pieces. 
    \item A spanning-2 surface $s_1\sqcup s_2$ is defined by an arbitrary intersection of two spanning-1 surfaces, such that $S$ is separated precisely into two disjoint pieces: $s_1$ and $s_2$. 
\end{itemize}
As a trivial example, for a disk with $n$ marked points on its boundary, all the curves separate it into two disjoint pieces, hence no spanning-1 surfaces. Here are the spanning surfaces for $S$ being an annulus with $n$ marked points on its boundary\footnote{To keep the analysis simple, we are restricting ourselves to half of the Schwinger region such that there is a single curve starting from an insertion and going around the puncture.}:
\begin{equation}
    \label{spanning surfaces for annulus}
    \begin{aligned}
        \text{spanning-1 surfaces:}\quad  \left\{ \ \ \begin{tikzpicture}[scale=0.5,baseline={([yshift=-.5ex]current bounding box.center)}]
    \draw (0,0) circle (1);
    \draw[fill=gray!50] (-0,0) circle (0.1);
    \node at (0,1) {\small $\times$};
    \node at (0.3,1.4) {$i$};
    \node at (-1.3,0.3) {$\vdots$};
    \node at (1.3,0.3) {$\vdots$};
    \node at (-0.707,-0.707) {\small $\times$};
    \node at (-1.05,-1.05) {$1$};
    \node at (0.707,-0.707) {\small $\times$};
    \node at (1.05,-1.05) {$n$};
    \draw (0,0) -- (0,1);
    \end{tikzpicture} \ \ \forall i\in [1,n] \ \ \right\} ~,\\
    \text{spanning-2 surfaces:}\ \  \left\{ \ \begin{tikzpicture}[scale=0.5,baseline={([yshift=-.5ex]current bounding box.center)}]
    \draw (0,0) circle (1);
    \draw[fill=gray!50] (-0,0) circle (0.1);
    \node at (0,1) {\small $\times$};
    \node at (0.4,1.5) {$i$};
    \node at (-1.3,0.3) {$\vdots$};
    \node at (1.3,0.3) {$\vdots$};
    \node at (-0,-1) {\small $\times$};
    \node at (-0.4,-1.5) {$j$};
    \draw (0,-1) -- (0,1);
    \end{tikzpicture} \ \ \begin{matrix}
        \forall\ (i,j) \in [1,n]\\ i\neq j
    \end{matrix} \ \right\} ~.
    \end{aligned}
\end{equation}
For the case of a planar two-loop with an arbitrary $n$ marked points on a boundary, the spanning surfaces are as follows\footnote{Similar to the one-loop case, we are considering a smaller region of Schwinger space such that only one curve joins a puncture to the external marked point, for a fixed region in MCG orbit. There is a sum over all the MCG inequivalent curves with the same end-points, which we are not writing explicitly here for brevity.}:

\begin{equation}
\begin{aligned}
        &\text{spanning-1 surfaces:}\quad \left\{
\begin{tikzpicture}[scale=0.6,baseline={([yshift=-.5ex]current bounding box.center)}]
    \draw (0,0) circle (1);
    \draw[fill=gray!50] (-0.4,0) circle (0.1);
    \draw[fill=gray!50] (0.4,0) circle (0.1);
    \node at (0,1) {\small $\times$};
    \node at (0,-1) {\small $\times$};
    \node at (0.3,1.4) {$i$};
    \node at (-0.3,-1.4) {$j$};
    \node at (-1.3,0.3) {$\vdots$};
    \node at (1.3,0.3) {$\vdots$};
    \draw (0,1) -- (-0.4,0) -- (0.4,0);
\end{tikzpicture} ~,~~
\begin{tikzpicture}[scale=0.6,baseline={([yshift=-.5ex]current bounding box.center)}]
    \draw (0,0) circle (1);
    \draw[fill=gray!50] (-0.4,0) circle (0.1);
    \draw[fill=gray!50] (0.4,0) circle (0.1);
    \node at (0,1) {\small $\times$};
    \node at (0,-1) {\small $\times$};
    \node at (0.3,1.4) {$i$};
    \node at (-0.3,-1.4) {$j$};
    \node at (-1.3,0.3) {$\vdots$};
    \node at (1.3,0.3) {$\vdots$}; 
    \draw (0,-1) -- (-0.4,0) -- (0.4,0);
\end{tikzpicture}~,~~
\begin{tikzpicture}[scale=0.6,baseline={([yshift=-.5ex]current bounding box.center)}]
    \draw (0,0) circle (1);
    \draw[fill=gray!50] (-0.4,0) circle (0.1);
    \draw[fill=gray!50] (0.4,0) circle (0.1);
    \node at (0,1) {\small $\times$};
    \node at (0,-1) {\small $\times$};
    \node at (0.3,1.4) {$i$};
    \node at (-0.3,-1.4) {$j$};
    \node at (-1.3,0.3) {$\vdots$};
    \node at (1.3,0.3) {$\vdots$}; 
    \draw (0,1) -- (-0.4,0);
    \draw (0.4,0) -- (0,-1);
\end{tikzpicture} \right\}  \\
&\text{spanning-2 surfaces:}\quad  \\
&\left\{
\begin{tikzpicture}[scale=0.6,baseline={([yshift=-.5ex]current bounding box.center)}]
    \draw (0,0) circle (1);
    \draw[fill=gray!50] (-0.4,0) circle (0.1);
    \draw[fill=gray!50] (0.4,0) circle (0.1);
    \node at (0,1) {\small $\times$};
    \node at (0,-1) {\small $\times$};
    \node at (0.3,1.4) {$i$};
    \node at (-0.3,-1.4) {$j$};
    \node at (-1.3,0.3) {$\vdots$};
    \node at (1.3,0.3) {$\vdots$}; 
    \node at (0.707,-0.707) {\small $\times$};
    \node at (1.05,-1.05) {$k$};
    \draw (0,1) -- (-0.4,0) -- (0,-1);
    \draw (0.4,0) -- (0.707,-0.707);
\end{tikzpicture} ~,~~
\begin{tikzpicture}[scale=0.6,baseline={([yshift=-.5ex]current bounding box.center)}]
    \draw (0,0) circle (1);
    \draw[fill=gray!50] (-0.4,0) circle (0.1);
    \draw[fill=gray!50] (0.4,0) circle (0.1);
    \node at (0,1) {\small $\times$};
    \node at (0,-1) {\small $\times$};
    \node at (0.3,1.4) {$i$};
    \node at (-0.3,-1.4) {$j$};
    \node at (-1.3,0.3) {$\vdots$};
    \node at (1.3,0.3) {$\vdots$}; 
    \node at (-0.707,-0.707) {\small $\times$};
    \node at (-1.05,-1.05) {$k$};
    \draw (0,1) -- (0.4,0) -- (0,-1);
    \draw (-0.4,0) -- (-0.707,-0.707);
\end{tikzpicture}~,~~
\begin{tikzpicture}[scale=0.6,baseline={([yshift=-.5ex]current bounding box.center)}]
    \draw (0,0) circle (1);
    \draw[fill=gray!50] (-0.4,0) circle (0.1);
    \draw[fill=gray!50] (0.4,0) circle (0.1);
    \node at (0,1) {\small $\times$};
    \node at (0,-1) {\small $\times$};
    \node at (0.3,1.4) {$i$};
    \node at (-0.3,-1.4) {$j$};
    \node at (-1.3,0.3) {$\vdots$};
    \node at (1.3,0.3) {$\vdots$}; 
    \draw (0,1) -- (0.4,0) -- (-0.4,0);
    \draw (0.4,0) -- (0,-1);
\end{tikzpicture}~,~~
\begin{tikzpicture}[scale=0.6,baseline={([yshift=-.5ex]current bounding box.center)}]
    \draw (0,0) circle (1);
    \draw[fill=gray!50] (-0.4,0) circle (0.1);
    \draw[fill=gray!50] (0.4,0) circle (0.1);
    \node at (0,1) {\small $\times$};
    \node at (0,-1) {\small $\times$};
    \node at (0.3,1.4) {$i$};
    \node at (-0.3,-1.4) {$j$};
    \node at (-1.3,0.3) {$\vdots$};
    \node at (1.3,0.3) {$\vdots$}; 
    \draw (0,1) -- (-0.4,0) -- (0.4,0);
    \draw (-0.4,0) -- (0,-1);
\end{tikzpicture}~,~~
\begin{tikzpicture}[scale=0.6,baseline={([yshift=-.5ex]current bounding box.center)}]
    \draw (0,0) circle (1);
    \draw[fill=gray!50] (-0.4,0) circle (0.1);
    \draw[fill=gray!50] (0.4,0) circle (0.1);
    \node at (0,1) {\small $\times$};
    \node at (0,-1) {\small $\times$};
    \node at (0.3,1.4) {$i$};
    \node at (-0.3,-1.4) {$j$};
    \node at (-1.3,0.3) {$\vdots$};
    \node at (1.3,0.3) {$\vdots$}; 
    \draw (0,1) -- (-0.4,0) -- (0.4,0) -- (0,-1);
\end{tikzpicture} \right\} 
\end{aligned}
\label{spanning surfaces for two loop planar}
\end{equation}

The surface Symanzik polynomials in terms of the spanning surfaces are given as follows:
\begin{align}
    \mathcal{U}_S &= \sum_{s_1\in\left\{\substack{\text{spanning-} 1 \\ \text{surfaces}}\right\}}\prod_{C_i\in S/s_1} \alpha_{C_i}~, \\
    \mathcal{F}_{S} &= \sum_{s_1\sqcup s_2\in\left\{\substack{\text{spanning-} 2 \\ \text{surfaces}}\right\}} p_{s_1}.p_{s_2} \prod_{C_i\in S/(s_1\sqcup s_2)}\alpha_{C_i} ~.
\end{align}
Using these definitions, we can readily write down the $\mathcal{U}$ and $\mathcal{F}$ for any surface without actually carrying out the loop integral. In both cases, for spanning-2 surfaces, the disjoint surfaces have incoming momenta in such a way that $p_{s_1}.p_{s_2}=X_{ij}~,$ the usual planar kinematic variable. Hence, for one-loop, we have\footnote{We have restricted ourselves to on-shell external particles, hence $n>3~,$ and $X_{ii}=X_{i,i+1}=0$~.}:
\begin{align}
    \mathcal{U}_{\text{planar},L=1} = \sum_{i=1}^n \alpha_{C_i'} ~, \qquad \mathcal{F}_{\text{planar},L=1} = \sum_{i+1<j}X_{ij} \,\alpha_{C_i'}\alpha_{C_j'}~.
\end{align}
For the two-loop, let us denote the two punctures as ${}_{a,b}$, and we have (terms arranged as in \eqref{spanning surfaces for two loop planar}):
\begin{equation}
    \begin{aligned}
        \mathcal{U}_{\text{pl},L=2} &= \alpha_{ab}\sum_{i=1}^n \alpha_{C_{ai}} + \alpha_{ab}\sum_{i=1}^n \alpha_{C_{bi}} + \sum_{i=1}^n \alpha_{C_{ai}}\sum_{i=1}^n \alpha_{C_{bi}} ~,  \\
        \mathcal{F}_{\text{pl},L=2} &= \sum_{i<j} X_{ij} \left(\sum_{k}\alpha_{ai}\alpha_{aj}\alpha_{bk} + \sum_{k}\alpha_{bi}\alpha_{bj}\alpha_{ak} + \alpha_{bi}\alpha_{bj}\alpha_{ab}  + \alpha_{ai}\alpha_{aj}\alpha_{ab} + \alpha_{ai}\alpha_{bj}\alpha_{ab} \right)
    \end{aligned}
\end{equation}
In both cases, $\mathcal{Z}$ contains the terms from curves with momenta independent of the loop momentum: $\mathcal{Z} = \sum_{i<j-1}\alpha_{ij}X_{ij}~.$ The whole Schwinger integrand is determined, and the external momenta appear only as $X_{ij}~.$

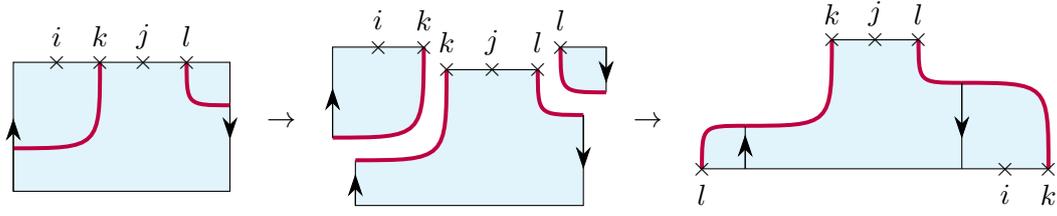
\begin{figure}[!t]
    \centering
    \begin{tikzpicture}[scale=0.57,baseline={([yshift=-1.9ex]current bounding box.center)}]
    \fill[cyan!10] (0,0) rectangle (5,3);
    \draw (0,0) rectangle (5,3);
    \draw[line width=1.5, purple] (0,1) .. controls (2,1) .. (2,3);
    \draw[line width=1.5, purple] (4,3) .. controls (4,2) .. (5,2);
    \node at (1,3) {\small $\times$ };
    \node at (2,3) {\small $\times$ };
    \node at (3,3) {\small $\times$ };
    \node at (4,3) {\small $\times$ };
    \node at (1,3.6) {$i$};  
    \node at (2,3.6) {$k$};  
    \node at (3,3.6) {$j$};  
    \node at (4,3.6) {$l$};  
    \draw[-{Stealth[length=3mm, width=2mm]}] (0,0) -- (0,1.7);
    \draw [-{Stealth[length=3mm, width=2mm]}] (5,2) -- (5,1.2);
\end{tikzpicture} $ \ \ \to \ \ $
\begin{tikzpicture}[scale=0.6,baseline={([yshift=-1.9ex]current bounding box.center)}]
    \fill[cyan!10] (0,0) -- (5,0) -- (5,2) .. controls (4,2) .. (4,3) -- (2,3) .. controls (2,1) .. (0,1) -- (0,0);
    \fill[cyan!10] (-0.5,1.5) -- (-0.5,3.5) -- (1.5,3.5) .. controls (1.5,1.5) .. (-0.5,1.5);
    \fill[cyan!10] (4.5,3.5) -- (5.5,3.5) -- (5.5,2.5) .. controls (4.5,2.5) .. (4.5,3.5);
    \draw (0,1) -- (0,0) -- (5,0) -- (5,2);
    \draw (5.5,2.5) -- (5.5,3.5) -- (4.5,3.5);
    \draw (2,3) -- (4,3);
    \draw (-0.5,1.5) -- (-0.5,3.5) -- (1.5,3.5);
    \draw[line width=1.5, purple] (0,1) .. controls (2,1) .. (2,3);
    \draw[line width=1.5, purple] (-0.5,1.5) .. controls (1.5,1.5) .. (1.5,3.5);
    \draw[line width=1.5, purple] (4,3) .. controls (4,2) .. (5,2);
    \draw[line width=1.5, purple] (4.5,3.5) .. controls (4.5,2.5) .. (5.5,2.5);
    \node at (0.5,3.5) {\small $\times$ };
    \node at (2,3) {\small $\times$ };
    \node at (3,3) {\small $\times$ };
    \node at (4,3) {\small $\times$ };
    \node at (1.5,3.5) {\small $\times$ };
    \node at (4.5,3.5) {\small $\times$ };
    \node at (0.5,4.1) {$i$};  
    \node at (2,3.6) {$k$};    
    \node at (1.5,4.1) {$k$};  
    \node at (3,3.6) {$j$};  
    \node at (4,3.6) {$l$};  
    \node at (4.5,4.1) {$l$};  
    \draw[-{Stealth[length=3mm, width=2mm]}] (0,0) -- (0,0.7);
    \draw [-{Stealth[length=3mm, width=2mm]}] (5,2) -- (5,0.8);
    \draw[-{Stealth[length=3mm, width=2mm]}] (-0.5,1.5) -- (-0.5,2.6);
    \draw [-{Stealth[length=3mm, width=2mm]}] (5.5,3.5) -- (5.5,2.7);
\end{tikzpicture} $\ \to \ $
\begin{tikzpicture}[scale=0.57,baseline={([yshift=-1.9ex]current bounding box.center)}]
    \fill[cyan!10] (-1,0) -- (7,0) .. controls (7,2) .. (5,2) .. controls (4,2) .. (4,3) -- (2,3) .. controls (2,1) .. (0,1) .. controls (-1,1) .. (-1,0);
    \draw[line width=1.5, purple] (0,1) .. controls (2,1) .. (2,3);
    \draw[line width=1.5, purple] (4,3) .. controls (4,2) .. (5,2);
    \draw[line width=1.5, purple] (0,1) .. controls (-1,1) .. (-1,0);
    \draw[line width=1.5, purple] (5,2) .. controls (7,2) .. (7,0);
    \node at (2,3) {\small $\times$ };
    \node at (3,3) {\small $\times$ };
    \node at (4,3) {\small $\times$ };
    \node at (2,3.6) {$k$};  
    \node at (3,3.6) {$j$};  
    \node at (4,3.6) {$l$};
    \draw (0,0) -- (0,1);
    \draw (5,0) -- (5,2);
    \draw[-{Stealth[length=3mm, width=2mm]}] (0,0) -- (0,0.8);
    \draw[-{Stealth[length=3mm, width=2mm]}] (5,2) -- (5,0.7);
    \node at (7,-0.6) {$k$};
    \node at (-1,-0.6) {$l$};
    \node at (6,-0.6) {$i$};
    \node at (7,0) {\small $\times$};
    \node at (-1,0) {\small $\times$};
    \node at (6,0) {\small $\times$};
    \draw (-1,0) -- (7,0);
    \draw (2,3) -- (4,3);
\end{tikzpicture}
    \caption{The curve $C^{\ot}_{kl}$ \emph{breaks open} an arbitrary Möbius strip with $n$ marked points into a disk with $n+2$ marked points.}
    \label{fig:curve Ckl breaks open the Möbius strip into a disk}
\end{figure}

\paragraph{Surface Symanzik for Möbius strip:} Let us use the spanning surfaces to construct the surface Symanzik for Möbius strip. The spanning-1 surfaces for Möbius strip are simply as follows:
\begin{align}
    \text{spanning-1 surfaces:}\ \  \left\{ \ \begin{tikzpicture}[scale=0.5,baseline={([yshift=-.5ex]current bounding box.center)}]
    \draw (0,0) circle (1);
    \node at (0,0) {\small $\otimes$};
    \node at (0,1) {\small $\times$};
    \node at (0.4,1.5) {$i$};
    \node at (-1.3,0.3) {$\vdots$};
    \node at (1.3,0.3) {$\vdots$};
    \node at (-0,-1) {\small $\times$};
    \node at (-0.4,-1.5) {$j$};
    \draw (0,-1) -- (0,1);
    \end{tikzpicture} \ \ \forall\ 1\leq i\leq j \leq n  \ \right\} ~. 
    \label{spanning 1 surfaces of Möbius}
\end{align}
The curves $C^{\ot}_{ik}$ cut open a Möbius strip into a disk as shown in Figure \ref{fig:curve Ckl breaks open the Möbius strip into a disk}. Since $C_{ij}^{\ot}$ is symmetric in its two indices, so to avoid double counting, we have $i\leq j~.$

Note that $C^{\ot}$ curve on a Möbius strip also does not separate it into two disjoint pieces, as shown in Figure \ref{fig:Cutting the Möbius strip along Cot}. It is a reflection of the fact that the momentum associated with $C^{\ot}$ is the loop momentum. Famously, cutting the Möbius strip along its length does not divide it into two parts, but rather turns it into an orientable surface, topologically a cylinder. To include the possibility of such curves, we defined our spanning-1 surfaces as subsurfaces with the inclusion of \emph{non-separating curves}, so that the resultant subsurface is still a single connected piece, not necessarily a disk. However, we restrict to the region in the global Schwinger space $t_0<0$ (with the tadpole (Figure \ref{fig:Schwinger parameters for generic n with tadpole as reference}) as the reference triangulation), so that the headlight $\alpha^{\ot}$ is \emph{turned off} for all $n$. We omit the quasi-triangulations coming from $C^{\ot}$, and as a result, we have:
\begin{align}
    \mathcal{U}_{\text{Möb}} = \sum_{i\leq k} \alpha^{\ot}_{ik}~.
\end{align}

\begin{figure}
    \centering
    \begin{tikzpicture}[scale=0.8,baseline={([yshift=-.5ex]current bounding box.center)}]
    \draw[thick] (0,0) circle (1);
    \node at (0,0) {\small $\otimes$};
    \node at (0,1) {\small $\times$};
    \node at (0,1.4) {$i$};
    \node at (-0,-1) {\small $\times$};
    \node at (-0,-1.4) {$j$};
    \node at (1,0) {\small $\times$};
    \node at (-1,0) {\small $\times$};
    \draw[line width=1.5, teal] (0,-1) -- (0,1);
    \draw[line width=1.5, purple] (-1,0) -- (1,0);
    \node at (1.4,0) {$k$};
    \node at (-1.4,0) {$l$};
    \node at (1,1) {$\ddots$};
    \node at (1,-0.8) {$\iddots$};
    \node at (-1,1) {$\iddots$};
    \node at (-1,-0.8) {$\ddots$};
    \end{tikzpicture} \ \ $\equiv$ \ \ 
\begin{tikzpicture}[scale=0.7,baseline={([yshift=-1.9ex]current bounding box.center)}]
    \fill[yellow!20] (0,2) .. controls (1,2) .. (1,3) -- (2,3) .. controls (2,1) .. (0,1) -- (0,2);
    \fill[yellow!20] (4,3) .. controls (4,2) .. (5,2) -- (5,1) .. controls (3,1) .. (3,3) -- (4,3);
    \fill[cyan!10] (4,3) .. controls (4,2) .. (5,2) -- (5,3) -- (4,3);
    \fill[cyan!10] (0,2) .. controls (1,2) .. (1,3) -- (0,3) -- (0,2);
    \fill[cyan!10] (3,3) .. controls (3,1) .. (5,1) -- (5,0) -- (0,0) -- (0,1)  .. controls (2,1) .. (2,3) -- (3,3);
    \draw (0,0) rectangle (5,3);
    \draw[line width=1.5, teal] (0,2) .. controls (1,2) .. (1,3); 
    \draw[line width=1.5, teal] (3,3) .. controls (3,1) .. (5,1);
    \draw[line width=1.5, purple] (0,1) .. controls (2,1) .. (2,3);
    \draw[line width=1.5, purple] (4,3) .. controls (4,2) .. (5,2);
    \node at (1,3) {\small $\times$ };
    \node at (2,3) {\small $\times$ };
    \node at (3,3) {\small $\times$ };
    \node at (4,3) {\small $\times$ };
    \node at (1,3.4) {$i$};  
    \node at (2,3.4) {$k$};  
    \node at (3,3.4) {$j$};  
    \node at (4,3.4) {$l$};  
    \draw[-{Stealth[length=3mm, width=2mm]}] (0,0) -- (0,1.7);
    \draw [-{Stealth[length=3mm, width=2mm]}] (5,2) -- (5,1.2);
    \draw [-{Stealth[length=3mm, width=2mm]}, line width=1.5] (1.5,3.9) -- (1.5,2.5);
    \node at (1.5,4.3) {$P_{ik}$};
    \draw [-{Stealth[length=3mm, width=2mm]}, line width=1.5] (2.5,3.9) -- (2.5,2.5);
    \node at (2.5,4.3) {$P_{kj}$};
    \draw [-{Stealth[length=3mm, width=2mm]}, line width=1.5] (3.5,3.9) -- (3.5,2.5);
    \node at (3.5,4.3) {$P_{jl}$};
    \draw [-{Stealth[length=3mm, width=2mm]}, line width=1.5] (4.5,3.9) -- (4.5,2.5);
    \node at (4.5,4.3) {$P_{li}$};
\end{tikzpicture} \ \ $\equiv$ \ \ 
\begin{tikzpicture}[scale=0.6,baseline={([yshift=-1.9ex]current bounding box.center)}]
    \fill[cyan!10] (3,3) .. controls (3,1) .. (5,1)  .. controls (6,1) .. (6,0) -- (-1,0) .. controls (-1,1) .. (0,1) .. controls (2,1) .. (2,3) -- (3,3);
    \fill[yellow!10] (3.5,3.5) .. controls (3.5,1.5) .. (5.5,1.5)  .. controls (6.5,1.5) .. (6.5,0.5) -- (7.5,0.5) .. controls (7.5,2.5) .. (5.5,2.5) .. controls (4.5,2.5) .. (4.5,3.5) -- (3.5,3.5);
    \draw[line width=1.5, purple] (0,1) .. controls (2,1) .. (2,3);
    \draw[line width=1.5, purple] (4.5,3.5) .. controls (4.5,2.5) .. (5.5,2.5) .. controls (7.5,2.5) .. (7.5,0.5);
    \draw[line width=1.5, purple] (0,1) .. controls (-1,1) .. (-1,0);
    \draw[line width=1.5, teal] (3,3) .. controls (3,1) .. (5,1)  .. controls (6,1) .. (6,0);
    \draw[line width=1.5, teal] (3.5,3.5) .. controls (3.5,1.5) .. (5.5,1.5)  .. controls (6.5,1.5) .. (6.5,0.5);
    \node at (2,3) {\small $\times$ };
    \node at (3,3) {\small $\times$ };
    \node at (3.5,3.5) {\small $\times$ };
    \node at (4.5,3.5) {\small $\times$ };
    \node at (2,3.5) {$k$};  
    \node at (3,3.5) {$j$};  
    \node at (3.5,4.1) {$j$};  
    \node at (4.5,4.1) {$l$};
    \draw (0,0) -- (0,1);
    \draw (5,0) -- (5,1);
    \draw (5.5,1.5) -- (5.5,2.5);
    \node at (7.5,-0.1) {$k$};
    \node at (6.5,-0.1) {$i$};
    \node at (-1,-0.6) {$l$};
    \node at (6,-0.6) {$i$};
    \node at (7.5,0.5) {\small $\times$};
    \node at (-1,0) {\small $\times$};
    \node at (6,0) {\small $\times$};
    \draw (-1,0) -- (6,0);
    \draw (6.5,0.5) -- (7.5,0.5);
    \draw (2,3) -- (3,3);
    \draw (3.5,3.5) -- (4.5,3.5);
    \draw [-{Stealth[length=3mm, width=2mm]}, line width=1.5] (2.5,3.9) -- (2.5,2.5);
    \node at (2.5,4.3) {$P_{kj}$};
    \draw [-{Stealth[length=3mm, width=2mm]}, line width=1.5] (2.5,-0.8) -- (2.5,0.5);
    \node at (2.5,-1.3) {$P_{li}$};
\end{tikzpicture}
    \caption{The curves $C_{ij}^{\ot}$ and $C_{kl}^{\ot}$ \emph{breaks open} the Möbius strip into two disks. Incoming momenta in the two disks are $P_{ik}+P_{jl}$ and $P_{kj}+P_{li}~,$ contributing $\alpha_{ij}^{\ot}\alpha_{kl}^{\ot}(P_{kj}+P_{li})^2$ to $\mathcal{F}_{\text{Mob}}~.$ }
    \label{fig: 2 curves breaking open Möbius strip into two disks}
\end{figure}
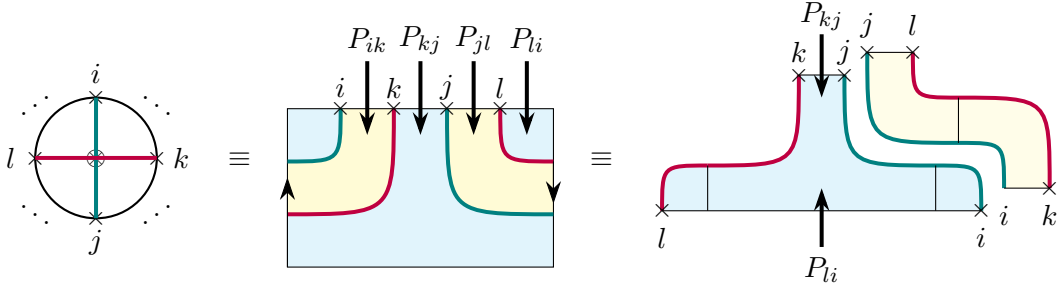

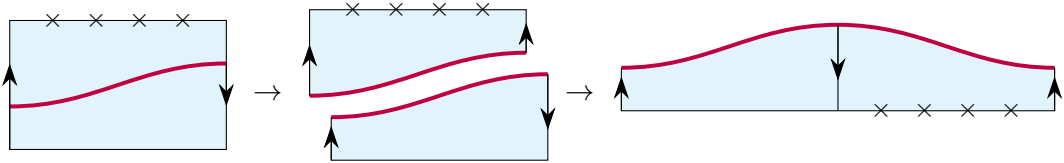
\begin{figure}
    \centering
\begin{tikzpicture}[scale=0.57,baseline={([yshift=-1.9ex]current bounding box.center)}]
    \fill[cyan!10] (0,0) rectangle (5,3);
    \draw (0,0) rectangle (5,3);
    \draw[line width=1.5, purple] (0,1) .. controls (2,1) and (3,2) .. (5,2);
    \node at (1,3) {\small $\times$ };
    \node at (2,3) {\small $\times$ };
    \node at (3,3) {\small $\times$ };
    \node at (4,3) {\small $\times$ };
    \draw[-{Stealth[length=3mm, width=2mm]}] (0,0) -- (0,2);
    \draw [-{Stealth[length=3mm, width=2mm]}] (5,2) -- (5,1);
\end{tikzpicture} $\ \to \ $
\begin{tikzpicture}[scale=0.57,baseline={([yshift=-1.9ex]current bounding box.center)}]
    \fill[cyan!10] (0,0) -- (0,1) .. controls (2,1) and (3,2) .. (5,2) -- (5,0) -- (0,0);
    \fill[cyan!10] (-0.5,1.5) -- (-0.5,3.5) -- (4.5,3.5) -- (4.5,2.5) ..  controls (2.5,2.5) and (1.5,1.5) .. (-0.5,1.5);
    \draw[line width=1.5, purple] (0,1) .. controls (2,1) and (3,2) .. (5,2);
    \draw[line width=1.5, purple] (4.5,2.5) ..  controls (2.5,2.5) and (1.5,1.5) .. (-0.5,1.5);
    \node at (0.5,3.5) {\small $\times$ };
    \node at (1.5,3.5) {\small $\times$ };
    \node at (2.5,3.5) {\small $\times$ };
    \node at (3.5,3.5) {\small $\times$ };
    \draw[-{Stealth[length=3mm, width=2mm]}] (0,0) -- (0,0.8);
    \draw [-{Stealth[length=3mm, width=2mm]}] (5,2) -- (5,0.7);
    \draw[-{Stealth[length=3mm, width=2mm]}] (-0.5,1.5) -- (-0.5,2.7);
    \draw [-{Stealth[length=3mm, width=2mm]}] (4.5,2.5) -- (4.5,3.2);
    \draw (0,1) -- (0,0) -- (5,0) -- (5,2);
    \draw (-0.5,1.5) -- (-0.5,3.5) -- (4.5,3.5) -- (4.5,2.5);
\end{tikzpicture}$\ \to \ $
\begin{tikzpicture}[scale=0.57,baseline={([yshift=-1.9ex]current bounding box.center)}]
    \fill[cyan!10] (0,0) -- (0,1) .. controls (2,1) and (3,2) .. (5,2) .. controls (7,2) and (8,1) .. (10,1) -- (10,0) -- (0,0);  
    \draw[line width=1.5, purple] (0,1) .. controls (2,1) and (3,2) .. (5,2)  .. controls (7,2) and (8,1) .. (10,1);
    \node at (6,0) {\small $\times$ };
    \node at (7,0) {\small $\times$ };
    \node at (8,0) {\small $\times$ };
    \node at (9,0) {\small $\times$ };
    \draw[-{Stealth[length=3mm, width=2mm]}] (0,0) -- (0,0.8);
    \draw [-{Stealth[length=3mm, width=2mm]}] (5,2) -- (5,0.7);
    \draw (0,1) -- (0,0) -- (10,0) -- (10,1);
    \draw (5,0) -- (5,2);
    \draw (10,0) -- (10,1);
    \draw[-{Stealth[length=3mm, width=2mm]}] (10,0) -- (10,0.8);
\end{tikzpicture}
    \caption{Cutting a Möbius strip along its length, i.e., along the curve $C^{\ot}$, turns it into a surface with two twists: topologically a cylinder, an orientable surface.}
    \label{fig:Cutting the Möbius strip along Cot}
\end{figure}

The spanning-2 surfaces for the Möbius strip are made up of two curves $C^{\ot}_{ij}$ and $C^{\ot}_{kl}~.$ 
\begin{align}
    \text{spanning-2 surfaces:}\ \  \left\{ \ \begin{tikzpicture}[scale=0.5,baseline={([yshift=-.5ex]current bounding box.center)}]
    \draw (0,0) circle (1);
    \node at (0,0) {\small $\otimes$};
    \node at (0,1) {\small $\times$};
    \node at (0.4,1.5) {$i$};
    \node at (-0,-1) {\small $\times$};
    \node at (-0.4,-1.5) {$j$};
    \node at (1,0) {\small $\times$};
    \node at (-1,0) {\small $\times$};
    \draw (0,-1) -- (0,1);
    \draw (-1,0) -- (1,0);
    \node at (1.5,-0.4) {$k$};
    \node at (-1.5,0.4) {$l$};
    \end{tikzpicture} \ \ \forall\ \ 1\leq i\leq k\leq j \leq l \leq n \ \right\} ~. 
    \label{spanning 2 surfaces of Möbius}
\end{align}
Figure \ref{fig: 2 curves breaking open Möbius strip into two disks} depicts how the two curves breaks open the Möbius strip into two disks. By convention $i\leq j$ and $k\leq l$~, and to avoid double counting, we choose $i\leq k~.$ With these constraints, the two curves $C_{ij}^{\ot}$ and $C_{kl}^{\ot}$ are compatible only if $i\leq k\leq j\leq l~.$ The momentum flowing inside the two disks, as shown in Figure \ref{fig: 2 curves breaking open Möbius strip into two disks} is $(P_{kj}+P_{li})~.$ Note that $P_{ik} \equiv p_i+p_{i+1}+\hdots + p_{k-1}~,$ and hence $P_{ik}+P_{kj}+P_{jl}+P_{li}=0~.$ Hence, the second Symanzik polynomial is as follows:
\begin{align}
    \mathcal{F}_{\text{Mob}} &= \sum_{i\leq k\leq j \leq l}\alpha_{ij}^{\ot}\alpha_{kl}^{\ot}(P_{kj}+P_{li})^2 \nonumber \\
    &= \sum_{i\leq k\leq j \leq l}\alpha_{ij}^{\ot}\alpha_{kl}^{\ot} \big(X_{ik}+X_{jl}+X_{il}+X_{kj}-X_{ij}-X_{kl}\big)~.
\end{align}
The following identities might be useful to obtain the last equality: $P_{ik}=P_{1k}-P_{1i}~,~~ 2P_{1i}.P_{1k}=X_{1i}+X_{1k}-X_{ik}~.$ The surface Symanzik polynomials obtained from the spanning surfaces match with the explicit momentum integral carried out in the Appendix \ref{appendix B details}: \eqref{explicit loop integration 1} - \eqref{explicit loop integration 3}~.  

\section{Two-loop non-orientable surface}
\label{section 5 two loop}
In this subsection, we briefly discuss how the constructions for non-orientable surfaces generalize to the higher genus surfaces. With a cross-cap and a puncture, we can construct higher genus surfaces. At two loops, along with the orientable surface, a disk with two punctures, we receive a contribution subdominant as $N^{-1}$ due to the following surface with a cross-cap and a puncture:
\begin{align}
\label{S2 surface}
    S_{(2)}\equiv \ \begin{tikzpicture}[scale=0.7,baseline={([yshift=-.5ex]current bounding box.center)}]
    \draw[fill=cyan!10] (0,0) circle (1);
    \draw[fill=white] (-0.4,0) circle (0.1);
    \node at (0.4,0) {$\ot$};
    \node at (0,1) {\small $\times$};
    \node at (0,-1) {\small $\times$};
    \node at (0.3,1.4) {$i$};
    \node at (-0.3,-1.4) {$j$};
    \node at (-1.3,0.3) {$\vdots$};
    \node at (1.3,0.3) {$\vdots$};
\end{tikzpicture} \ \ \equiv \ \ \begin{tikzpicture}[scale=0.45,baseline={([yshift=-1.9ex]current bounding box.center)}]
    \fill[cyan!10] (0,0) rectangle (5,3);
    \draw (0,0) rectangle (5,3);
    \node at (1,3) {\small $\times$ };
    \node at (2,3.5) {$\hdots$};
    \node at (3,3) {\small $\times$ };
    \node at (4,3.5) {$\hdots$ };
    \node at (1,3.6) {$i$};  
    \node at (3,3.6) {$j$};  
    \draw[-{Stealth[length=2.5mm, width=1.5mm]}] (0,0) -- (0,1.7);
    \draw [-{Stealth[length=2.5mm, width=1.5mm]}] (5,2) -- (5,1.2);
    \draw[fill=white] (2.5,1.5) circle (0.3);
\end{tikzpicture} 
\end{align}
The surface $S_{(2)}$ is obtained from cutting out a small disk from the Möbius strip. We shall discuss the curves and triangulations of this surface below. 

What about a surface with two cross-caps? A sphere with two cross-caps is a Klein bottle, and a disk with two cross-caps is a Klein bottle with a disk removed. The \emph{handle} due to the Klein bottle represents the propagation of an uncolored particle, a closed string propagation. However, in our current model, we do not have uncolored particles, and the $\Phi$ excitations can not propagate in the Klein bottle. So, in the current setup, a disk with cross-caps does not contribute due to the same reason that a 2-torus with a disk removed does not contribute. One can imagine a different setup with uncolored particles, and such surfaces will contribute. For instance:
\begin{align}
    \mathcal{L} \supset -\frac{1}{2}(\partial_{\mu} \sigma)^2 -\frac{1}{2} \tr(\partial \Phi.\partial \Phi) -\frac{1}{3} \tr(\Phi^3) -\frac{1}{3} \sigma\tr(\Phi^2) ~.  
\end{align}
We do not discuss such cases with uncolored fields. 

\paragraph{The curves and their momenta: }
The surface $S_{(2)}$ will have the following possible curves:
\begin{equation}
\begin{aligned}
    C_{0i}:&= \ \begin{tikzpicture}[scale=0.7,baseline={([yshift=-.5ex]current bounding box.center)}]
    \draw[fill=cyan!10] (0,0) circle (1);
    \node at (0.4,0) {$\ot$};
    \node at (0,1) {\small $\times$};
    \node at (0,-1) {\small $\times$};
    \node at (0.3,1.4) {$i$};
    \node at (-0.3,-1.4) {$j$};
    \node at (-1.3,0.3) {$\vdots$};
    \node at (1.3,0.3) {$\vdots$};
    \draw[line width=1.3, purple] (-0.4,0) -- (0,1);
    \draw[fill=white] (-0.4,0) circle (0.1);
\end{tikzpicture} ~, & \hspace{0.4cm} C_{ij}^{\ot} :&= \ \begin{tikzpicture}[scale=0.7,baseline={([yshift=-.5ex]current bounding box.center)}]
    \draw[fill=cyan!10] (0,0) circle (1);
    \node at (0.4,0) {$\ot$};
    \node at (0,1) {\small $\times$};
    \node at (0,-1) {\small $\times$};
    \node at (0.3,1.4) {$i$};
    \node at (-0.3,-1.4) {$j$};
    \node at (-1.3,0.3) {$\vdots$};
    \node at (1.3,0.3) {$\vdots$};
    \draw[line width=1.3, teal] (0,1) .. controls (0.55,0.25) and (0.55,-0.25) .. (0,-1);
    \draw[fill=white] (-0.4,0) circle (0.1);
\end{tikzpicture}~, & \hspace{0.4cm} C_{0j}^{\ot} :&= \ \begin{tikzpicture}[scale=0.7,baseline={([yshift=-.5ex]current bounding box.center)}]
    \draw[fill=cyan!10] (0,0) circle (1);
    \node at (0.4,0) {$\ot$};
    \node at (0,1) {\small $\times$};
    \node at (0,-1) {\small $\times$};
    \node at (0.3,1.4) {$i$};
    \node at (-0.3,-1.4) {$j$};
    \node at (-1.3,0.3) {$\vdots$};
    \node at (1.3,0.3) {$\vdots$};
    \draw[line width=1.3, teal!50!purple] (-0.4,0) .. controls (0.45,0.5) and (0.8,-0.25) .. (0,-1);
    \draw[fill=white] (-0.4,0) circle (0.1);
\end{tikzpicture}  \\
C_{00}^{\ot}:&= \ \begin{tikzpicture}[scale=0.7,baseline={([yshift=-.5ex]current bounding box.center)}]
    \draw[fill=cyan!10] (0,0) circle (1);
    \node at (0.4,0) {$\ot$};
    \node at (0,1) {\small $\times$};
    \node at (0,-1) {\small $\times$};
    \node at (0.3,1.4) {$i$};
    \node at (-0.3,-1.4) {$j$};
    \node at (-1.3,0.3) {$\vdots$};
    \node at (1.3,0.3) {$\vdots$};
    \draw[line width=1.3, blue!50!cyan] (0,0) circle (0.4);
    \draw[fill=white] (-0.4,0) circle (0.1);
\end{tikzpicture} ~, & \hspace{0.4cm} C^{\ot} :&= \ \begin{tikzpicture}[scale=0.7,baseline={([yshift=-.5ex]current bounding box.center)}]
    \draw[fill=cyan!10] (0,0) circle (1);
    \node at (0.4,0) {$\ot$};
    \node at (0,1) {\small $\times$};
    \node at (0,-1) {\small $\times$};
    \node at (0.3,1.4) {$i$};
    \node at (-0.3,-1.4) {$j$};
    \node at (-1.3,0.3) {$\vdots$};
    \node at (1.3,0.3) {$\vdots$};
    \draw[line width=1.3, gray!50!black] (0.4,-0.2) circle (0.2);
    \draw[fill=white] (-0.4,0) circle (0.1);
\end{tikzpicture}~, & \hspace{0.4cm} C_{(0)}^{\ot} :&= \ \begin{tikzpicture}[scale=0.7,baseline={([yshift=-.5ex]current bounding box.center)}]
    \draw[fill=cyan!10] (0,0) circle (1);
    \node at (0.4,0) {$\ot$};
    \node at (0,1) {\small $\times$};
    \node at (0,-1) {\small $\times$};
    \node at (0.3,1.4) {$i$};
    \node at (-0.3,-1.4) {$j$};
    \node at (-1.3,0.3) {$\vdots$};
    \node at (1.3,0.3) {$\vdots$};
    \draw[line width=1.3, gray!50!black] (-0.2,0) circle (0.6);
    \draw[fill=white] (-0.4,0) circle (0.1);
\end{tikzpicture}  \\
C_{00(\ot)}:&= \ \begin{tikzpicture}[scale=0.7,baseline={([yshift=-.5ex]current bounding box.center)}]
    \draw[fill=cyan!10] (0,0) circle (1);
    \node at (0.4,0) {$\ot$};
    \node at (0,1) {\small $\times$};
    \node at (0,-1) {\small $\times$};
    \node at (0.3,1.4) {$i$};
    \node at (-0.3,-1.4) {$j$};
    \node at (-1.3,0.3) {$\vdots$};
    \node at (1.3,0.3) {$\vdots$};
    \draw[line width=1.3, gray!50!black] (0.2,0) circle (0.6);
    \draw[fill=white] (-0.4,0) circle (0.1);
\end{tikzpicture} ~, & \hspace{0.4cm} C_{ij} :&= \ \begin{tikzpicture}[scale=0.7,baseline={([yshift=-.5ex]current bounding box.center)}]
    \draw[fill=cyan!10] (0,0) circle (1);
    \node at (0.4,0) {$\ot$};
    \node at (0,1) {\small $\times$};
    \node at (0,-1) {\small $\times$};
    \node at (0.3,1.4) {$i$};
    \node at (-0.3,-1.4) {$j$};
    \node at (-1.3,0.3) {$\vdots$};
    \node at (1.3,0.3) {$\vdots$};
    \draw[line width=1.3, gray!50!black] (0,-1) -- (0,1);
    \draw[fill=white] (-0.4,0) circle (0.1);
\end{tikzpicture}~ . & & 
\end{aligned}
\end{equation}
The surface $S^{(2)}$ has a non-trivial Mapping Class Group (MCG) due to the Dehn-twist: \begin{tikzpicture}[scale=0.6,baseline={([yshift=-.5ex]current bounding box.center)}]
    \draw[fill=cyan!10] (0,0) circle (1);
    \node at (0.4,0) {$\ot$};
    \node at (0,1) {\small $\times$};
    \node at (0,-1) {\small $\times$};
    \draw[thick,-{Stealth[length=1mm, width=1.2mm]}] (0.35,0.3) arc (20:160:0.4);
    \draw[thick,-{Stealth[length=1mm, width=1.2mm]}] (-0.35,-0.3) arc (200:340:0.4);
    \draw[fill=white] (-0.4,0) circle (0.1);
\end{tikzpicture}
The resulting quasi-cluster algebra and hence the Teichmüller space is infinite-dimensional. There are infinitely many distinct copies of curves $C_{0i}~, C_{ik}^{\ot}~, C_{0i}$ and $C_{ik}$ based on how they wind around the puncture and cross-cap. There are also three copies of $C_{ik}:$ \begin{tikzpicture}[scale=0.6,baseline={([yshift=-.5ex]current bounding box.center)}]
    \draw[fill=cyan!10] (0,0) circle (1);
    \node at (0.4,0) {$\ot$};
    \node at (0,1) {\small $\times$};
    \node at (0,-1) {\small $\times$};
    \draw[line width=1.3, gray!50!black] (0,-1) -- (0,1);    
    \draw[line width=1.3, gray!50!black] (0,-1) .. controls (-1,-0.3) and (-1,0.3) .. (0,1);
    \draw[line width=1.3, gray!50!black] (0,-1) .. controls (1.1,-0.3) and (1.1,0.3) .. (0,1);
    \draw[fill=white] (-0.4,0) circle (0.1);
\end{tikzpicture}~. These carry the same momenta, and while specifying a particular triangulation, we need to specify the particular curve. We omit this difference to keep the notations clean. So, each of these curves carries an extra index, with the information of windings and their path, suppressed here. The combinatorial polytope obtained from the mutations never closes onto itself and is an infinite structure. Analogous to the Faddeev-Popov gauge fixing, one can fix the MCG by inserting a \emph{Mirzakhani kernel} (equivalent to the factor of $1/(\text{Vol MCG})$): an appropriate set of headlight functions, in the curve integral. It picks up a finite region in the Teichmüller space, and gives us the moduli space of the surface. 

The momentum assignment to a curve is the same for all its MCG variants. The momentum assignment to the following curves is straightforward:
\begin{align}
    C_{ij}\leftrightarrow P_{ij} = p_i+\hdots +p_{j-1}~, \qquad C_{0i} \leftrightarrow P_{0i} = l_1 + P_{1i}~, \qquad C_{ik}^{\ot} \leftrightarrow P_{ik}^{\ot} = l_2 + P_{1i} + P_{1k}~.
\end{align}
We avoid the cumbersome procedure of drawing the doubled orientable surface, and determine the momentum for some of the curves by looking at subsurfaces as follows:
\begin{align}
    \begin{tikzpicture}[scale=0.8,baseline={([yshift=-.5ex]current bounding box.center)}]
    \draw[fill=cyan!10] (0,0) circle (1);
    \node at (0.4,0) {$\ot$};
    \node at (0,1) {\small $\times$};
    \node at (0,-1) {\small $\times$};
    \node at (0.3,1.4) {$i$};
    \node at (-0.3,-1.4) {$j$};\draw[thick, -{Stealth[length=2mm, width=1.5mm]}] (1.5,0) -- (0.9,0);
    \node at (2,0) {$P_{ij}$};
    \draw[line width=1.3, purple] (-0.4,0) -- (0,1);
    \draw[line width=1.3, teal!50!purple] (-0.4,0) .. controls (0.45,0.5) and (0.8,-0.25) .. (0,-1);
    \draw[fill=white] (-0.4,0) circle (0.1);
\end{tikzpicture} \ \ \simeq \ \ \begin{tikzpicture}[scale=0.8,baseline={([yshift=-.5ex]current bounding box.center)}]
    \draw[fill=cyan!10] (0,0) circle (1);
    \node at (0,0) {$\ot$};
    \node at (0,1) {\small $\times$};
    \node at (0,-1) {\small $\times$};
    \node at ({cos(45)},{sin(45)}) {\small $\times$};
    \node at ({cos(135)},{sin(135)}) {\small $\times$};
    \node at (0,1.5) {$0$};
    \node at (-0,-1.5) {$j$};
    \draw[thick, -{Stealth[length=2mm, width=1.5mm]}] (1.5,0) -- (0.5,0); 
    \node at (2,0) {$P_{ij}$};
    \draw[thick, -{Stealth[length=2mm, width=1.5mm]}] ({1.8*cos(67)},{1.8*sin(67)}) -- ({0.5*cos(67)},{0.5*sin(67)});
    \node at ({2.2*cos(67)},{2.2*sin(67)}) {$P_{0i}$};
    \draw[thick, -{Stealth[length=2mm, width=1.5mm]}] ({1.8*cos(113)},{1.8*sin(113)}) -- ({0.5*cos(113)},{0.5*sin(113)});
    \node at ({2.2*cos(113)},{2.2*sin(113)}) {$-P_{0i}$};
    \node at (0.9,1) {$i$};
    \node at (-0.9,1) {$i'$};
    \draw[line width=1.3, purple] ({cos(45)},{sin(45)}) arc (45:135:1);
    \draw[line width=1.3, teal!50!purple] (0,1) -- (0,-1);
\end{tikzpicture} \ \ \ \begin{matrix}
    \Rightarrow  \ \ P^{\ot}_{0j} = l_2 + (p_1+\hdots +p_{i-1} -P_{0i}) \\  + (p_1+\hdots + p_{j-1}) \\ \quad \\ 
    \Rightarrow  \ \ P^{\ot}_{0j} = l_2 - l_1 + P_{1j}~.
\end{matrix}
\end{align}
\begin{align}
    \begin{tikzpicture}[scale=0.7,baseline={([yshift=-.5ex]current bounding box.center)}]
    \draw[fill=cyan!10] (0,0) circle (1);
    \node at (0.4,0) {$\ot$};
    \node at (0,1) {\small $\times$};
    \node at (0.3,1.4) {$i$};
    \node at (1.3,0.3) {$\vdots$};
    \node at (-1.3,0.3) {$\vdots$};
    \draw[line width=1.3, blue!50!cyan] (0,0) circle (0.4);
    \draw[line width=1.3, purple] (0,1) .. controls (-0.5,0.3) .. (-0.4,0);
    \draw[fill=white] (-0.4,0) circle (0.1);
\end{tikzpicture} \ \ \simeq \ \ \begin{tikzpicture}[scale=0.8,baseline={([yshift=-.5ex]current bounding box.center)}]
    \draw[fill=cyan!10] (0,0) circle (1);
    \node at (0,0) {$\ot$};
    \node at (0,1) {\small $\times$};
    \node at ({cos(45)},{sin(45)}) {\small $\times$};
    \node at ({cos(135)},{sin(135)}) {\small $\times$};
    \node at (0,1.4) {$0$};
    \draw[thick, -{Stealth[length=2mm, width=1.5mm]}] ({1.8*cos(67)},{1.8*sin(67)}) -- ({0.9*cos(67)},{0.9*sin(67)});
    \node at ({2.2*cos(67)},{2.2*sin(67)}) {$P_{0i}$};
    \draw[thick, -{Stealth[length=2mm, width=1.5mm]}] ({1.8*cos(113)},{1.8*sin(113)}) -- ({0.9*cos(113)},{0.9*sin(113)});
    \node at ({2.2*cos(113)},{2.2*sin(113)}) {$-P_{0i}$};
    \node at (1,0.9) {$i$};
    \node at (-1,0.9) {$i'$};
    \draw[line width=1.3, purple] ({cos(45)},{sin(45)}) arc (45:135:1);
    \draw[line width=1.3, blue!50!cyan] (0,1) .. controls (-0.6,0.15) .. (0,0) .. controls (0.6,0.15) .. (0,1);
\end{tikzpicture}  \ \ \ \begin{matrix}
    \Rightarrow  \ \ P^{\ot}_{00} = l_2 + 2(p_1+\hdots +p_{i-1} -P_{0i}) \\ \quad \\ 
    \Rightarrow  \ \ P^{\ot}_{00} = l_2 - 2l_1~.
\end{matrix}
\end{align}
We are left with three curves: $C^{\ot}~, C^{\ot}_{(0)}$ and $C_{00(\ot)}~.$ Analogous to the Möbius strip case, $C^{\ot}$ and $C_{(0)}^{\ot}$ are the tadpole loop curves. In the string moduli space, they correspond to a (unoriented) closed string insertion ($\tau\to 0$). Just like the Möbius strip case, we choose not to include such degeneration and graphs. The last curve, $C_{00(\ot)}$, is a curve appearing exclusively in tadpoles, and carries zero momentum. 

We have enumerated all the curves and their associated momenta. One can construct explicit $g$-vector fans and the dual headlight functions. Though no simpler way other than analysing the curves on the doubled orientable surface is known.
 
\paragraph{Surface Symanzik for two-loop non-orientable surface: } We construct the spanning-1 and 2 subsurfaces of the $S_{(2)}$ surface below, and consequently write down the surface Symanzik polynomials. 
\begin{align}
    &\text{Spanning-1 sub-surfaces of }\, S_{(2)}: \nonumber \\ 
    &\left\{
    \begin{tikzpicture}[scale=0.7,baseline={([yshift=-.5ex]current bounding box.center)}]
    \draw[fill=cyan!10] (0,0) circle (1);
    \node at (0.4,0) {$\ot$};
    \node at (0,-1) {\small $\times$};
    \node at ({cos(60)},{sin(60)}) {\small $\times$};
    \node at ({cos(120)},{sin(120)}) {\small $\times$};
    \node at (0.9,1.1) {$k$};
    \node at (-0.9,1.1) {$i$};
    \node at (-0.3,-1.4) {$j$};
    \node at (-1.3,0.3) {$\vdots$};
    \node at (1.3,0.3) {$\vdots$};
    \draw[line width=1.3, purple] (-0.4,0) -- ({cos(120)},{sin(120)});
    \draw[line width=1.3, teal] ({cos(60)},{sin(60)}) .. controls (0.45,0.25) and (0.45,-0.25) .. (0,-1);
    \draw[fill=white] (-0.4,0) circle (0.1);
\end{tikzpicture}, \ 
\begin{tikzpicture}[scale=0.7,baseline={([yshift=-.5ex]current bounding box.center)}]
    \draw[fill=cyan!10] (0,0) circle (1);
    \node at (0.4,0) {$\ot$};
    \node at (0,1) {\small $\times$};
    \node at (0,-1) {\small $\times$};
    \node at (0.3,1.4) {$i$};
    \node at (-0.3,-1.4) {$j$};
    \node at (-1.3,0.3) {$\vdots$};
    \node at (1.3,0.3) {$\vdots$};
    \draw[line width=1.3, purple] (-0.4,0) -- (0,1);
    \draw[line width=1.3, teal!50!purple] (-0.4,0) .. controls (0.45,0.5) and (0.8,-0.25) .. (0,-1);
    \draw[fill=white] (-0.4,0) circle (0.1);
\end{tikzpicture}, \  
\begin{tikzpicture}[scale=0.7,baseline={([yshift=-.5ex]current bounding box.center)}]
    \draw[fill=cyan!10] (0,0) circle (1);
    \node at (0.4,0) {$\ot$};
    \node at (0,1) {\small $\times$};
    \node at (0,-1) {\small $\times$};
    \node at (0.3,1.4) {$i$};
    \node at (-0.3,-1.4) {$j$};
    \node at (1.3,0.3) {$\vdots$};
    \node at (-1.3,0.3) {$\vdots$};
    \draw[line width=1.3, blue!50!cyan] (0,0) circle (0.4);
    \draw[line width=1.3, purple] (0,1) .. controls (-0.5,0.3) .. (-0.4,0);
    \draw[fill=white] (-0.4,0) circle (0.1);
\end{tikzpicture} , \ 
\begin{tikzpicture}[scale=0.7,baseline={([yshift=-.5ex]current bounding box.center)}]
    \draw[fill=cyan!10] (0,0) circle (1);
    \node at (0.4,0) {$\ot$};
    \node at (0,1) {\small $\times$};
    \node at (0,-1) {\small $\times$};
    \node at (1,0) {\small $\times$};
    \node at (0.3,1.4) {$i$};
    \node at (-0.3,-1.4) {$j$};
    \node at (1.4,0) {$k$};
    \node at (-1.3,0.3) {$\vdots$};
    \draw[line width=1.3, blue!50!cyan] (0,0) circle (0.4);
    \draw[line width=1.3, teal!50!purple] (-0.4,0) -- (1,0);
    \draw[fill=white] (-0.4,0) circle (0.1);
\end{tikzpicture}, \
\begin{tikzpicture}[scale=0.7,baseline={([yshift=-.5ex]current bounding box.center)}]
    \draw[fill=cyan!10] (0,0) circle (1);
    \node at (0.4,0) {$\ot$};
    \node at (0,1) {\small $\times$};
    \node at (0,-1) {\small $\times$};
    \node at (1,0) {\small $\times$};
    \node at (0.3,1.4) {$i$};
    \node at (-0.3,-1.4) {$j$};
    \node at (1.4,0) {$k$};
    \node at (-1.3,0.3) {$\vdots$};
    \draw[line width=1.3, teal] (0,1) .. controls (0.55,0.25) and (0.55,-0.25) .. (0,-1);
    \draw[line width=1.3, teal!50!purple] (-0.4,0) -- (1,0);
    \draw[fill=white] (-0.4,0) circle (0.1);
\end{tikzpicture}
\right\} \\
&\Rightarrow \quad \mathcal{U}_{S_{(2)}} = \sum_{i} \alpha_{i0}\left(\sum_{j\leq k} \alpha_{jk}^{\ot} + \alpha_{0j}^{\ot} + \alpha_{00}^{\ot}\right) + \sum_{k}\alpha_{0k}^{\ot}\left(\alpha_{00}^{\ot} + \sum_{i\leq j}\alpha_{ij}^{\ot} \right)~.  
\end{align}

\begin{align}
    &\text{Spanning-2 sub-surfaces of }\, S_{(2)}: \nonumber \\ 
    &\left\{ \begin{tikzpicture}[scale=0.7,baseline={([yshift=-.5ex]current bounding box.center)}]
    \draw[fill=cyan!10] (0,0) circle (1);
    \node at (0.4,0) {$\ot$};
    \node at (0,1) {\small $\times$};
    \node at (0,-1) {\small $\times$};
    \node at (1,0) {\small $\times$};
    \node at (0.25,1.4) {$i$};
    \node at (-0.25,-1.4) {$k$};
    \node at (1.4,0) {$j$};
    \node at (-1.3,0.3) {$\vdots$};
    \draw[line width=1.3, purple] (0,-1) -- (-0.4,0) -- (0,1);
    \draw[line width=1.3, teal!50!purple] (-0.4,0) -- (1,0);
    \draw[fill=white] (-0.4,0) circle (0.1);
\end{tikzpicture}~, \ 
\begin{tikzpicture}[scale=0.7,baseline={([yshift=-.5ex]current bounding box.center)}]
    \draw[fill=cyan!10] (0,0) circle (1);
    \node at (0.4,0) {$\ot$};
    \node at ({cos(60)},{sin(60)}) {\small $\times$};
    \node at ({cos(120)},{sin(120)}) {\small $\times$};
    \node at ({cos(-60)},{sin(-60)}) {\small $\times$};
    \node at ({cos(-120)},{sin(-120)}) {\small $\times$};
    \node at (0.9,1.1) {$j$};
    \node at (-0.9,1.1) {$i$};
    \node at (0.9,-1.1) {$l$};
    \node at (-0.9,-1.1) {$k$};
    \node at (-1.3,0.3) {$\vdots$};
    \node at (1.3,0.3) {$\vdots$};
    \draw[line width=1.3, purple] ({cos(120)},{sin(120)}) -- (-0.4,0) -- ({cos(-120)},{sin(-120)});
    \draw[line width=1.3, teal] ({cos(60)},{sin(60)}) .. controls (0.45,0.25) and (0.45,-0.25) .. ({cos(-60)},{sin(-60)});
    \draw[fill=white] (-0.4,0) circle (0.1);
\end{tikzpicture}~, \ 
\begin{tikzpicture}[scale=0.7,baseline={([yshift=-.5ex]current bounding box.center)}]
    \draw[fill=cyan!10] (0,0) circle (1);
    \node at (0.4,0) {$\ot$};
    \node at ({cos(120)},{sin(120)}) {\small $\times$};
    \node at ({cos(-120)},{sin(-120)}) {\small $\times$};
    \node at (-0.9,1.1) {$i$};
    \node at (-0.9,-1.1) {$k$};
    \node at (-1.3,0.3) {$\vdots$};
    \node at (1.3,0.3) {$\vdots$};
    \draw[line width=1.3, purple] ({cos(120)},{sin(120)}) -- (-0.4,0) -- ({cos(-120)},{sin(-120)});   
    \draw[line width=1.3, blue!50!cyan] (0,0) circle (0.4);
    \draw[fill=white] (-0.4,0) circle (0.1);
\end{tikzpicture}~, \ \begin{tikzpicture}[scale=0.7,baseline={([yshift=-.5ex]current bounding box.center)}]
    \draw[fill=cyan!10] (0,0) circle (1);
    \node at (0.4,0) {$\ot$};
    \node at (0,1) {\small $\times$};
    \node at (0,-1) {\small $\times$};
    \node at (0.3,1.4) {$i$};
    \node at (-0.3,-1.4) {$k$};
    \node at (-1.3,0.3) {$\vdots$};
    \node at (1.3,0.3) {$\vdots$};
    \draw[line width=1.3, purple] (0,1) .. controls (-0.5,0.3) .. (-0.4,0);
    \draw[line width=1.3, teal!50!purple] (-0.4,0) .. controls (0.45,0.5) and (0.8,-0.25) .. (0,-1);    \draw[line width=1.3, blue!50!cyan] (0,0) circle (0.4);
    \draw[fill=white] (-0.4,0) circle (0.1);
\end{tikzpicture}~, \ \begin{tikzpicture}[scale=0.7,baseline={([yshift=-.5ex]current bounding box.center)}]
    \draw[fill=cyan!10] (0,0) circle (1);
    \node at (0.4,0) {$\ot$};
    \node at (0,1) {\small $\times$};
    \node at (0,-1) {\small $\times$};
    \node at (0.3,1.4) {$i$};
    \node at (-0.3,-1.4) {$k$};
    \node at (-1.3,0.3) {$\vdots$};
    \node at (1.3,0.3) {$\vdots$};
    \draw[line width=1.3, teal!50!purple] (-0.4,0) .. controls (0.45,-0.5) and (0.8,0.25) .. (0,1); 
    \draw[line width=1.3, teal!50!purple] (-0.4,0) .. controls (0.45,0.5) and (0.8,-0.25) .. (0,-1);    \draw[line width=1.3, blue!50!cyan] (0,0) circle (0.4);
    \draw[fill=white] (-0.4,0) circle (0.1);
\end{tikzpicture}~, \right. \nonumber \\
&\hspace{-0.5cm} \left. \begin{tikzpicture}[scale=0.7,baseline={([yshift=-.5ex]current bounding box.center)}]
    \draw[fill=cyan!10] (0,0) circle (1);
    \node at (0.4,0) {$\ot$};
    \node at (0,1) {\small $\times$};
    \node at (0,-1) {\small $\times$};
    \node at (0.3,1.4) {$i$};
    \node at (-0.3,-1.4) {$k$};
    \node at (-1,0) {\small $\times$};
    \node at (-1.4,0) {$j$};
    \node at (1.3,0.3) {$\vdots$};
    \draw[line width=1.3, teal!50!purple] (-0.4,0) .. controls (0.45,-0.5) and (0.8,0.25) .. (0,1); 
    \draw[line width=1.3, teal!50!purple] (-0.4,0) .. controls (0.45,0.5) and (0.8,-0.25) .. (0,-1);    \draw[line width=1.3, purple] (-1,0) -- (-0.4,0);
    \draw[fill=white] (-0.4,0) circle (0.1);
\end{tikzpicture} ~, \ 
\begin{tikzpicture}[scale=0.7,baseline={([yshift=-.5ex]current bounding box.center)}]
    \draw[fill=cyan!10] (0,0) circle (1);
    \node at (0.4,0) {$\ot$};
    \node at (0,1) {\small $\times$};
    \node at (0,-1) {\small $\times$};
    \node at (0.3,1.4) {$i$};
    \node at (-0.3,-1.4) {$k$};
    \node at ({cos(150)},{sin(150)}) {\small $\times$};
    \node at ({cos(-150)},{sin(-150)}) {\small $\times$};
    \node at (-1.1,1) {$j$};
    \node at (-1.1,-1) {$l$};
    \node at (1.3,0.3) {$\vdots$};
    \draw[line width=1.3, teal!50!purple] (-0.4,0) .. controls (0.45,-0.5) and (0.8,0.25) .. (0,1); 
    \draw[line width=1.3, teal!50!purple] (-0.4,0) .. controls (0.45,0.5) and (0.8,-0.25) .. (0,-1);    \draw[line width=1.3, teal] ({cos(150)},{sin(150)}) .. controls (0.3,0.4) .. (0.4,0) .. controls (0.3,-0.4) .. ({cos(-150)},{sin(-150)});
    \draw[fill=white] (-0.4,0) circle (0.1);
\end{tikzpicture} \right\} \cup \left\{ \begin{tikzpicture}[scale=0.7,baseline={([yshift=-.5ex]current bounding box.center)}]
    \draw[fill=cyan!10] (0,0) circle (1);
    \node at (0.4,0) {$\ot$};
    \node at (0,-1) {\small $\times$};
    \node at (1,0) {\small $\times$};
    \node at ({cos(60)},{sin(60)}) {\small $\times$};
    \node at ({cos(120)},{sin(120)}) {\small $\times$};
    \node at (0.9,1.1) {$k$};
    \node at (-0.9,1.1) {$i$};
    \node at (-0.3,-1.4) {$l$};
    \node at (1.3,-0.3) {$j$};
    \node at (-1.3,0.3) {$\vdots$};
    \draw[line width=1.3, purple] (-0.4,0) -- ({cos(120)},{sin(120)});
    \draw[line width=1.3, teal] ({cos(60)},{sin(60)}) .. controls (0.45,0.25) and (0.45,-0.25) .. (0,-1);
    \draw[line width=1.3, teal!50!purple]  (-0.4,0) -- (1,0);
    \draw[fill=white] (-0.4,0) circle (0.1);
\end{tikzpicture}, \ \begin{tikzpicture}[scale=0.7,baseline={([yshift=-.5ex]current bounding box.center)}]
    \draw[fill=cyan!10] (0,0) circle (1);
    \node at (0.4,0) {$\ot$};
    \node at (-1,0) {\small $\times$};
    \node at ({cos(60)},{sin(60)}) {\small $\times$};
    \node at ({cos(-60)},{sin(-60)}) {\small $\times$};
    \node at ({cos(120)},{sin(120)}) {\small $\times$};
    \node at ({cos(-120)},{sin(-120)}) {\small $\times$};
    \node at (0.9,1.1) {$k$};
    \node at (-0.9,1.1) {$i$};
    \node at (0.9,-1.1) {$j$};
    \node at (-0.9,-1.1) {$l$};
    \node at (-1.4,-0.3) {$m$};
    \node at (1.3,0.3) {$\vdots$};
    \draw[line width=1.3, teal] ({cos(60)},{sin(60)}) .. controls (0.65,0.35) and (0,-0.9) .. ({cos(-120)},{sin(-120)});
    \draw[line width=1.3, teal] ({cos(-60)},{sin(-60)}) .. controls (0.65,-0.35) and (0,0.9) .. ({cos(120)},{sin(120)});
    \draw[line width=1.3, purple]  (-0.4,0) -- (-1,0);
    \draw[fill=white] (-0.4,0) circle (0.1);
\end{tikzpicture}, \ \begin{tikzpicture}[scale=0.7,baseline={([yshift=-.5ex]current bounding box.center)}]
    \draw[fill=cyan!10] (0,0) circle (1);
    \node at (0.4,0) {$\ot$};
    \node at (1,0) {\small $\times$};
    \node at ({cos(60)},{sin(60)}) {\small $\times$};
    \node at ({cos(-60)},{sin(-60)}) {\small $\times$};
    \node at ({cos(120)},{sin(120)}) {\small $\times$};
    \node at ({cos(-120)},{sin(-120)}) {\small $\times$};
    \node at (0.9,1.1) {$k$};
    \node at (-0.9,1.1) {$i$};
    \node at (0.9,-1.1) {$j$};
    \node at (-0.9,-1.1) {$l$};
    \node at (1.4,-0.3) {$m$};
    \node at (-1.3,0.3) {$\vdots$};
    \draw[line width=1.3, teal] ({cos(60)},{sin(60)}) .. controls (0.65,0.35) and (0,-0.9) .. ({cos(-120)},{sin(-120)});
    \draw[line width=1.3, teal] ({cos(-60)},{sin(-60)}) .. controls (0.65,-0.35) and (0,0.9) .. ({cos(120)},{sin(120)});
    \draw[line width=1.3, teal!50!purple]  (-0.4,0) -- (1,0);
    \draw[fill=white] (-0.4,0) circle (0.1);
\end{tikzpicture} \right\} \\
&\hspace{-0.5cm} \Rightarrow \quad \mathcal{F}_{S_{(2)}} = \sum_{i<{k-1}}X_{ik} \left[\alpha_{i0}\,\alpha_{k0} \left( \sum_{i\leq j\leq k}\alpha_{j0}^{\ot} + \sum_{i\leq j\leq l\leq k}\alpha_{jl}^{\ot} + \alpha_{00}^{\ot}\right) + \alpha_{00}^{\ot}\,\alpha_{i0}\,\alpha_{k0}^{\ot} \right. \nonumber \\
&\left. \hspace{4cm} + \alpha^{\ot}_{i0}\,\alpha_{k0}^{\ot}\left(\alpha_{00}^{\ot} + \sum_{k\leq j\leq i}\alpha_{j0} + \sum_{k\leq l\leq j\leq i}\alpha_{jl}^{\ot}\right) \right] \nonumber \\
& \hspace{1.5cm} + \sum_{i\leq k\leq j \leq l}(P_{kj}+P_{li})^2\alpha_{kl}^{\ot}\left[\alpha_{i0}\,\alpha_{j0}^{\ot} +  \alpha_{ij}^{\ot}\left(\sum_{l\leq m\leq i}\alpha_{m0} + \sum_{k\leq m \leq j}\alpha_{m0}^{\ot}\right) \right] ~.
\end{align}
The third surface Symanzik polynomial $\mathcal{Z}$ is given by the following:
\begin{align}
    \mathcal{Z}_{S_{(2)}} = \sum_{i<j-1} \alpha_{ij}X_{ij}~. 
\end{align}
With this information, we can write down the scattering amplitude as follows:
\begin{align}
    A_{n}^{S_{(2)}} = \int_{0}^{\infty} \dd t_0\,\dd t_0'\int_{-\infty}^{\infty} \dd^{n+1}\vec{t} \ \mathcal{K} \,\frac{1}{\mathcal{U}^{d/2}}\,e^{-\mathcal{F}/\mathcal{U}- \mathcal{Z}}~.
\end{align}

\section{Tropicalization of Möbius strip string amplitude}
\label{section 6 string amplitudes}
In this section, we revisit the field theory limit of the type-I superstring one-loop amplitudes with four massless particles. It is famously known that the field theory limit of type-I open strings gives the $N=4$ super Yang-Mills amplitudes \cite{Green:1982sw, Green:1987mn}. Our objective in this analysis is to see the combinatorial polytope $M_n$ appearing out of the degeneration of the Möbius strip moduli space. We check that the field theory limit of the four-particle amplitude gives us the expected eight box diagrams from the polytope $M_4$ with the appropriate momenta. This serves as a check for our curve integral construction, and for $P_{ij}^{\ot} = l+P_{1i}+P_{1j}~.$

\subsection{Warmup: tropicalization of annulus string amplitude}
\begin{figure}[H]
    \centering
\begin{tikzpicture}[scale=0.7,baseline={([yshift=-.5ex]current bounding box.center)}]
        \draw[gray, <->] (-1,0) -- (4.5,0);
        \draw[gray, <->] (0,-1) -- (0,4.5);
        \draw[white,fill=teal!10] (0,0) -- (3.5,0) -- (3.5,3.5) -- (0,3.5) -- (0,0);
        \draw[very thick, purple!20!black] (0,0) -- (0,3.5);
        \draw[very thick, purple!80!black] (3.5,0) -- (3.5,3.5);
        \draw[thick, dashed, -{Stealth[length=3mm, width=2mm]}] (3.5,0) -- (1.65,0);
        \draw[thick, dashed] (0,0) -- (1.7,0);
        \draw[thick, dashed] (0,3.5) -- (1.85,3.5);
        \draw[thick, dashed, {Stealth[length=3mm, width=2mm]}-] (1.7,3.5) -- (3.5,3.5);
        \node at (0,5.1) {Im$(\zeta)$};
        \node at (5.4,0) {Re$(\zeta)$};
        \draw (4.5,4.9) -- (4.5,4.5) -- (4.9,4.5);
        \node at (4.8,4.8) {$\zeta$};
        \node at (3.5,-0.55) {$1/2$};
        \node at (-0.55,3.5) {$it$};
        \node at (4.45,3.5) {$\frac{1}{2}+it$};
        \node at (-0.35,-0.35) {0};
        \draw[fill=purple!50!black] (0,0) circle (0.07);
        \draw[fill=purple!50!black] (3.5,0) circle (0.07);
        \draw[fill=purple!50!black] (3.5,3.5) circle (0.07);
        \draw[fill=purple!50!black] (0,3.5) circle (0.07);
    \end{tikzpicture}
    \qquad 
    \begin{tikzpicture}[scale=0.7,baseline={([yshift=-.5ex]current bounding box.center)}]
        \draw[white,fill=teal!10] (4.4,0) arc (0:180:4.4) -- (-2,0) -- (-2,0) arc (180:0:2) -- (4.4,0); 
        \draw[gray, <->] (-5,0) -- (5,0);
        \draw[gray, <->] (0,-1) -- (0,5);
        \node at (5.9,0) {Re$(\rho)$};
        \node at (0,5.6) {Im$(\rho)$};
        \draw (4.4,4.7) -- (4.4,4.3) -- (4.8,4.3);
        \node at (5.7,4.8) {$\rho=e^{2\pi i \zeta}$}; 
        \node at (5.7,3.5) {$\rho \sim w\rho$};
        \draw[thick, dashed] (4.4,0) arc (0:95:4.4);
        \draw[thick, dashed, {Stealth[length=3mm, width=2mm]}-] (0,4.4) arc (90:180:4.4);
        \draw[thick, dashed] (2,0) arc (0:95:2);
        \draw[thick, dashed, {Stealth[length=3mm, width=2mm]}-] (0,2) arc (90:180:2);
        \node at (4.4,-0.4) {$1$};
        \node at (-4.4,-0.4) {$-1$};
        \node at (-0.4,-0.4) {$0$};
        \node at (2,-0.4) {$w$};
        \node at (-2,-0.4) {$-w$};
        \node at (2.3,-0.9) {$=e^{-2\pi t}$};
        \draw[very thick, purple!20!black] (2,0) -- (4.4,0);
        \draw[very thick, purple!80!black] (-2,0) -- (-4.4,0);
        \draw[fill=purple!50!black] (4.4,0) circle (0.07);
        \draw[fill=purple!50!black] (2,0) circle (0.07);
        \draw[fill=purple!50!black] (-2,0) circle (0.07);
        \draw[fill=purple!50!black] (-4.4,0) circle (0.07);
    \end{tikzpicture}
    \caption{Annulus obtained by identifying $\rho$ uppper half plane as $\rho \sim w{\rho}~.$ The fundamental region under identification is shaded, and the thick lines mark the two distinct boundaries of the annulus.}
    \label{fig:Annulus obtained from identification}
\end{figure}
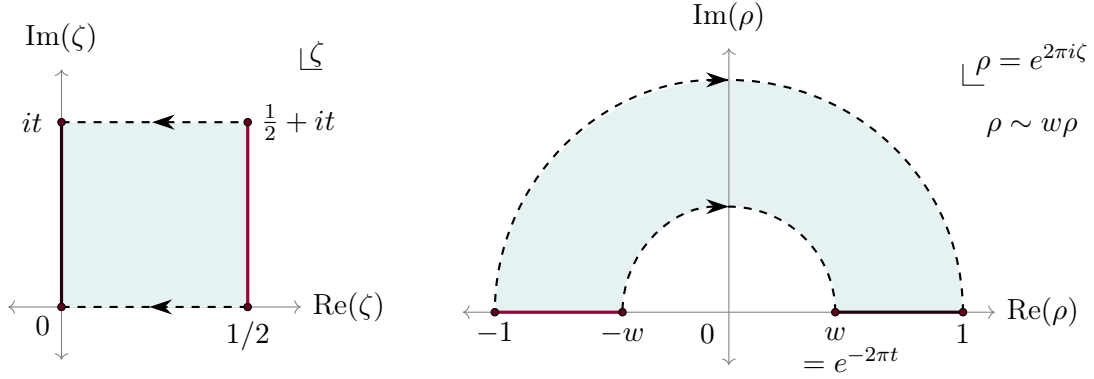

Consider a complex half plane, parameterized by $\rho$. One can obtain an annulus by the following identification: $\rho \sim w\rho~.$ Let us express $\rho = e^{2\pi i \zeta}~,$ and we have $\zeta \sim \zeta + it~.$ Figure \ref{fig:Annulus obtained from identification} depicts the fundamental region under this identification. There are two boundaries: $\rho \in [w,1]$ and $\rho \in [-w,-1]~.$ 
Let us have four marked insertions on the boundary $[w,1]$ with a fixed ordering. We use the translation on the boundary to fix $\zeta_4 = it~:$
\begin{align}
    \zeta_i \equiv it\nu_i~, \qquad \nu_1\leq \nu_2 \leq \nu_3 \leq \nu_4 = 1~. \label{ordering in nu_i}
\end{align}

\paragraph{Superstring amplitude:} The open string amplitude, due to annulus worldsheet for four massless external particles in type I superstring theory is as follows\footnote{Changing the integration variable to $1/w$ in place of $w$, and using appropriate modular properties of $\theta_1$, one can rewrite the amplitude in a neater expression, as given in \cite{Eberhardt:2023xck}.} \cite{Green:1987mn}:
\begin{align}
    \mathcal{A}^{\text{annulus}}[1234] = N\,g_s^4 \,\tr\big(t_{a_1}t_{a_2}t_{a_3}t_{a_4}\big)\,t_8 \int_0^1\frac{\dd w}{w}\,\left(\frac{-2\pi}{\log w}\right)^5\int\prod_{r=1}^3\frac{\dd \rho_r}{\rho_r}\prod_{r<s}\big(\psi_{rs}\big)^{\alpha'\,k_r.k_s}~.
\end{align}
{  Since $\rho_r = e^{2\pi i \zeta_r} = e^{-2\pi t \nu_r}$, and $\nu_r$ are ordered as \eqref{ordering in nu_i}, the integration over $\rho_r$ in the above (color ordered) amplitude is for a fixed ordering: $1\geq \rho_1\geq \rho_2\geq \rho_3 \geq \rho_4 = e^{-2\pi t}~.$} Note that $k_i^2=0$~. The factor $t_8$ contains the polarization vectors and kinematic factors corresponding to various scattering states in the supermultiplet. For $k_i^2=0$, the factor $\prod \psi$ is given by the following:
\begin{align}
    \prod_{r<s}\big(\psi_{rs}\big)^{k_r.k_s} = \prod_{\substack{r,s=1 \\ r<s}}^4e^{\pi (\alpha'\,k_r.k_s)\zeta_{rs}^2/t} \ \theta_1\left(\zeta_{rs}\,\big|\,it\right)^{\alpha'\,k_r.k_s}~, \qquad\quad \big(\zeta_{rs} = \zeta_s-\zeta_r\big)~.
\end{align}
We are interested in the moduli space integral, so let us drop the overall factor $N \, g_s^4 \tr()$ for now. Also, let us change the integration variables to $t$ and $\zeta_r$ {  (keeping the same ordering of $\zeta_r$)}:
\begin{align}
    \mathcal{A}^{\text{Ann}} \equiv \int_0^{\infty}\frac{\dd t}{t^5}\,\int_{0}^{it}\prod_{r=1}^3\dd \zeta_r\ \prod_{r<s}^4e^{\pi (\alpha'\,k_r.k_s)\zeta_{rs}^2/t}\ \prod_{r<s} \theta_1\left(\zeta_{rs}\,\big|\,it\right)^{\alpha'\,k_r.k_s}~.
\end{align}

We introduce a Gaussian integral over the loop momentum to facilitate the comparison with the QFT one-loop amplitude, as follows ($d=10$):
\begin{align}
    \int \dd^dl\,e^{-\pi t\alpha' \left(l^2 - 2 \sum_r\nu_r\,l.k_r\right)} &= \int \dd^dl\,e^{-\pi t\alpha' \left(l - \sum_r\nu_rk_r\right)^2}\ e^{\pi t\alpha'\left(\sum\nu_rk_r\right)^2} \nonumber \nonumber \\
    &= \frac{1}{t^{d/2}}\,e^{\pi\alpha' \sum_{r<s}k_r.k_s\,\zeta_{rs}^2/t}~. \label{fake loop integral}
\end{align}
Plugging it back into the amplitude, we have:
\begin{align}
    \mathcal{A}^{\text{Ann.}} &= \int_0^{\infty}\dd t\,\int_{0}^{it}\prod_{r=1}^3\dd \zeta_r \int \dd^dl\,e^{-\pi t\alpha' \left(l^2 - 2 \sum_r\nu_r\,l.k_r \right)}    \ \prod_{r<s} \theta_1\left(\zeta_{rs}\,\big|\,it\right)^{\alpha'\,k_r.k_s}~. \label{annulus string amplitude usable}
\end{align}

\paragraph{Obtaining the box integral:}
We have four external massless particles, and a loop momentum $l$. We have the following scalar kinematic variables, constructed out of $k_i$ and $l$, motivated by the triangulations of an annulus:
\begin{align}
    X_{13} = (k_1+k_2)^2~, \qquad X_{24} = (k_2+k_3)^2 ~, \qquad X_{0i} = (l+k_1+\hdots + k_{i-1})^2~.
\end{align}
One can express $l^2$ and $2l.k_i$ in terms of these as follows:
\begin{equation}
\begin{aligned}
    l^2 = X_{10}~, \qquad 2l.p_1 &= X_{20}-X_{10}~, \qquad \qquad & 
    2l.p_2 &= X_{30}-X_{20}-X_{13}~, \\
    2l.p_3 &= X_{40}-X_{30}+X_{13}~, \qquad& \qquad 
    2l.p_4 &= X_{10}-X_{40}~.
\end{aligned}
\end{equation}
We have,
\begin{align*}
    l^2 -2\sum_r\nu_rl.k_r = X_{10}(\nu_1) + X_{20}(\nu_2-\nu_1)+ X_{30}(\nu_3-\nu_2) + X_{40}(1-\nu_3) + X_{13}(\nu_2-\nu_3)
\end{align*}
Let us change to the following $\eta_I$ variables. $I$ refers to the chords/propagators in the triangulations $\{X_{10},~X_{20},~X_{30},~X_{40}\}~.$ 
\begin{align}
\begin{tikzpicture}[scale=0.7,baseline={([yshift=-.5ex]current bounding box.center)}]
        \draw[gray, <->] (-1,0) -- (4.5,0);
        \draw[gray, <->] (0,-1) -- (0,5);
        \draw[white,fill=teal!10] (0,0) -- (3.5,0) -- (3.5,4) -- (0,4) -- (0,0);
        \draw[very thick, purple!20!black] (0,0) -- (0,4);
        \draw[very thick, purple!80!black] (3.5,0) -- (3.5,4);
        \draw[thick, dashed, -{Stealth[length=3mm, width=2mm]}] (0,0) -- (1.95,0);
        \draw[thick, dashed] (1.6,0) -- (3.5,0);
        \draw[thick, dashed] (0,4) -- (1.85,4);
        \draw[thick, dashed, {Stealth[length=3mm, width=2mm]}-] (1.7,4) -- (3.5,4);
        \draw (4.5,4.9) -- (4.5,4.5) -- (4.9,4.5);
        \node at (4.8,4.8) {$\zeta$};
        \node at (3.5,-0.55) {$1/2$};
        \node at (-1.2,4.2) {$\zeta_4=it$};
        \node at (-0.5,3) {$\zeta_3$};
        \node at (-0.5,2) {$\zeta_2$};
        \node at (-0.5,1) {$\zeta_1$};
        \node at (0,3.9) {$*$};
        \node at (0,2.9) {$*$};
        \node at (0,1.9) {$*$};
        \node at (0,0.9) {$*$};
        \node at (-0.35,-0.35) {0};
        \draw[fill=purple!50!black] (0,0) circle (0.07);
        \draw[fill=purple!50!black] (3.5,0) circle (0.07);
        \draw[fill=purple!50!black] (3.5,4) circle (0.07);
        \draw[fill=purple!50!black] (0,4) circle (0.07);
        \draw[<->, dashed] (-1.5,0) -- (-1.5,1);
        \draw[<->, dashed] (-1.5,1) -- (-1.5,2);
        \draw[<->, dashed] (-1.5,2) -- (-1.5,3);
        \draw[<->, dashed] (-1.5,3) -- (-1.5,4);
        \node at (-2.5,0.5) {$i\eta_{10}$};
        \node at (-2.6,1.5) {$i\eta_{20}$};
        \node at (-2.5,2.5) {$i\eta_{30}$};
        \node at (-2.5,3.5) {$i\eta_{40}$};
    \end{tikzpicture}    
    \hspace{1.5cm} \begin{matrix}
        \eta_{10}=t(\nu_1) \\
        \eta_{20}=t(\nu_2-\nu_1) \\
        \eta_{30} = t(\nu_3-\nu_2) \\
        \eta_{40}=t(1-\nu_3)
    \end{matrix} ~, \qquad \sum_{I\in T}\eta_I=t~.
\end{align}
The string amplitude \eqref{annulus string amplitude usable} expressed in terms of the kinematic variables $X_{I}$ and the worldsheet coordinates $\eta_I$, is as follows:
\begin{align}
    \mathcal{A}^{\text{Ann}} &= \int_0^{\infty}\prod_{i=1}^4\dd \eta_{i0} \int \dd^dl\ e^{-\pi \alpha'\sum_I\eta_IX_{I}}  \nonumber \\
    &\hspace{1.5cm} \times \left(\frac{\theta_1(\zeta_{12})\,\theta_1(\zeta_{34})}{\theta_1(\zeta_{13})\,\theta_1(\zeta_{24})}\,e^{2\pi t (\nu_3-\nu_2)}\right)^{\alpha'X_{13}/2}\left(\frac{\theta_1(\zeta_{23})\,\theta_1(\zeta_{14})}{\theta_1(\zeta_{13})\,\theta_1(\zeta_{24})}\right)^{\alpha'X_{24}/2}~.  \label{annulus case just before taking the tropical limit}
\end{align}
{  We have changed the integration variables from $\{t,\zeta_1,\zeta_2,\zeta_3\}$ to $\{\eta_I\}~.$ The change of the variables can be understood as follows: 
\begin{align}
    \int_0^{\infty}\dd  t\int \prod_{r=1}^3\dd\zeta_r = \int_0^\infty \dd t\int \prod_{I} \dd \eta_I \ \delta\left(t-\sum \eta_I\right) = \int_0^\infty \prod_{I} \dd \eta_I~. \label{change of variables from t zeta to eta}
\end{align}
Finally, the four $\eta_I$s are to be considered as independent variables, and the condition $\sum \eta_I =t$ does not play any role.}

We take the tropical limit of the integral \eqref{annulus case just before taking the tropical limit} as follows. Consider the high string tension limit $\alpha'\to 0~,$ but scaling the worldsheet coordinates such that the whole worldsheet does not collapse to a point: $\eta_I\to \eta_I/\alpha'~.$ In this limit, the worldsheet degenerates into a worldline, and the worldsheet coordinates $\eta_I$ become the Schwinger parameters. With the limit $t\to t/\alpha'~,$ one can check the following using the infinite product expansion of $\theta_1$ \footnote{The Jacobi theta function $\theta_1$ has the following product representation:
\begin{align}
    \theta_1(\zeta|\tau) = 2e^{\pi i \tau/4}\,\sin(\pi \zeta)\prod_{m=1}^\infty(1-e^{2\pi im \tau})(1-e^{2\pi i(m \tau+\zeta)})(1-e^{2\pi i(m \tau-\zeta)})~.
\end{align} In the limit $\tau \to i\infty$, and $\zeta$ not compensating for it, the infinite product can be replaced by 1, and we have:
\begin{align}
    \frac{\theta_1(\zeta_{12})\,\theta_1(\zeta_{34})}{\theta_1(\zeta_{13})\,\theta_1(\zeta_{24})} \xrightarrow{\ \ \tau \to i\infty \ \ } \frac{\sin(\pi \zeta_{12})\,\sin(\pi \zeta_{34})}{\sin(\pi \zeta_{13})\, \sin(\pi \zeta_{24})} \to e^{\pi t(\nu_{12}+\nu_{34}-\nu_{13}-\nu_{24})} = e^{2\pi t (\nu_{2}-\nu_3)}~.  
\end{align}
}:
\begin{align}
    \left(\frac{\theta_1(\zeta_{12})\,\theta_1(\zeta_{34})}{\theta_1(\zeta_{13})\,\theta_1(\zeta_{24})}\,e^{2\pi t (\nu_3-\nu_2)}\right) ~, ~~ \left(\frac{\theta_1(\zeta_{23})\,\theta_1(\zeta_{14})}{\theta_1(\zeta_{13})\,\theta_1(\zeta_{24})}\right)\xrightarrow{\ \ \ t\to \infty \ \ \ \ } 1~. \label{thetas not contributing}
\end{align}
With all the $\eta_I$ large, one can convince oneself that the worldsheet has degenerated into a box diagram with four propagators: $X_{i0}~,$ and the amplitude becomes: 
\begin{align}
    \mathcal{A}^{\text{Ann}} \to \int_0^{\infty}\prod_{i=1}^4\dd \eta_{i0} \int \dd^dl\ e^{-\pi \sum_I\eta_IX_{I}} ~. 
\end{align}
The worldsheet coordinates $\eta_I$ have become the Schwinger parameters for the Feynman diagram. This box diagram corresponds to one of the vertices in the $\hat{D}_4$ combinatorial polytope. The triangles and bubbles in the combinatorial polytope correspond to the region in the moduli space where certain $\eta_I\to 0 ~.$ The moduli space allows for triangles, bubbles, and tadpoles, though the field theory limit of superstrings is special, and famously there are no triangles or bubbles in the one-loop amplitude of $\mathcal{N}=4$ SYM. The above analysis is sufficient for us to study the field theory limit of Möbius strip string amplitude, and we revisit the triangles and bubbles later in more detail.  

\subsection{Field theory limit of Möbius string }
The open string amplitude with Möbius strip topology is an integral over the moduli space of the Möbius strip with $n$ marked points on its boundary. We argue that in the high string tension limit, the moduli space degenerates into the combinatorial polytope $M_n$, and the string amplitude, i.e., integration over the worldsheet moduli space, turns into the curve integral formula discussed earlier. For concreteness, we discuss a four-particle amplitude. Analysis for arbitrary $n$ follows in the same manner. 

\begin{figure}[!ht]
    \centering
\begin{tikzpicture}[scale=0.7,baseline={([yshift=-.5ex]current bounding box.center)}]
        \draw[gray, <->] (-1,0) -- (4.5,0);
        \draw[gray, <->] (0,-1) -- (0,4.5);
        \draw[white,fill=teal!10] (0,0) -- (3.5,0) -- (3.5,3.5) -- (0,3.5) -- (0,0);
        \draw[very thick, purple!50!black] (0,0) -- (0,3.5);
        \draw[very thick, purple!50!black] (3.5,0) -- (3.5,3.5);
        \draw[thick, dashed, -{Stealth[length=3mm, width=2mm]}] (0,0) -- (1.95,0);
        \draw[thick, dashed] (1.6,0) -- (3.5,0);
        \draw[thick, dashed] (0,3.5) -- (1.85,3.5);
        \draw[thick, dashed, {Stealth[length=3mm, width=2mm]}-] (1.7,3.5) -- (3.5,3.5);
        \node at (0,5.1) {Im$(\zeta)$};
        \node at (5.4,0) {Re$(\zeta)$};
        \draw (4.5,4.9) -- (4.5,4.5) -- (4.9,4.5);
        \node at (4.8,4.8) {$\zeta$};
        \node at (3.5,-0.55) {$1/2$};
        \node at (3.5,-1.2) {$\sim it$};
        \node at (-0.55,3.5) {$it$};
        \node at (4.45,3.5) {$\frac{1}{2}+it$};
        \node at (4.45,2.8) {$\sim 0$};
        \node at (-0.35,-0.35) {0};
        \draw[fill=purple!50!black] (0,0) circle (0.07);
        \draw[fill=purple!50!black] (3.5,0) circle (0.07);
        \draw[fill=purple!50!black] (3.5,3.5) circle (0.07);
        \draw[fill=purple!50!black] (0,3.5) circle (0.07);
    \end{tikzpicture}
    \qquad 
    \begin{tikzpicture}[scale=0.7,baseline={([yshift=-.5ex]current bounding box.center)}]
        \draw[white,fill=teal!10] (4.4,0) arc (0:180:4.4) -- (-2,0) -- (-2,0) arc (180:0:2) -- (4.4,0); 
        \draw[gray, <->] (-5,0) -- (5,0);
        \draw[gray, <->] (0,-1) -- (0,5);
        \node at (5.9,0) {Re$(\rho)$};
        \node at (0,5.6) {Im$(\rho)$};
        \draw (4.4,4.7) -- (4.4,4.3) -- (4.8,4.3);
        \node at (5.7,4.8) {$\rho=e^{2\pi i \zeta}$};  
        \draw[thick, dashed, -{Stealth[length=3mm, width=2mm]}] (4.4,0) arc (0:90:4.4);
        \draw[thick, dashed] (-4.4,0) arc (180:85:4.4);
        \draw[thick, dashed] (2,0) arc (0:95:2);
        \draw[thick, dashed, {Stealth[length=3mm, width=2mm]}-] (0,2) arc (90:180:2);
        \node at (4.4,-0.4) {$1$};
        \node at (-4.4,-0.4) {$-1$};
        \node at (-0.4,-0.4) {$0$};
        \node at (2,-0.4) {$w$};
        \node at (-2,-0.4) {$-w$};
        \node at (-2,-0.9) {$\sim 1$};
        \node at (-4.4,-0.9) {$\sim w$};
        \node at (2.3,-0.9) {$=e^{-2\pi t}$};
        \draw[very thick, purple!50!black] (2,0) -- (4.4,0);
        \draw[very thick, purple!50!black] (-2,0) -- (-4.4,0);
        \draw[fill=purple!50!black] (4.4,0) circle (0.07);
        \draw[fill=purple!50!black] (2,0) circle (0.07);
        \draw[fill=purple!50!black] (-2,0) circle (0.07);
        \draw[fill=purple!50!black] (-4.4,0) circle (0.07);
    \end{tikzpicture}
    \caption{Möbius strip obtained by identifying $\rho$ uppper half plane as $\rho \sim -w\overline{\rho}~.$ The fundamental region under identification is shaded, and the thick line marks the single boundary of the Möbius strip.}
    \label{fig:Möbius strip obtained from identification}
\end{figure}
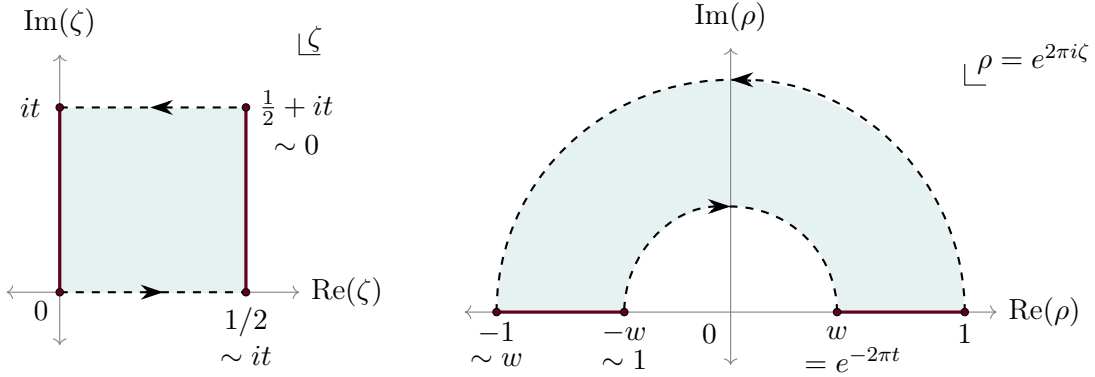

\paragraph{Constructing and parameterizing a Möbius strip:} 
Consider a complex half plane, parameterized by $\rho$. We obtain a Möbius strip by identifying the complex plane as follows: $    \rho \sim -\overline{\rho} \,w~.$ The fundamental region under this identification is depicted in Figure \ref{fig:Möbius strip obtained from identification}. Let us express $\rho$ as $\rho = e^{2\pi i \zeta}$. The identification for $\zeta$ is as follows:
\begin{align}
    \rho \sim -w\,\overline{\rho} ~, \qquad \quad \zeta \sim \frac{1}{2}+it-\overline{\zeta}~.
\end{align}
The factor $w=e^{-2\pi t}$ parameterizes the thickness of the Möbius strip. In the $\rho$ plane, the boundary of the Möbius strip is $[w,1]\cup [-1,-w]~.$ The single boundary appears divided into two parts in this parameterization. Hence, it is convenient to parameterize the boundary equivalently as $\rho \in [w^2,1]$ or $\zeta \in [0,2it]~.$ The region $\rho\in[w^2,w]$ is equivalent to $\rho\in[-w,-1]$ under the identification $\rho\sim -w\overline{\rho}~.$ With a given color ordering, the marked insertions will lie on this boundary in the following order, and we use the translation on the boundary to fix $\zeta_n=2it~:$ 
\begin{align}
    \zeta_i \equiv it\nu_i~, \quad \nu_1\leq \nu_2\leq \hdots \leq \nu_n=2~. 
\end{align}

\paragraph{Superstring amplitude:}
The open string amplitude, due to Möbius strip worldsheet, for massless external particles in type I superstring theory is given by the following \cite{Green:1987mn}:
\begin{align}
    \mathcal{A}^{\text{Mob}}\big[1234\big] = g_s^4 \,\tr\big(t_{a_1}t_{a_2}t_{a_3}t_{a_4}\big)\,t_8 \int_0^1\frac{\dd w}{w}\,\left(\frac{-2\pi}{\log w}\right)^5\int\prod_{r=1}^3\frac{\dd \rho_r}{\rho_r}\prod_{r<s}\big(\psi^N_{rs}\big)^{\alpha'\,k_r.k_s}~.
\end{align}
Note that $k_i^2=0$~. For $k_i^2=0$, the factor $\prod \psi^N$ is given by the following:
\begin{align}
    \prod_{r<s}\big(\psi^N_{rs}\big)^{k_r.k_s} = \prod_{\substack{r,s=1 \\ r<s}}^4e^{\pi (\alpha'\,k_r.k_s)\zeta_{rs}^2/t} \theta_1\left(\zeta_{rs}\,\bigg|\,\frac{1}{2}+it\right)^{\alpha'\,k_r.k_s}~, \qquad\quad \big(\zeta_{rs} = \zeta_s-\zeta_r\big)~.
\end{align}
Analogous to the annulus case, let us drop the prefactors $g_s^4 \,\tr(t_{a_1}\hdots)\,t_8$, change the integration variables to $t$ and $\zeta_r$, and introduce a gaussian loop integral \eqref{fake loop integral}, to obtain the following:
Plugging it back into the amplitude, we have:
\begin{align}
    \mathcal{A}^{\text{Mob}} &= \int_0^{\infty}\dd t\,\int_{0}^{2it}\prod_{r=1}^3\dd \zeta_r \int \dd^dl\,e^{-\pi t\alpha' \left(l^2 - 2 \sum_r\nu_r\,l.k_r \right)}    \ \prod_{r<s} \theta_1\left(\zeta_{rs}\,\bigg|\,\frac{1}{2}+it\right)^{\alpha'\,k_r.k_s}~. \label{Möbius string amplitude usable}
\end{align}

\paragraph{Obtaining box diagrams:}
We had constructed the invariants out of loop momenta $X_{ik}^{\ot} = (l+(p_1+\hdots + p_{i-1})+ (p_1+\hdots + p_{k-1}))^2$ while studying the curve integral formula. Let us trade off the $l^2~, 2l.k_r$ with $X_{ik}^{\ot}~.$ However, not all $X_{ik}^{\ot}$ are independent, and we have to choose a set of linearly independent $X_{ik}^{\ot}$'s. Let us choose the following four, to begin with: $X_{11}^{\ot}~, X_{12}^{\ot}~, X_{13}^{\ot}~, X_{14}^{\ot}~. $ We have the following identity:
\begin{align*}
    l^2 - 2 \sum_r\nu_r\,l.k_r = (\nu_1-1)X_{11}^{\ot} + (\nu_2-\nu_1)X_{12}^{\ot}+ (\nu_3-\nu_2)X_{13}^{\ot} + (2-\nu_3)X_{14}^{\ot} - (\nu_3-\nu_2)X_{13}~.
\end{align*}
We express \eqref{Möbius string amplitude usable} in terms of $X_{ik}^{\ot}$ as follows:
\begin{align}
    \mathcal{A}^{\text{Mob}} &= \int_0^{\infty}\dd t\,\int_{0}^{2it}\prod_{r=1}^3\dd \zeta_r \int \dd^dl\,e^{-\pi t\alpha' \left\{(\nu_1-1)X_{11}^{\ot} + (\nu_2-\nu_1)X_{12}^{\ot}+ (\nu_3-\nu_2)X_{13}^{\ot} + (2-\nu_3)X_{14}^{\ot} \right\}}  \nonumber \\
    &\hspace{1.5cm} \times \left(\frac{\theta_1(\zeta_{12})\,\theta_1(\zeta_{34})}{\theta_1(\zeta_{13})\,\theta_1(\zeta_{24})}\,e^{2\pi t (\nu_3-\nu_2)}\right)^{\alpha'X_{13}/2}\left(\frac{\theta_1(\zeta_{23})\,\theta_1(\zeta_{14})}{\theta_1(\zeta_{13})\,\theta_1(\zeta_{24})}\right)^{\alpha'X_{24}/2}~. 
\end{align}
Let us restrict ourselves to a particular region in the moduli space with $\nu_1>1~,$ so that we have the following configuration of the insertions on the boundary: 
\begin{align}
\label{configuration 11 12 13 14}
\begin{tikzpicture}[scale=0.7,baseline={([yshift=-.5ex]current bounding box.center)}]
        \draw[gray, <->] (-1,0) -- (4.5,0);
        \draw[gray, <->] (0,-1) -- (0,5);
        \draw[white,fill=teal!10] (0,0) -- (3.5,0) -- (3.5,4) -- (0,4) -- (0,0);
        \draw[very thick, purple!50!black] (0,0) -- (0,4);
        \draw[very thick, purple!50!black] (3.5,0) -- (3.5,4);
        \draw[thick, dashed, -{Stealth[length=3mm, width=2mm]}] (0,0) -- (1.95,0);
        \draw[thick, dashed] (1.6,0) -- (3.5,0);
        \draw[thick, dashed] (0,4) -- (1.85,4);
        \draw[thick, dashed, {Stealth[length=3mm, width=2mm]}-] (1.7,4) -- (3.5,4);
        \draw (4.5,4.9) -- (4.5,4.5) -- (4.9,4.5);
        \node at (4.8,4.8) {$\zeta$};
        \node at (3.5,-0.55) {$it$};
        \node at (-0.55,3.5) {$it$};
        \node at (4.7,4) {$\zeta_4=2it$};
        \node at (4,3) {$\zeta_3$};
        \node at (4,2) {$\zeta_2$};
        \node at (4,1) {$\zeta_1$};
        \node at (3.5,3.9) {$*$};
        \node at (3.5,2.9) {$*$};
        \node at (3.5,1.9) {$*$};
        \node at (3.5,0.9) {$*$};
        \node at (-0.35,-0.35) {0};
        \draw[fill=purple!50!black] (0,0) circle (0.07);
        \draw[fill=purple!50!black] (3.5,0) circle (0.07);
        \draw[fill=purple!50!black] (3.5,4) circle (0.07);
        \draw[fill=purple!50!black] (0,4) circle (0.07);
        \draw[<->, dashed] (6,0) -- (6,1);
        \draw[<->, dashed] (6,1) -- (6,2);
        \draw[<->, dashed] (6,2) -- (6,3);
        \draw[<->, dashed] (6,3) -- (6,4);
        \node at (6.6,0.5) {$i\eta_{11}$};
        \node at (6.6,1.5) {$i\eta_{12}$};
        \node at (6.6,2.5) {$i\eta_{13}$};
        \node at (6.6,3.5) {$i\eta_{14}$};
    \end{tikzpicture}    
    \hspace{2cm} \begin{matrix}
        \eta_{14}=t(2-\nu_3) \\
        \eta_{13}=t(\nu_3-\nu_2) \\
        \eta_{12} = t(\nu_2-\nu_1) \\
        \eta_{11}=t(\nu_1-1)
    \end{matrix} ~, \qquad \sum_{I\in T}\eta_I=t~.
\end{align}
The newly introduced variables $\eta_{I}\in [0,t]$ add up to $t~.$ The index ${}_I$ runs over ${(11),(12),(13),(14)}$, the chords in the particular triangulation $(T)$ chosen. {  We can change the integration variables to $\eta_I$, following \eqref{change of variables from t zeta to eta}, and the string integral written in terms of these variables is as follows: }
\begin{align}
    \mathcal{A}^{\text{Mob}} &\supset  \int_0^{\infty}\prod_{I\in T}\dd \eta_I\int \dd^dl\,e^{-\pi\alpha'\sum_I(\eta_IX_I^{\ot})} \nonumber \\
    &\hspace{1.5cm} \times\left(\frac{\theta_1(\zeta_{12})\,\theta_1(\zeta_{34})}{\theta_1(\zeta_{13})\,\theta_1(\zeta_{24})}\,e^{2\pi t (\nu_3-\nu_2)}\right)^{\alpha'X_{13}/2} \left(\frac{\theta_1(\zeta_{23})\,\theta_1(\zeta_{14})}{\theta_1(\zeta_{13})\,\theta_1(\zeta_{24})}\right)^{\alpha'X_{24}/2}~.
    \label{amplitude contribution from 11 12 13 14}
\end{align}
We wish to take the tropical limit of this integral. It is exactly the same as the annulus case \eqref{annulus case just before taking the tropical limit}. We proceed in the same manner: $\eta_I \to \eta_I/\alpha'$, and the factors with $X_{13}$ and $X_{24}$ go to 1 \eqref{thetas not contributing}. So, in the tropical limit of the region of the moduli space given by \eqref{configuration 11 12 13 14}, we have:
\begin{align}
    \mathcal{A}^{\text{Mob}} \supset \int_0^{\infty} \prod_{I\in T}\dd \eta_I\int \dd^dl\ e^{-\pi\sum_I(\eta_IX_I^{\ot})}~.
\end{align}
We have obtained the box diagram dual to complete triangulation $\{X_{11}^{\ot}\,,\ X_{12}^{\ot}\,,\ X_{13}^{\ot}\,,\ X_{14}^{\ot}\}~.$ 

In the above analysis, the marked points on the boundary of the Möbius strip were arranged such that $\zeta_1$ and $\zeta_4$ were far separated from each other, compared to the other pairs. With this, we argue that the worldsheet has degenerated into the following Feynman diagram:
\begin{align}
\begin{tikzpicture}[scale=0.45,baseline={([yshift=-.5ex]current bounding box.center)}]
    \fill[teal!10!white] (0,0) circle (2);
    \draw (0,0) circle (2);
    \node at (0,0) {$\ot$};
    \node at ({2*cos(-36)},{2*sin(-36)}) {$\times$};
    \node at ({2*cos(-72)},{2*sin(-72)}) {$\times$};
    \node at ({2*cos(-108)},{2*sin(-108)}) {$\times$};
    \node at ({2*cos(-144)},{2*sin(-144)}) {$\times$};
    \node at ({2.7*cos(-36)},{2.7*sin(-36)}) {$1$};
    \node at ({2.7*cos(-72)},{2.7*sin(-72)}) {$2$};
    \node at ({2.7*cos(-108)},{2.7*sin(-108)}) {$3$};
    \node at ({2.7*cos(-144)},{2.7*sin(-144)}) {$4$};
\end{tikzpicture} \xrightarrow{\text{  Tropicalization  }}
    \begin{tikzpicture}[scale=0.45,baseline={([yshift=-.5ex]current bounding box.center)}]
    \begin{scope}[rotate=0]
        \fill[teal!20!white] (-1,1) .. controls (-\change*2.5,1+\qq*0.2) .. (0,1+\qq*0.5) .. controls (-\change*2.5,1+\qq*0.8) .. (-1-\ror,1+\qq) -- (-1-\ror-0.8,1.8+\qq) -- (-\qq-1.8,1+\ror+0.8) --   (-1-\qq,1+\ror)  --   (-1-\qq,-1-\ror) -- (-\qq-1.8,-\ror-1.8) -- (-1-\ror-0.8,-\qq-1.8) --   (-1-\ror,-1-\qq) --   (1+\ror,-1-\qq) -- (1+\ror+0.8,-\qq-1.8) -- (1+\qq+0.8,-\ror-1.8) -- (1+\qq,-1-\ror) -- (1+\qq,1+\ror) -- (1+\qq+0.8,1+\ror+0.8) -- (1+\ror+0.8,1+\qq+0.8) -- (1+\ror,1+\qq) .. controls (\change*2.5,1+\qq*0.8) .. (0,1+\qq*0.5) .. controls (\change*2.5,1+\qq*0.2) .. (1,1) -- (1,-1) -- (-1,-1) -- (-1,1);
        \draw (-1,1) .. controls (-\change*2.5,1+\qq*0.2) .. (-\change*1.5,1+\qq*0.3);
        \draw (1+\ror,1+\qq) .. controls (\change*2.5,1+\qq*0.8) .. (\change*1.5,1+\qq*0.7);
        \draw (-1-\ror,1+\qq) .. controls (-\change,1+\qq) and (\change,1) .. (1,1);
		\draw (-1,-1) --  (1,-1) -- (1,1);
        \draw (-1,1) --  (-1,-1);
		\draw (1+\ror+0.8,-\qq-1.8) --   (1+\ror,-1-\qq) --   (-1-\ror,-1-\qq) --   (-1-\ror-0.8,-\qq-1.8);
		\draw (-\qq-1.8,-\ror-1.8) --   (-1-\qq,-1-\ror) --   (-1-\qq,1+\ror) --   (-\qq-1.8,1+\ror+0.8);
        \draw (-\ror-1.8,1+\qq+0.8) --   (-1-\ror,1+\qq);
        \draw (1+\ror,1+\qq) -- (1+\ror+0.8,1+\qq+0.8);
        \draw (1+\qq+0.8,1+\ror+0.8) -- (1+\qq,1+\ror) -- (1+\qq,-1-\ror) -- (1+\qq+0.8,-\ror-1.8);
    \end{scope}
    \node at (2.2,2.2) {${}_1$};
	\node at (2.2,-2.2) {${}_2$};
	\node at (-2.2,-2.2) {${}_3$};
	\node at (-2.2,2.2) {${}_4$};
    \draw[thick,-{Stealth[length=2mm, width=1.5mm]}] (2.07,0.5) -- (2.07,-0.5);
    \node at (3.7,0) {$l+p_1$};
\end{tikzpicture}
\end{align}

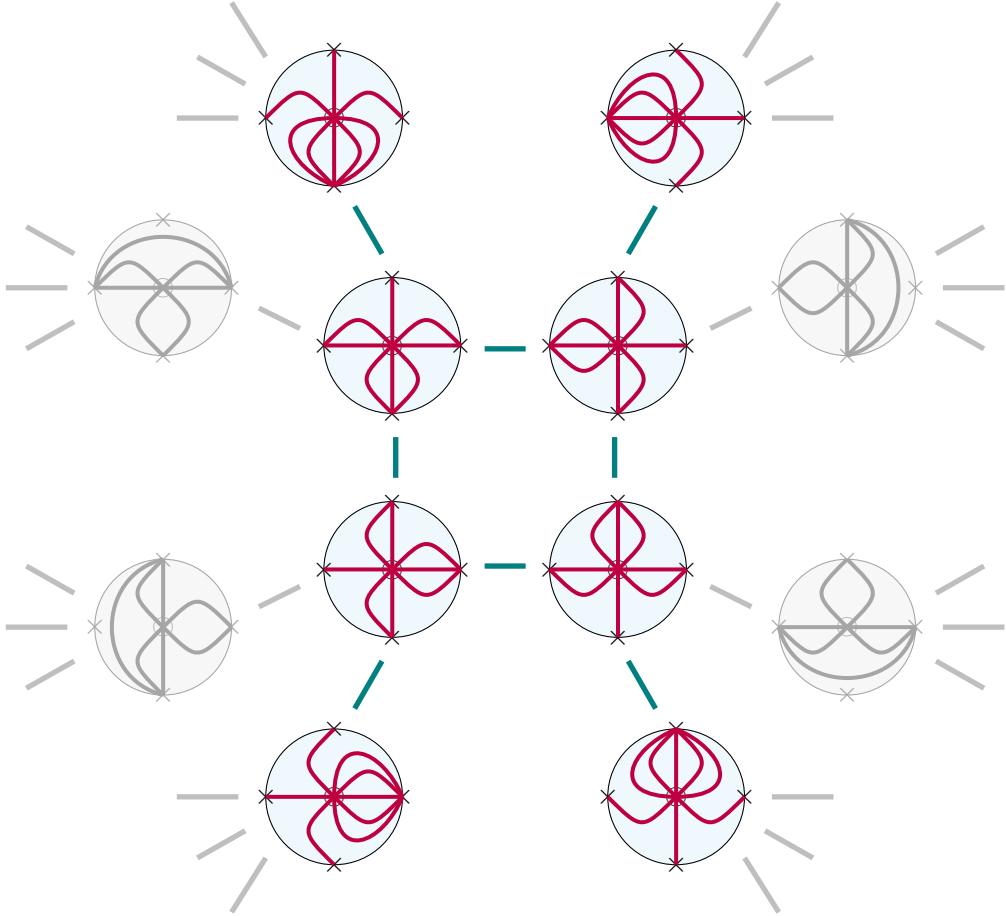
\begin{figure}[!t]
    \centering
\begin{tikzpicture}[scale=0.9,baseline={([yshift=-.5ex]current bounding box.center)}]
    \draw[line width=2,teal] (-0.3,1.6) -- (0.3,1.6);    
    \draw[line width=2,teal] (-0.3,-1.6) -- (0.3,-1.6);
    \draw[line width=2,teal] (1.6,-0.3) -- (1.6,0.3);  
    \draw[line width=2,teal] (-1.6,-0.3) -- (-1.6,0.3);
    \draw[line width=2,teal] (-1.8,3) -- (-2.2,3.7);
    \draw[line width=2,teal] (-1.8,-3) -- (-2.2,-3.7);
    \draw[line width=2,teal] (1.8,3) -- (2.2,3.7);
    \draw[line width=2,teal] (1.8,-3) -- (2.2,-3.7);
    \draw[line width=2,gray!50!white] (3,1.9) -- (3.6,2.2);
    \draw[line width=2,gray!50!white] (-3,1.9) -- (-3.6,2.2);
    \draw[line width=2,gray!50!white] (3,-1.9) -- (3.6,-2.2);
    \draw[line width=2,gray!50!white] (-3,-1.9) -- (-3.6,-2.2);
    \draw[line width=2,gray!50!white] (-6.3,3) -- (-7,3.4);
    \draw[line width=2,gray!50!white] (-6.4,2.5) -- (-7.3,2.5);
    \draw[line width=2,gray!50!white] (-6.3,2) -- (-7,1.6);
    \draw[line width=2,gray!50!white] (6.3,3) -- (7,3.4);
    \draw[line width=2,gray!50!white] (6.4,2.5) -- (7.3,2.5);
    \draw[line width=2,gray!50!white] (6.3,2) -- (7,1.6);
    \draw[line width=2,gray!50!white] (-6.3,-3) -- (-7,-3.4);
    \draw[line width=2,gray!50!white] (-6.4,-2.5) -- (-7.3,-2.5);
    \draw[line width=2,gray!50!white] (-6.3,-2) -- (-7,-1.6);
    \draw[line width=2,gray!50!white] (6.3,-3) -- (7,-3.4);
    \draw[line width=2,gray!50!white] (6.4,-2.5) -- (7.3,-2.5);
    \draw[line width=2,gray!50!white] (6.3,-2) -- (7,-1.6);
    \draw[line width=2,gray!50!white] (-3.8,-5.5) -- (-4.5,-5.9);
    \draw[line width=2,gray!50!white] (-3.9,-5) -- (-4.8,-5);
    \draw[line width=2,gray!50!white] (-3.5,-5.9) -- (-4,-6.7);
    \draw[line width=2,gray!50!white] (3.8,-5.5) -- (4.5,-5.9);
    \draw[line width=2,gray!50!white] (3.9,-5) -- (4.8,-5);
    \draw[line width=2,gray!50!white] (3.5,-5.9) -- (4,-6.7);
    \draw[line width=2,gray!50!white] (-3.8,5.5) -- (-4.5,5.9);
    \draw[line width=2,gray!50!white] (-3.9,5) -- (-4.8,5);
    \draw[line width=2,gray!50!white] (-3.5,5.9) -- (-4,6.7);
    \draw[line width=2,gray!50!white] (3.8,5.5) -- (4.5,5.9);
    \draw[line width=2,gray!50!white] (3.9,5) -- (4.8,5);
    \draw[line width=2,gray!50!white] (3.5,5.9) -- (4,6.7);
    \begin{scope}[scale=0.5,xshift=-3.3cm, yshift=3.3cm]
        \fill[cyan!6!white] (0,0) circle (2);
        \draw (0,0) circle (2);
        \node at (0,0) {$\ot$};
        \node at (0,2) {$\times$};
        \node at (0,-2) {$\times$};
        \node at (2,0) {$\times$};
        \node at (-2,0) {$\times$};
        \draw[line width=1.5, purple] (-2,0) -- (2,0);
        \draw[line width=1.5, purple] (0,-2) -- (0,2);
        \draw[line width=1.5, purple] (-2,0) .. controls (-1,1) .. (0,0) .. controls (1,-1) .. (0,-2);            
        \draw[line width=1.5, purple] (2,0) .. controls (1,1) .. (0,0) .. controls (-1,-1) .. (0,-2); 
    \end{scope}
	\begin{scope}[scale=0.5,xshift=3.3cm, yshift=3.3cm]
        \fill[cyan!6!white] (0,0) circle (2);
        \draw (0,0) circle (2);
        \node at (0,0) {$\ot$};
        \node at (0,2) {$\times$};
        \node at (0,-2) {$\times$};
        \node at (2,0) {$\times$};
        \node at (-2,0) {$\times$};
        \draw[line width=1.5, purple] (-2,0) -- (2,0);
        \draw[line width=1.5, purple] (0,-2) -- (0,2);
        \draw[line width=1.5, purple] (-2,0) .. controls (-1,1) .. (0,0) .. controls (1,-1) .. (0,-2);            
        \draw[line width=1.5, purple] (0,2) .. controls (1,1) .. (0,0) .. controls (-1,-1) .. (-2,0); 
    \end{scope}
	\begin{scope}[scale=0.5,xshift=-3.3cm, yshift=-3.3cm]
        \fill[cyan!6!white] (0,0) circle (2);
        \draw (0,0) circle (2);
        \node at (0,0) {$\ot$};
        \node at (0,2) {$\times$};
        \node at (0,-2) {$\times$};
        \node at (2,0) {$\times$};
        \node at (-2,0) {$\times$};
        \draw[line width=1.5, purple] (-2,0) -- (2,0);
        \draw[line width=1.5, purple] (0,-2) -- (0,2);
        \draw[line width=1.5, purple] (2,0) .. controls (1,-1) .. (0,0) .. controls (-1,1) .. (0,2);           
        \draw[line width=1.5, purple] (2,0) .. controls (1,1) .. (0,0) .. controls (-1,-1) .. (0,-2); 
    \end{scope}
	\begin{scope}[scale=0.5,xshift=3.3cm, yshift=-3.3cm]
        \fill[cyan!6!white] (0,0) circle (2);
        \draw (0,0) circle (2);
        \node at (0,0) {$\ot$};
        \node at (0,2) {$\times$};
        \node at (0,-2) {$\times$};
        \node at (2,0) {$\times$};
        \node at (-2,0) {$\times$};
        \draw[line width=1.5, purple] (-2,0) -- (2,0);
        \draw[line width=1.5, purple] (0,-2) -- (0,2);
        \draw[line width=1.5, purple] (2,0) .. controls (1,-1) .. (0,0) .. controls (-1,1) .. (0,2);           
        \draw[line width=1.5, purple] (-2,0) .. controls (-1,-1) .. (0,0) .. controls (1,1) .. (0,2); 
    \end{scope}
    \begin{scope}[scale=0.5,xshift=-5cm, yshift=10cm]
        \fill[cyan!6!white] (0,0) circle (2);
        \draw (0,0) circle (2);
        \node at (0,0) {$\ot$};
        \node at (0,2) {$\times$};
        \node at (0,-2) {$\times$};
        \node at (2,0) {$\times$};
        \node at (-2,0) {$\times$};
        \draw[line width=1.5, purple] (0,-2) .. controls (-1.4,-1.4) and (-2,0) .. (0,0) .. controls (2,0) and (1.4,-1.4) .. (0,-2);
        \draw[line width=1.5, purple] (0,-2) -- (0,2);
        \draw[line width=1.5, purple] (-2,0) .. controls (-1,1) .. (0,0) .. controls (1,-1) .. (0,-2);            
        \draw[line width=1.5, purple] (2,0) .. controls (1,1) .. (0,0) .. controls (-1,-1) .. (0,-2); 
    \end{scope}
	\begin{scope}[scale=0.5,xshift=5cm, yshift=10cm]
        \fill[cyan!6!white] (0,0) circle (2);
        \draw (0,0) circle (2);
        \node at (0,0) {$\ot$};
        \node at (0,2) {$\times$};
        \node at (0,-2) {$\times$};
        \node at (2,0) {$\times$};
        \node at (-2,0) {$\times$};
        \draw[line width=1.5, purple] (-2,0) -- (2,0);
        \draw[line width=1.5, purple] (-2,0) .. controls (-1.4,-1.4) and (0,-2) .. (0,0) .. controls (0,2) and (-1.4,1.4) .. (-2,0);
        \draw[line width=1.5, purple] (-2,0) .. controls (-1,1) .. (0,0) .. controls (1,-1) .. (0,-2);            
        \draw[line width=1.5, purple] (0,2) .. controls (1,1) .. (0,0) .. controls (-1,-1) .. (-2,0); 
    \end{scope}
    \begin{scope}[scale=0.5,xshift=-5cm, yshift=-10cm]
        \fill[cyan!6!white] (0,0) circle (2);
        \draw (0,0) circle (2);
        \node at (0,0) {$\ot$};
        \node at (0,2) {$\times$};
        \node at (0,-2) {$\times$};
        \node at (2,0) {$\times$};
        \node at (-2,0) {$\times$};
        \draw[line width=1.5, purple] (-2,0) -- (2,0);
        \draw[line width=1.5, purple] (2,0) .. controls (1.4,-1.4) and (0,-2) .. (0,0) .. controls (0,2) and (1.4,1.4) .. (2,0);
        \draw[line width=1.5, purple] (2,0) .. controls (1,-1) .. (0,0) .. controls (-1,1) .. (0,2);           
        \draw[line width=1.5, purple] (2,0) .. controls (1,1) .. (0,0) .. controls (-1,-1) .. (0,-2); 
    \end{scope}
	\begin{scope}[scale=0.5,xshift=5cm, yshift=-10cm]
        \fill[cyan!6!white] (0,0) circle (2);
        \draw (0,0) circle (2);
        \node at (0,0) {$\ot$};
        \node at (0,2) {$\times$};
        \node at (0,-2) {$\times$};
        \node at (2,0) {$\times$};
        \node at (-2,0) {$\times$};
        \draw[line width=1.5, purple] (0,2) .. controls (-1.4,1.4) and (-2,0) .. (0,0) .. controls (2,0) and (1.4,1.4) .. (0,2);
        \draw[line width=1.5, purple] (0,-2) -- (0,2);
        \draw[line width=1.5, purple] (2,0) .. controls (1,-1) .. (0,0) .. controls (-1,1) .. (0,2);           
        \draw[line width=1.5, purple] (-2,0) .. controls (-1,-1) .. (0,0) .. controls (1,1) .. (0,2); 
    \end{scope}
    \begin{scope}[scale=0.5,xshift=-10cm, yshift=5cm, gray!70!white]
        \fill[gray!6!white] (0,0) circle (2);
        \draw (0,0) circle (2);
        \node at (0,0) {$\ot$};
        \node at (0,2) {$\times$};
        \node at (0,-2) {$\times$};
        \node at (2,0) {$\times$};
        \node at (-2,0) {$\times$};
        \draw[line width=1.5] (-2,0) -- (2,0);
        \draw[line width=1.5] (-2,0) .. controls (-1.5,2) and (1.5,2) .. (2,0);
        \draw[line width=1.5] (-2,0) .. controls (-1,1) .. (0,0) .. controls (1,-1) .. (0,-2);            
        \draw[line width=1.5] (2,0) .. controls (1,1) .. (0,0) .. controls (-1,-1) .. (0,-2); 
    \end{scope}
	\begin{scope}[scale=0.5,xshift=10cm, yshift=5cm, gray!70!white]
        \fill[gray!6!white] (0,0) circle (2);
        \draw (0,0) circle (2);
        \node at (0,0) {$\ot$};
        \node at (0,2) {$\times$};
        \node at (0,-2) {$\times$};
        \node at (2,0) {$\times$};
        \node at (-2,0) {$\times$};
        \draw[line width=1.5] (0,2) .. controls (2,1.5) and (2,-1.5) .. (0,-2);
        \draw[line width=1.5] (0,-2) -- (0,2);
        \draw[line width=1.5] (-2,0) .. controls (-1,1) .. (0,0) .. controls (1,-1) .. (0,-2);            
        \draw[line width=1.5] (0,2) .. controls (1,1) .. (0,0) .. controls (-1,-1) .. (-2,0); 
    \end{scope}
	\begin{scope}[scale=0.5,xshift=-10cm, yshift=-5cm, gray!70!white]
        \fill[gray!6!white] (0,0) circle (2);
        \draw (0,0) circle (2);
        \node at (0,0) {$\ot$};
        \node at (0,2) {$\times$};
        \node at (0,-2) {$\times$};
        \node at (2,0) {$\times$};
        \node at (-2,0) {$\times$};
        \draw[line width=1.5] (0,2) .. controls (-2,1.5) and (-2,-1.5) .. (0,-2);
        \draw[line width=1.5] (0,-2) -- (0,2);
        \draw[line width=1.5] (2,0) .. controls (1,-1) .. (0,0) .. controls (-1,1) .. (0,2);           
        \draw[line width=1.5] (2,0) .. controls (1,1) .. (0,0) .. controls (-1,-1) .. (0,-2); 
    \end{scope}
	\begin{scope}[scale=0.5,xshift=10cm, yshift=-5cm, gray!70!white]
        \fill[gray!6!white] (0,0) circle (2);
        \draw (0,0) circle (2);
        \node at (0,0) {$\ot$};
        \node at (0,2) {$\times$};
        \node at (0,-2) {$\times$};
        \node at (2,0) {$\times$};
        \node at (-2,0) {$\times$};
        \draw[line width=1.5] (-2,0) -- (2,0);
        \draw[line width=1.5] (2,0) .. controls (1.5,-2) and (-1.5,-2) .. (-2,0);
        \draw[line width=1.5] (2,0) .. controls (1,-1) .. (0,0) .. controls (-1,1) .. (0,2);           
        \draw[line width=1.5] (-2,0) .. controls (-1,-1) .. (0,0) .. controls (1,1) .. (0,2); 
    \end{scope}
\end{tikzpicture}
    \caption{A portion of the combinatorial polytope $M_4$ showing the \emph{eight} vertices (complete triangulations) corresponding to eight \emph{box} diagrams in the moduli space of a Möbius strip with four marked points, as shown in Figure 8.13 of \cite{Green:1987mn}. The four gray triangulations are triangle graphs, and more triangulations are omitted. Look at Table \ref{tab:eight box diagrams coming from Möbius4} to see dual box diagrams.}
    \label{fig:the eight box diagrams for Möbius 4 from Green Schwarz Witten}
\end{figure}

We chose a particular linear independent set $\{X_{11}^{\ot}, ~X_{12}^{\ot}, ~X_{13}^{\ot}, ~X_{14}^{\ot}\}$ earlier, demanded that the worldsheet coordinates $\eta_I$ in exponent in the amplitude were positive, and obtained a particular box diagram via tropicalization. There are eight such linearly independent sets of propagators $\{X_I\}$ corresponding to eight box diagrams in the combinatorial polytope $M_4$, as depicted in Figure \ref{fig:the eight box diagrams for Möbius 4 from Green Schwarz Witten}. Repeating the same exercise with any such set, and restricting to a particular region in the moduli space yields a particular box diagram. The list of all eight diagrams appearing in the different regions of the moduli space is given in Table \ref{tab:eight box diagrams coming from Möbius4}~. We illustrate the idea via one more detailed example: $\{X_{24}^{\ot},~X_{12}^{\ot},~X_{13}^{\ot},~X_{14}^{\ot}\}~.$ The string amplitude \eqref{Möbius string amplitude usable} written in the basis $\{X_{24}^{\ot},~X_{12}^{\ot},~X_{13}^{\ot},~X_{14}^{\ot}\}$ is as follows:
\begin{align}
    \mathcal{A}^{\text{Mob}} &= \int_0^{\infty}\dd t\,\int_{0}^{2it}\prod_{r=1}^3\dd \zeta_r \int \dd^dl\,e^{-\pi t\alpha' \left\{(1-\nu_1)X_{24}^{\ot} + (\nu_2-1)X_{12}^{\ot}+ (\nu_3-\nu_2)X_{13}^{\ot} + (1+\nu_1-\nu_3)X_{14}^{\ot} \right\}}  \nonumber \\
    &\hspace{0cm} \times \left(\frac{\theta_1(\zeta_{12})\,\theta_1(\zeta_{34})}{\theta_1(\zeta_{13})\,\theta_1(\zeta_{24})}\,e^{2\pi t (\nu_3-\nu_2)}\right)^{\alpha'X_{13}/2}\left(\frac{\theta_1(\zeta_{23})\,\theta_1(\zeta_{14})}{\theta_1(\zeta_{13})\,\theta_1(\zeta_{24})}\,e^{-2\pi t(1-\nu_1)}\right)^{\alpha'X_{24}/2}~. 
\end{align}
Let us restrict ourselves to the following configuration of the marked points, such that the coefficients of $X_{ik}^{\ot}$ in the exponent are positive, and eventually turn into Schwinger parameters:

\newpage
\begin{table}[H]
    \centering
\begin{tabular}{c c c c c c c}
          \begin{tikzpicture}[scale=0.399,baseline={([yshift=-.5ex]current bounding box.center)}]
    \fill[cyan!6!white] (0,0) circle (2);
        \draw (0,0) circle (2);
        \node at (0,0) {$\ot$};
        \node at (0,2) {$\times$};
        \node at (0,-2) {$\times$};
        \node at (2,0) {$\times$};
        \node at (-2,0) {$\times$};
        \draw[line width=1.2, purple] (0,2) .. controls (-1.4,1.4) and (-2,0) .. (0,0) .. controls (2,0) and (1.4,1.4) .. (0,2);
        \draw[line width=1.2, purple] (0,-2) -- (0,2);
        \draw[line width=1.2, purple] (2,0) .. controls (1,-1) .. (0,0) .. controls (-1,1) .. (0,2);     
        \draw[line width=1.2, purple] (-2,0) .. controls (-1,-1) .. (0,0) .. controls (1,1) .. (0,2);
        \node at (0.7,2.6) {$1'$};
        \node at (2.8,0) {$2'$};
        \node at (-0.7,-2.6) {$3'$};
        \node at (-2.8,0) {$4'$};
        \end{tikzpicture} 
        & \qquad
        & $\begin{matrix}
      (\nu_1-1)X_{11}^{\ot} \\ +(\nu_2-\nu_1)X_{12}^{\ot} \\ + (\nu_3-\nu_2)X_{13}^{\ot} \\ + (2-\nu_3)X_{14}^{\ot}
 \end{matrix}$ 
        &\qquad 
        & \begin{tikzpicture}[scale=0.5,baseline={([yshift=-.5ex]current bounding box.center)}]
    \fill[teal!6!white] (0,0) rectangle (3,4);
    \draw[line width=1] (0,0) -- (0,4);
    \draw[dashed,-{Stealth[length=2mm, width=1.5mm]}] (0,0) -- (1.7,0);
    \draw[dashed,-{Stealth[length=2mm, width=1.5mm]}] (3,4) -- (1.3,4);
    \draw[dashed] (1.6,0) -- (3,0);
    \draw[dashed] (0,4) -- (1.4,4);
    \draw[line width=1] (3,0) -- (3,4);
    \node at (3,1-0.1) {\textbf{*}};
    \node at (3,2-0.1) {\textbf{*}};
    \node at (3,3-0.1) {\textbf{*}};
    \node at (3,4-0.1) {\textbf{*}};
    \node at (3.75,1) {$\nu_1$};
    \node at (3.75,2) {$\nu_2$};
    \node at (3.75,3) {$\nu_3$};
    \node at (3.6,4.6) {$\nu_4$};
    \node at (4,3.9) {\small (=2)};
    \node at (-0.6,0) {0};
\end{tikzpicture}
        & \qquad 
        & \begin{tikzpicture}[scale=0.45,baseline={([yshift=-.5ex]current bounding box.center)}]
    \begin{scope}[rotate=0]
        \fill[teal!20!white] (-1,1) .. controls (-\change*2.5,1+\qq*0.2) .. (0,1+\qq*0.5) .. controls (-\change*2.5,1+\qq*0.8) .. (-1-\ror,1+\qq) -- (-1-\ror-0.8,1.8+\qq) -- (-\qq-1.8,1+\ror+0.8) --   (-1-\qq,1+\ror)  --   (-1-\qq,-1-\ror) -- (-\qq-1.8,-\ror-1.8) -- (-1-\ror-0.8,-\qq-1.8) --   (-1-\ror,-1-\qq) --   (1+\ror,-1-\qq) -- (1+\ror+0.8,-\qq-1.8) -- (1+\qq+0.8,-\ror-1.8) -- (1+\qq,-1-\ror) -- (1+\qq,1+\ror) -- (1+\qq+0.8,1+\ror+0.8) -- (1+\ror+0.8,1+\qq+0.8) -- (1+\ror,1+\qq) .. controls (\change*2.5,1+\qq*0.8) .. (0,1+\qq*0.5) .. controls (\change*2.5,1+\qq*0.2) .. (1,1) -- (1,-1) -- (-1,-1) -- (-1,1);
        \draw (-1,1) .. controls (-\change*2.5,1+\qq*0.2) .. (-\change*1.5,1+\qq*0.3);
        \draw (1+\ror,1+\qq) .. controls (\change*2.5,1+\qq*0.8) .. (\change*1.5,1+\qq*0.7);
        \draw (-1-\ror,1+\qq) .. controls (-\change,1+\qq) and (\change,1) .. (1,1);
		\draw (-1,-1) --  (1,-1) -- (1,1);
        \draw (-1,1) --  (-1,-1);
		\draw (1+\ror+0.8,-\qq-1.8) --   (1+\ror,-1-\qq) --   (-1-\ror,-1-\qq) --   (-1-\ror-0.8,-\qq-1.8);
		\draw (-\qq-1.8,-\ror-1.8) --   (-1-\qq,-1-\ror) --   (-1-\qq,1+\ror) --   (-\qq-1.8,1+\ror+0.8);
        \draw (-\ror-1.8,1+\qq+0.8) --   (-1-\ror,1+\qq);
        \draw (1+\ror,1+\qq) -- (1+\ror+0.8,1+\qq+0.8);
        \draw (1+\qq+0.8,1+\ror+0.8) -- (1+\qq,1+\ror) -- (1+\qq,-1-\ror) -- (1+\qq+0.8,-\ror-1.8);
    \end{scope}
    \node at (2.2,2.2) {${}_1$};
	\node at (2.2,-2.2) {${}_2$};
	\node at (-2.2,-2.2) {${}_3$};
	\node at (-2.2,2.2) {${}_4$};
    \draw[thick,-{Stealth[length=2mm, width=1.5mm]}] (2.07,0.5) -- (2.07,-0.5);
    \node at (3,0) {$P_{12}^{\ot}$};
\end{tikzpicture} \\
\end{tabular}
    \caption{Feynman diagrams appearing out of moduli space of Möbius${}_4$. Second column is the exponent in the amplitude \eqref{Möbius string amplitude usable} expressed in terms of $X_{ik}^{\ot}$ in the triangulation. Third coulmn shows the configuration of marked points on the Möbius strip boundary. Fourth column shows the resultant box Feynman diagram with all external momenta incoming, and $P_{ik}^{\ot}=l+(p_1+\hdots+p_{i-1})+(p_1+\hdots+p_{k-1})~.$}
    \label{tab:eight box diagrams coming from Möbius4}
\end{table}

\newpage

\begin{align}
\label{configuration 24 12 13 14}
\begin{tikzpicture}[scale=0.7,baseline={([yshift=-.5ex]current bounding box.center)}]
        \draw[gray, <->] (-1,0) -- (4.5,0);
        \draw[gray, <->] (0,-1) -- (0,5);
        \draw[white,fill=teal!10] (0,0) -- (3.5,0) -- (3.5,4) -- (0,4) -- (0,0);
        \draw[very thick, purple!50!black] (0,0) -- (0,4);
        \draw[very thick, purple!50!black] (3.5,0) -- (3.5,4);
        \draw[thick, dashed, -{Stealth[length=3mm, width=2mm]}] (0,0) -- (1.95,0);
        \draw[thick, dashed] (1.6,0) -- (3.5,0);
        \draw[thick, dashed] (0,4) -- (1.85,4);
        \draw[thick, dashed, {Stealth[length=3mm, width=2mm]}-] (1.7,4) -- (3.5,4);
        \draw (4.5,4.9) -- (4.5,4.5) -- (4.9,4.5);
        \node at (4.8,4.8) {$\zeta$};
        \node at (3.5,-0.55) {$it$};
        \node at (-0.55,4.2) {$it$};
        \node at (4.7,4) {$\zeta_4=2it$};
        \node at (-0.5,3) {$\zeta_1$};
        \node at (4,2.4) {$\zeta_3$};
        \node at (4,1) {$\zeta_2$};
        \node at (3.5,3.9) {$*$};
        \node at (0,2.9) {$*$};
        \node at (3.5,1.96) {$*$};
        \node at (3.5,0.9) {$*$};
        \node at (-0.35,-0.35) {0};
        \draw[fill=purple!50!black] (0,0) circle (0.07);
        \draw[fill=purple!50!black] (3.5,0) circle (0.07);
        \draw[fill=purple!50!black] (3.5,4) circle (0.07);
        \draw[fill=purple!50!black] (0,4) circle (0.07);
        \draw[<->, dashed] (5,0) -- (5,1);
        \draw[<->, dashed] (5,1) -- (5,2);
        \draw[<->, dashed] (-1.3,2) -- (-1.3,3);
        \draw[<->, dashed] (-1.3,3) -- (-1.3,4);
        \node at (5.6,0.5) {$i\eta'_{12}$};
        \node at (5.6,1.5) {$i\eta'_{13}$};
        \node at (-1.9,2.5) {$i\eta'_{14}$};
        \node at (-1.9,3.5) {$i\eta'_{24}$};
        \draw[thin, dotted] (-1.5,2) -- (5.5,2);
    \end{tikzpicture}    
    \hspace{2cm} \begin{matrix}
        \eta'_{24}=t(1-\nu_1) \\
        \eta'_{14}=t(1+\nu_1-\nu_3) \\
        \eta'_{13} = t(\nu_3-\nu_2) \\
        \eta'_{12}=t(\nu_2-1)
    \end{matrix} ~, \qquad \sum_{I\in T}\eta_I=t~.
\end{align}

Changing the integration variables to $\eta'_I$, proceeding just like the earlier example, we have:
\begin{align}
    \mathcal{A}^{\text{Mob}} &\supset \int_0^{\infty}\prod_{I\in T'}\dd \eta'_I\int \dd^dl\,e^{-\pi\alpha'\sum_I(\eta'_IX_I^{\ot})} \times\left(\frac{\theta_1(\zeta_{12})\,\theta_1(\zeta_{34})}{\theta_1(\zeta_{13})\,\theta_1(\zeta_{24})}\,e^{2\pi t (\nu_3-\nu_2)}\right)^{\alpha'X_{13}/2}\nonumber \\
    &\hspace{5cm} \times\left(\frac{\theta_1(\zeta_{23})\,\theta_1(\zeta_{14})}{\theta_1(\zeta_{13})\,\theta_1(\zeta_{24})}\,e^{-2\pi t(1-\nu_1)}\right)^{\alpha'X_{24}/2}~.
\end{align}
Taking the tropical limit: $\eta'_I\to \eta'_I/\alpha'$ and $\alpha'\to 0~,$ the factors involving $X_{13}$ and $X_{24}$ drops out\footnote{Keeping in the mind the identification $\zeta=2it\sim 0$, in the $\{X_{24}^{\ot},~X_{12}^{\ot},~X_{13}^{\ot},~X_{14}^{\ot}\}$ configuration, we have $\zeta_{23}=2it-i\eta'_{13}~, \zeta_{14}=i(t+\eta'_{24})~, \zeta_{13}=i(t+\eta'_{14})~, \zeta_{24}=i(t+i\eta_{12}')~,$ and so the limit of the ratio of theta functions is modified. For instance, 
\begin{align}
    \frac{\theta_1(\zeta_{23})\,\theta_1(\zeta_{14})}{\theta_1(\zeta_{13})\,\theta_1(\zeta_{24})} \to \frac{\sin(\pi\zeta_{23})\,\sin(\pi\zeta_{14})}{\sin(\pi\zeta_{13})\,\sin(\pi\zeta_{24})} \to \frac{e^{\pi(2t- \eta'_{13})}\, e^{\pi (t+\eta_{24}') } }{ e^{\pi (t+\eta_{14}') }\, e^{\pi (t+\eta_{12}') } }  \to e^{2\pi\eta'_{24}} = e^{2\pi t(1-\nu_1)}~.
\end{align}}, and we are left with the box diagram: 
\begin{align}
    \mathcal{A}^{\text{Mob}} &\supset \int_0^{\infty}\prod_{I\in T'}\dd \eta'_I\int \dd^dl\,e^{-\pi\alpha'\sum_I(\eta'_IX_I^{\ot})}~.
\end{align}
In this case, one can interpret the degenerated worldsheet as a box diagram with twists in three propagators, as shown in Table \ref{tab:eight box diagrams coming from Möbius4}.

\paragraph{Obtaining triangles and bubbles:}
The triangle and bubble graphs correspond to the regions in the moduli space where two marked points come close to each other. One can parameterize the moduli space integral appropriately to make such graphs manifest. However, we have been analyzing type-I superstring amplitudes and their field theory limit is the $\mathcal{N}=4$ SYM, and famously, the one-loop amplitudes of $\mathcal{N}=4$ SYM do not have any triangle or box contributions. We illustrate the idea via an example, and then briefly discuss how the triangles and bubbles appear in non-supersymmetric one-loop string amplitudes. 

Consider the region of the moduli space of Möbius strip with four insertions such that $1\leq \nu_1\leq \nu_2\leq \nu_3\leq \nu_4=2~,$ as depicted in \eqref{configuration 11 12 13 14}. The contribution to the amplitude from this region is given by \eqref{amplitude contribution from 11 12 13 14}:
\begin{align}
    \mathcal{A}^{\text{Mob}}[1234] &\supset \int_0^{\infty}\prod_{I\in T}\dd \eta_I\int \dd^dl\,e^{-\pi\alpha'\sum_I(\eta_IX_I^{\ot})} \, \mathfrak{F}(\eta_I)~, \nonumber \\
    \mathfrak{F}(\eta_I) &= \left(\frac{\theta_1(\zeta_{12})\,\theta_1(\zeta_{34})}{\theta_1(\zeta_{13})\,\theta_1(\zeta_{24})}\,e^{2\pi t (\nu_3-\nu_2)}\right)^{\alpha'X_{13}/2} \left(\frac{\theta_1(\zeta_{23})\,\theta_1(\zeta_{14})}{\theta_1(\zeta_{13})\,\theta_1(\zeta_{24})}\right)^{\alpha'X_{24}/2}~.
\end{align}
Recall that $I\in\{11,12,13,14\}~.$ Let us change variables appropriately such that the triangle graph $\{X^{\ot}_{11},~X^{\ot}_{12},~X^{\ot}_{13},~X_{13}\}$ is made manifest as well:
\begin{align}
    e^{-\pi \eta_{14}} \equiv \frac{e^{-\vp}}{1+e^{-\varphi}} ~, \qquad \Rightarrow \quad \int_0^{\infty}\dd \eta_{14} = \frac{1}{\pi}\int_{-\infty}^{\infty}\dd \vp \,\frac{1}{1+e^{-\vp}}~. 
\end{align}

Taking the tropical limit for $\eta_{11}~,\eta_{12}~,\eta_{13}$, but keeping $\eta_{14}$ arbitrary, one obtains $\mathfrak{F}$ to be dependent only on $\eta_{14}$: 
\begin{align}
    \mathfrak{F}(\eta_{14}) = \left(\frac{1+2e^{-\vp}}{(1+e^{-\vp})^2}\right)^{\alpha'X_{13}/2}~. 
\end{align}
Note that in the tropical limit of $\eta_{14}~,$ we revert back to $\mathfrak{F}(\eta_{14}\to \infty) \to 1~.$ So, the contribution to the amplitude is as follows:
\begin{align}
    \mathcal{A}^{\text{Mob}}[1234] &\supset \int_0^{\infty}\prod_{I\in T}\dd \eta_I\int \dd^dl\,e^{-\pi\alpha'(\eta_{11}X_{11}^{\ot} + \eta_{12}X_{12}^{\ot} + \eta_{13}X_{13}^{\ot})}\nonumber \\
    &\hspace{1.5cm} \times\frac{1}{\pi}\int_{-\infty}^{\infty}\frac{\dd \vp}{(1+e^{-\vp})} \left(\frac{e^{-\vp}}{1+e^{-\vp}}\right)^{\alpha' X^{\ot}_{14}} \left(\frac{1+2e^{-\vp}}{(1+e^{-\vp})^2}\right)^{\alpha'X_{13}/2}
\end{align}
In these coordinates, the region $\vp\to \infty$ has the integrand $e^{-\vp X_{14}^{\ot}}$, and the $\mathfrak{F}$ does not contribute, and we have the box diagram. In the region $\vp \to -\infty$, we have $e^{+\vp X_{13}}$ contributing, while the $X_{14}^{\ot}$ term becomes 1. In this tropical limit $\vp \to \vp/\alpha'$ and $\alpha'\to 0~,$ the leading exponential survives and the exact factors of $\pm 2$ become irrelevant. This is precisely how the piecewise linear headlight functions appear as the tropical limit of the string amplitudes. In the integration region $\vp\to \infty$, with the $X_{13}$ propagator turned on, the two insertions $\zeta_3$ and $\zeta_4$ are coming close, and we are to get a triangle diagram. However, the integration measure contains $1/(1+e^{-\vp})$, and it makes the triangle contribution zero. If we were dealing with a bosonic open string theory with tachyons as external particles, the extra factor in the measure would be absorbed into the exponent containing $X_{13}$. In that case, we obtain triangles and bubbles in the similar manner via tropicalization and isolating the leading exponent. 

\section{Discussion}
\label{section 7 discussion}
We have studied the scattering amplitudes for scalars transforming as the adjoint of $so(n)$ or $sp(n),$ leading to the non-orientable surfaces. We constructed the usual surfaceology structures by embedding the non-orientable surface into a doubled orientable surface, and projecting the information from the doubled surface. We saw that, analogous to the orientable surfaces, we can construct the appropriate $g$-vectors, headlight functions, and associate momenta to the curves. Similar constructions are possible for higher loops as well. We have explicitly worked out the two-loop case, enumerated the possible curves, associated momenta, and written down the surface Symanzik polynomials. The generalization of spanning trees for surfaces allows us to write down the Symanzik polynomials for higher loop surfaces directly. There are many possible open questions that we point out in this section. 

A major question is: what exactly is the curve integral formula calculating? It is calculating an entity, which generically has bubbles and tadpoles on its external legs. In QFT, one has to perform the LSZ reduction to obtain the on-shell scattering amplitude from such an object. The bubbles and tadpoles on the external legs lead to the mass renormalization for QFT and string theory (unless the mass is protected due to supersymmetry or gauge invariance). In string theory, for generic amplitudes receiving mass renormalization, one can use the string field theory framework, decompose the moduli space into different 'Feynman diagrams,' and deal with the divergent diagrams separately \cite{Pius:2013sca, Pius:2014iaa}. In the curve integral, we do not wish to divide up the moduli space of graphs into a sum over graphs, because that would be equivalent to the usual QFT route. We can, however, put an appropriate factor in the curve integral, similar to the MCG fixing kernel, which \emph{turns off} all the cones with bubbles and tadpoles on the external legs, similar in spirit to the decapitation of tadpoles in \cite{Banerjee:2025hnz}. We can argue that the external legs in the curve integral have physical renormalized masses, have no tadpoles or bubbles on external legs, resulting in sensible QFT amplitudes. However, further investigations regarding the renormalization of the curve integral are needed. For instance, see \cite{Banerjee:2025hnz}.  

In Section \ref{section 6 string amplitudes}, we took a field theory limit of the Möbius strip superstring amplitude, and obtained the $1/N$ subleading contribution to the one-loop amplitude of $SO(N)$ $\mathcal{N}=4$ SYM. It is a sum of eight diagrams, and one can identify the same loop momentum across different non-planar Feynman diagrams using the technology developed. It reinforces the idea about the curve integral, facilitating the choice of a common loop momentum across Feynman diagrams for non-planar processes. The curve integral has potential for efficient writing of the full loop integrand for non-planar amplitudes in $\mathcal{N}=4$ SYM. We will explore these in future works. 

In this article, we considered the field theory limit of the string amplitudes, but did not construct explicit binary geometry coordinates $u_I$ for the worldsheet, as done in \cite{Arkani-Hamed:2019_binary_geometry, Arkani-Hamed:2024pzc}. It is still an open question whether the complete string loop amplitudes can be computed using the surfaceology and combinatorics. In \cite{Arkani-Hamed:2024pzc}, the authors have defined a surface function $G$ closely related to the matrix model correlators, and it will be interesting to construct such surface functions for non-orientable surfaces and to see if one can discover some novel features. 

One can spin up the scalar global Schwinger integral into a gluon scattering amplitude, using the Corolla differential operators \cite{Laddha:2024qtn}. In \cite{Arkani-Hamed:2023-HiddenZeros, Arkani-Hamed:2023-ScalarScaffolding, Arkani-Hamed:2024nhp}, it was discovered that certain deformations of the $\tr \phi^3$ amplitudes yield NLSM and YM amplitudes. It would be interesting to see if similar constructions work for the non-orientable surfaces. We will explore this in future works. 

\subsection*{Acknowledgements} The author is grateful to Pinaki Banerjee, Abhijit Gadde, Harsh, Shiraz Minwalla, Shruti Paranjape, Onkar Parrikar, and Adarsh Sudhakar for discussions. The author thanks Nima Arkani-Hamed for comments on a preliminary version of this manuscript. The author is especially grateful to Sujay Ashok and Alok Laddha for discussions, encouragement, and comments on the manuscript. The research is partly funded by the Infosys Endowment for the study of the Quantum Structure of Spacetime. 

\appendix

\section{Review of combinatorics of the curve integral formula}
\label{appendix A review}
In this section, we briefly review the combinatorics of the curve integral formula. Refer to \cite{Arkani-Hamed:2023CurveIntegral, Arkani-Hamed:2023Multiplicity} for details. Let $S$ be a surface with at least one boundary component and $n$ marked points on the boundaries. We can triangulate the surface $S$ by drawing curves that either end at the marked points, wind around the boundary components with no marked points, or are closed. Note that we are referring to the equivalence class of homotopically equivalent paths on $S$ as the \emph{curve/chord}. Below, we describe how to associate a fixed momentum $P_C$, $\vec{g}_C$, and the headlight function $\alpha_C$ to each curve $C$. 

Let us choose an arbitrary triangulation $T_0$ of $S$ as the reference. Let us draw the dual trivalent graph $\Gamma$, label the internal legs as $\{t_i\}$, and label the momenta on all its legs. For higher genus surfaces, certain momenta are loop momenta. Let us denote the path of a curve $C$ on the graph $\Gamma$ as a series of left and right turns. For instance, \eqref{C_14 curve on the five point} depicts the path of curve $C_{14}$ on a reference triangulation of the disk with five marked points.
\begin{align}
    \label{C_14 curve on the five point}
    \begin{tikzpicture}[scale=0.9,baseline={([yshift=-.5ex]current bounding box.center)}]
        \fill[cyan!10!white] (-1,0) -- (0,0) -- (0,1) -- (0.5,1) -- (0.5,0) -- (1.5,0) -- (1.5,1) -- (2,1) -- (2,0) -- (3,0) -- (3,1) -- (3.5,1) -- (3.5,0) -- (4.5,0) -- (4.5,-0.5) -- (-1,-0.5);
        \draw (-1,-0.5) -- (4.5,-0.5);
        \draw (-1,0) -- (0,0) -- (0,1);
        \draw (0.5,1) -- (0.5,0) -- (1.5,0) -- (1.5,1);
        \draw (2,1) -- (2,0) -- (3,0) -- (3,1);
        \draw (3.5,1) -- (3.5,0) -- (4.5,0);
        \draw[-{Stealth[length=2mm, width=1.5mm]}, thick] (-1.6,-0.25) -- (-1.1,-0.25);
        \draw[-{Stealth[length=2mm, width=1.5mm]}, thick] (5.1,-0.25) -- (4.6,-0.25);
        \node at (-1.9,-0.25) {$p_5$};
        \node at (5.4,-0.25) {$p_1$};
        \draw[-{Stealth[length=2mm, width=1.5mm]}, thick] (0.25,1.6) -- (0.25,1.1);
        \node at (-0.1,1.5) {$p_4$};
        \draw[-{Stealth[length=2mm, width=1.5mm]}, thick] (1.75,1.6) -- (1.75,1.1);
        \node at (1.4,1.5) {$p_3$};
        \draw[-{Stealth[length=2mm, width=1.5mm]}, thick] (3.25,1.6) -- (3.25,1.1);
        \node at (2.9,1.5) {$p_2$};
        \draw[very thick, purple] (0.25,1.05) -- (0.25,-0.25) -- (4.55,-0.25);
        \node at (1,-0.8) {$t_2$};
        \node at (2.5,-0.8) {$t_1$};
    \end{tikzpicture}
    \qquad C_{14}: \ \ 1\,L\,t_1\,L\,t_2\,R\,4 
\end{align}

\paragraph{Momentum:} Using a reference triangulation $T_0$, we can associate the following momentum to every curve $C$:
\begin{align}
    P_C = P_{\text{initial}} + \sum_{\text{right turns}} P_{\text{from left}}~. 
\end{align}
For instance, for $C_{14}$ in \eqref{C_14 curve on the five point}, considering $p_1$ as the initial point, $P_{14} = p_1 + p_2+p_3~.$ Considering $p_4$ as the starting point, we obtain an overall negative sign. One can check that for $S$ being a disk, the momentum squared associated to a curve $C_{ij}$ is $X_{ij} = P_{ij}^2 = (p_i+\hdots + p_{j-1})^2~.$ The momentum associated with a curve is independent of the reference triangulation $T_0~.$ 

\paragraph{g vectors: } Let $E$ be the number of chords used to completely triangulate $S$. One can associate a unique $g$-vector $g_C\in \mathbb{R}^E$ with every curve $C$ by drawing oriented quiver paths inside the reference triangulation, as reviewed in \cite{Laddha:2024qtn}. We review the combinatorics described in \cite{Arkani-Hamed:2023CurveIntegral}. Let $\{t_1,\,\hdots,\,t_E\}$ be the labels of internal edges of the graph dual to $T_0$. For every instance of $L\,t_i\,R$ in the path traversed by $C$, we include $+1$ in the $t_i$ entry of the $E$-dimensional $g$-vector, and for every instance of $R\,t_i\,L$, we include $-1$. For instance, the curve $C_{14}$ in \eqref{C_14 curve on the five point} has a single $L\,t_2\,R$, and so we have $\vec{g}_{14}=(0,1).$ The five curves on the disk with five marked points, and their resultant $g$-vectors are as follows:
\begin{equation}
\begin{aligned}
    C_{13}&: \ \ 1\,L\,t_1\,R\,3 \qquad & \qquad \vec{g}_{13} &= (1,0) ~,\\ 
    C_{14}&: \ \ 1\,L\,t_1\,L\,t_2\,R\,4 \qquad & \qquad \vec{g}_{14} &= (0,1) ~,\\
    C_{24}&: \ \ 2\,R\,t_1\,L\,t_2\,R\,4 \qquad & \qquad \vec{g}_{24} &= (-1,1) ~,\\
    C_{25}&: \ \ 2\,R\,t_1\,L\,t_2\,L\,5 \qquad & \qquad \vec{g}_{25} &= (-1,0)~, \\
    C_{35}&: \ \ 3\,R\,t_2\,L\,5 \qquad & \qquad \vec{g}_{35} &= (0,-1) ~.
\end{aligned}
\label{evaluation of g-vectors for tree 5 points}
\end{equation}
The result is the $g$-vector fan shown in Figure \ref{fig:headlight functions for tree level 5 points}. Note that the $g$-vectors depend on the reference triangulation $T_0~.$

\paragraph{Headlight functions:} The headlight functions $\alpha_C$ are defined to be dual to the $g$-vectors as $\alpha_C(\vec{g}_{C'})=\delta_{CC'}~.$ Let $\{y_1,\hdots,y_E\}$ be variables associated with each internal edge of the graph dual to $T_0.$ Given the path traversed by a curve $C$, let us omit the end-points, and construct the following matrix $M_C$ using the following replacement:
\begin{align}
    L \to \left(\begin{matrix}
        1 & 0 \\ 1 & 1
    \end{matrix}\right). ~, \qquad R \to \left(\begin{matrix}
        1 & 1 \\ 0 & 1
    \end{matrix}\right)  ~, \qquad t_i \to \left(\begin{matrix}
        1 & 0 \\ 0 & y_i
    \end{matrix}\right)~. 
\end{align}
For instance, the curve $C_{13}$ on the disk with five marked points leads to:
\begin{align}
    C_{13} : \ \ 1\,L\,t_1\,R\,3 \qquad \to \qquad M_{13} = \left(\begin{matrix}
        1 & 0 \\ 1 & 1
    \end{matrix}\right).\left(\begin{matrix}
        1 & 0 \\ 0 & y_1
    \end{matrix}\right).\left(\begin{matrix}
        1 & 1 \\ 0 & 1
    \end{matrix}\right) = \left(\begin{matrix}
        1 & 1 \\ 1 & 1+y_1
    \end{matrix}\right) 
\end{align}
Given the matrix $M_C = \left(\begin{matrix}
    M_{C11} & M_{C12} \\ M_{C21} & M_{C22} 
\end{matrix}\right)$, one obtains the headlight functions via tropicalization as follows:
\begin{align}
    \alpha_C = -\text{Trop}\left(\frac{M_{C12}M_{C21}}{M_{C11}M_{C22}}\right)~.
\end{align}
Note that Trop($f$) picks up the leading exponential. For instance, Trop($1+e^{x}+e^{x+y}$) $ = \mxx{0,x,x+y}~.$ In the example of $C_{13}$, introducing $y_i=e^{t_i}$, we obtain:
\begin{align}
    \alpha_{13} = -\text{Trop}\left(\frac{1}{1+y_{1}}\right) = \text{Trop}(0+y_1) = \mxx{0,t_1}~.
\end{align}
Following \eqref{evaluation of g-vectors for tree 5 points}, we see that $\alpha_{13}(\vec{g}_{13})= 1~,$ and $\alpha_{13}(\vec{g}_{C \neq 13}) = 0~.$

{ \paragraph{Non-planar example:}
In the main text, we have realized the Mobius strip with $n$ marked points, as a certain projection of an annulus with $n$ marked points on each of the two boundaries. Here, we review the $g$-vectors for the curves for the simplest case of an annulus with a single marked point on each of its boundary \cite{Arkani-Hamed:2023CurveIntegral}. Let us consider the reference triangulation dual to the following trivalent graph:  
\begin{align}
    \begin{tikzpicture}[scale=0.41,baseline={([yshift=-.5ex]current bounding box.center)}]
    \def \rin {1.7};
    \def \rout {2.5};
    \def \rone {1.9667};
    \def \rtwo {2.2333};
    \def \r {2.1};
    \def \deltaout {9};
    \def \deltain {13};
    \def \lenout {0.6};
    \def \lenin {0.6};     
    \draw ({\rout * cos(90+\deltaout)},{\rout * sin(90+\deltaout) + \lenout}) -- ({\rout * cos(90+\deltaout) },{\rout * sin(90+\deltaout)}) arc (90+\deltaout : 360+90-\deltaout : \rout) -- ({\rout * cos(90-\deltaout)},{\rout * sin(90-\deltaout) + \lenout});
    \draw ({\rin * cos(270+\deltain)},{\rin * sin(270+\deltain) + \lenin}) -- ({\rin * cos(270+\deltain)},{\rin * sin(270+\deltain)}) arc (270+\deltain : 360+270-\deltain : \rin) -- ({\rin * cos(270-\deltain)},{\rin * sin(270-\deltain) + \lenin});
    \node at (0.7,\rout +0.6) {$1$};
    \node at (0,{-\rin+1.2}) {$1'$};
    \node at (\rout+0.5, 0) {$t$};
    \node at (-\rout-0.5, 0) {$t'$};
\end{tikzpicture}
\end{align}
All the curves on this surface are listed below:
\begin{align}
    \textcolor{purple}{C_n := 1Rt'(RtLt')^nL1'}~, \quad \textcolor{teal}{C_n' := 1Lt(Lt'Rt)^nR1'} ~, \qquad  n \geq 0~. 
\end{align}
The curves \textcolor{purple}{$C_0$} and \textcolor{teal}{$C_0'$} are drawn below:
\begin{align}
    \begin{tikzpicture}[scale=0.46,baseline={([yshift=-.5ex]current bounding box.center)}]
    \def \rin {1.7};
    \def \rout {2.5};
    \def \rone {1.9667};
    \def \rtwo {2.2333};
    \def \r {2.1};
    \def \deltaout {9};
    \def \deltain {13};
    \def \lenout {0.6};
    \def \lenin {0.6};     
    \draw ({\rout * cos(90+\deltaout)},{\rout * sin(90+\deltaout) + \lenout}) -- ({\rout * cos(90+\deltaout) },{\rout * sin(90+\deltaout)}) arc (90+\deltaout : 360+90-\deltaout : \rout) -- ({\rout * cos(90-\deltaout)},{\rout * sin(90-\deltaout) + \lenout});
    \draw ({\rin * cos(270+\deltain)},{\rin * sin(270+\deltain) + \lenin}) -- ({\rin * cos(270+\deltain)},{\rin * sin(270+\deltain)}) arc (270+\deltain : 360+270-\deltain : \rin) -- ({\rin * cos(270-\deltain)},{\rin * sin(270-\deltain) + \lenin});
    \node at (0.7,\rout +0.6) {$1$};
    \node at (0.7,{-\rin+0.8}) {$1'$};
    \node at (\rout+0.5, 0) {$t$};
    \node at (-\rout-0.5, 0) {$t'$};
    \draw[line width=1.6, purple] ({\r * cos(90+0.33*\deltaout)},{\r * sin(90+0.33*\deltaout) + 2*\lenout}) -- ({\r * cos(90+0.33*\deltaout)},{\r * sin(90+0.33*\deltaout)}) arc (90+0.33*\deltaout : 270-0.33*\deltaout : \r) -- ({\r * cos(270-0.33*\deltaout)},{\r * sin(270-0.33*\deltaout) + 2*\lenout});
    \draw[line width=1.6, teal] ({\r * cos(90-0.33*\deltaout)},{\r * sin(90-0.33*\deltaout) + 2*\lenout}) -- ({\r * cos(90-0.33*\deltaout)},{\r * sin(90-0.33*\deltaout)}) arc (90-0.33*\deltaout : -90+0.33*\deltaout : \r) -- ({\r * cos(270+0.33*\deltaout)},{\r * sin(270+0.33*\deltaout) + 2*\lenout}); 
\end{tikzpicture}  \qquad 
\begin{matrix}
    \textcolor{purple}{C_0 := 1Rt'L1} ~, \quad \vec{g}_{C_0} = (0,-1) ~; \\ \quad \\ \textcolor{teal}{C_0':= 1LtR1'}~, \quad \vec{g}_{C'_0} = (1,0) ~.
\end{matrix} 
\end{align}
For the curves with generic winding, we have:
\begin{align}
    \vec{g}_{C_n} = (0,-1) + n(-1,1)~, \qquad \vec{g}_{C_n'} = (1,0) + n(-1,1)~.  
\end{align}
Hence, the $g$-vector fan is as follows:
\begin{align}
    \begin{tikzpicture}[scale=1.2,baseline={([yshift=-.5ex]current bounding box.center)}]
        \draw[<->] (-1.2,0) -- (1.2,0);
        \draw[<->] (0,-1.2) -- (0,1.2);
        \draw[thick,teal,-{Stealth[length=2mm, width=1.5mm]}] (0,0) -- (1,0);
        \draw[thick,teal,-{Stealth[length=2mm, width=1.5mm]}] (0,0) -- (0,1);
        \draw[thick,teal,-{Stealth[length=2mm, width=1.5mm]}] (0,0) -- (-1,2);
        \draw[thick,teal,-{Stealth[length=2mm, width=1.5mm]}] (0,0) -- (-2,3);
        \draw[thick,purple,-{Stealth[length=2mm, width=1.5mm]}] (0,0) -- (0,-1);
        \draw[thick,purple,-{Stealth[length=2mm, width=1.5mm]}] (0,0) -- (-1,0);
        \draw[thick,purple,-{Stealth[length=2mm, width=1.5mm]}] (0,0) -- (-2,1);
        \draw[thick,purple,-{Stealth[length=2mm, width=1.5mm]}] (0,0) -- (-3,2);        
        \node[purple, rotate=120] at (-2.5,2.1) {$\hdots$};
        \node[teal, rotate=-20] at (-2.1,2.5) {$\hdots$};
        \draw[thick,-{Stealth[length=2mm, width=1.5mm]}, dashed] (0,0) -- (-1,1);
        \node at (1.3,0.2) {$t$};
        \node at (0.2,1.3) {$t'$};
        \node[teal] at (0.9,-0.3) {$C_0'$};
        \node[teal] at (0.3,0.9) {$C_1'$};
        \node[teal] at (-0.6,1.9) {$C_2'$};
        \node[teal] at (-1.5,2.9) {$C_3'$};
        \node[purple] at (0.3,-0.9) {$C_0$};
        \node[purple] at (-0.9,-0.3) {$C_1$};
        \node[purple] at (-1.9,0.6) {$C_2$};
        \node[purple] at (-2.9,1.6) {$C_3$};
        \node at (-1.2,1.2) {$\Delta$};
    \end{tikzpicture}
    \label{fan for annulus with 2 marked points}
\end{align}
The set of vectors $\vec{g}_{C_n}$ and $\vec{g}_{C'_n}$ do not span the entire $\mathbb{R}^2$, so we need include the (dashed) vector $\vec{g}_\Delta = (-1,1)~. $ It corresponds to the closed curve $C_\Delta \equiv (RtLt')^\infty~.$ 

We have similar constructions for the higher points generalizations. For example, the consider the case of annulus with $n$ marked points on either of the two boundaries and consider the two curves depicted in Figure \ref{fig:Schwinger parameters for generic n with tadpole as reference}. The path for the two is as follows:
\begin{equation}
    \begin{aligned}
    \textcolor{purple}{ C_{i'j}^{\text{aux}}:= i'Rt'_{i-1}L\hdots L t'_{n-1}L t_0 Rt_{n-1}L\hdots Lt_{j-1}Rj}~~,  \\
    \textcolor{teal}{ C_{ij'}^{\text{aux}}:= iLt_{i-1}R\hdots R t_{n-1} R t_0' Lt_{n-1}'R\hdots R t'_{j-1}Lj' } ~~. 
\end{aligned}
\label{curves from figure 9}
\end{equation}
Thus, we have the $g$-vectors as follows:
\begin{equation}
    \begin{aligned}
    \textcolor{purple}{ \vec{g}^{\text{aux}}_{i'j} =\big(\stackrel{t_0}{1},\hdots, \stackrel{t_{j-1}}{1},\hdots,\stackrel{t_{n-1}}{-1}~;~\hdots,\stackrel{t'_{i-1}}{-1},\hdots\big) } ~~,\\ 
    \textcolor{teal}{ \vec{g}^{\text{aux}}_{ij'} = \big(\hdots,\stackrel{t_{i-1}}{1},\hdots~;~\stackrel{t'_0}{-1},\hdots,\stackrel{t'_{j-1}}{-1},\hdots,\stackrel{t'_{n-1}}{1}\big)} ~~.
\end{aligned}
\end{equation}
We project these $g$-vectors to obtain the $g$-vectors for curves on the Mobius strip in \eqref{projection of random two curves}. Note that the closed curve with reference triangulation shown in Figure \ref{fig:Schwinger parameters for generic n with tadpole as reference} is $(Rt_0Lt_0')^\infty$ for generic $n$.
}

\section{Details omitted in the main text}
\label{appendix B details}

In this section, we present some detailed calculations omitted in the main text. 
\subsection*{Explicit loop momentum integral for Möbius strip}
Let us restrict ourselves to the $t_0>0$ region, and have consistently $\alpha^{\ot}=0$. The curve integral for Möbius${}_n$ is as follows:
\begin{align}
    A_n^{\text{Mob}}(\{p_i\}) = \int_{t_0>0}\dd^n \vec{t}\int \dd^D l \ e^{-\sum_{i\leq j}\alpha^{\ot}_{ij}(l+P_{1i}+P_{1j})^2 + \sum_{i,k}\alpha_{ik}X_{ik}}~.
    \label{explicit loop integration 1}
\end{align}
Let us carry out the loop integration:
\begin{align}
    \int \dd^D l \ e^{-\sum_{i\leq j}\alpha^{\ot}_{ij}(l+P_{1i}+P_{1k})^2} &= \int \dd^D l\ e^{-\sum_{i\leq j}\alpha_{ij}^{\ot}\big(l^2 + 2l.(P_{1i}+P_{1j}) + (P_{1i}+P_{1j})^2 \big) } \nonumber \\
    &\equiv \int \dd^D l \ e^{-\mathcal{U}\left[l^2 + 2l.\frac{1}{\mathcal{U}}\sum_{i\leq j}(P_{1i}+P_{1j})\alpha_{ij}^{\ot} + \frac{1}{\mathcal{U}}\sum_{i\leq j}(P_{1i}+P_{1j})^2\alpha_{ij}^{\ot} \right]} \nonumber \\
    &=\int \dd^D l \ e^{ -\mathcal{U}\left[ \tilde{l}^2 - \left(\frac{1}{\mathcal{U}}\sum_{i\leq j}(P_{1i}+P_{1j})\alpha_{ij}^{\ot}\right)^2 + \frac{1}{\mathcal{U}}\sum_{i\leq j}(P_{1i}+P_{1j})^2\alpha_{ij}^{\ot} \right]} \nonumber \\
    &= \frac{1}{\mathcal{U}^{D/2}} \ e^{+\frac{1}{\mathcal{U}}\left(\sum_{i\leq j}(P_{1i}+P_{1j})\alpha_{ij}^{\ot}\right)^2 - \sum_{i\leq j}(P_{1i}+P_{1j})^2\alpha_{ij}^{\ot}} \nonumber \\
    &\equiv \frac{1}{\mathcal{U}^{D/2}} \ e^{-\mathcal{F}/\mathcal{U}}~.
\end{align}
We introduced $\mathcal{U}=\sum_{i\leq j}\alpha_{ij}^{\ot}$ in the second equality, and $\mathcal{F}$ in the last equality:
\begin{align}
    \mathcal{F} &= -\left(\sum_{i\leq j}(P_{1i}+P_{1j})\alpha_{ij}^{\ot}\right)^2 + \mathcal{U}\sum_{i\leq j}(P_{1i}+P_{1j})^2\alpha_{ij}^{\ot} \nonumber \\
    &= -\sum_{i\leq j} \sum_{k\leq l} (P_{1i}+P_{1j}).  (P_{1k}+P_{1l})\alpha_{ij}^{\ot}\,\alpha_{kl}^{\ot} + \sum_{i\leq j} \sum_{k\leq l} (P_{1i}+P_{1j})^2 \alpha_{ij}^{\ot}\alpha_{kl}^{\ot} \nonumber \\
    &= \sum_{i\leq j} \sum_{k\leq l} \alpha_{ij}^{\ot}\alpha_{kl}^{\ot} \big((P_{1i}+P_{1j})^2 -(P_{1i}+P_{1j}).  (P_{1k}+P_{1l}) \big) \nonumber \\
    &= \sum_{i\leq j} \sum_{k\leq l} \alpha_{ij}^{\ot}\alpha_{kl}^{\ot} \left(-X_{ij}+X_{1i}+X_{1j}-X_{1k}-X_{1l}+\frac{1}{2}\big(X_{ik}+X_{jk}+X_{il}+X_{jl}\big)\right)~.
\end{align}
In the last expression, the coefficient of $\alpha_{ij}^{\ot}\alpha_{kl}^{\ot}$ is not yet symmetric under $(ij)\leftrightarrow (kl)~.$ For instance, there are two terms that contribute to $\alpha_{13}^{\ot}\alpha_{24}^{\ot}$, namely $(ij)=(13)~,~(kl)=(24)$ and $(ij)=(24)~,~(kl)=(13)$. We simplify further by choosing $i\leq k~,$ so that we have:
\begin{align}
    &\mathcal{F} = \nonumber \\
    &\!\!\sum_{\substack{i\leq k\leq j \\ k\leq l}}\!\!\alpha_{ij}^{\ot}\alpha_{kl}^{\ot} \bigg[ \left(-X_{ij}+X_{1i}+X_{1j}-X_{1k}-X_{1l}+\frac{1}{2}\big(X_{ik}+X_{jk}+X_{il}+X_{jl}\big)\right) + \bigg((ij)\leftrightarrow (kl)\bigg) \bigg] \nonumber \\
    &= \sum_{\substack{i\leq k\leq j \\ k\leq l}}\alpha_{ij}^{\ot}\alpha_{kl}^{\ot} \big( -X_{ij} -X_{kl} + X_{ik}+X_{jk}+X_{il}+X_{jl} \big)~.
\end{align}
With the constraints $i\leq k\leq j$ and $k\leq l~,$ there are two possibilities $i\leq k\leq j\leq l$ and $i\leq j\leq k\leq l~.$ For the latter case, the curves $C_{ij}^{\ot}$ and $C_{kl}^{\ot}$ are not compatible:
\begin{align}
    \alpha_{ij}^{\ot}\alpha_{kl}^{\ot} = 0 ~, \quad \forall\quad   i\leq j\leq k\leq l~.
\end{align}
So, we obtain the following formula for $\mathcal{F},$ with the limits of $i,j,k,l$ updated as follows:
\begin{align}
    \mathcal{F}^{\text{Mob}} = \sum_{i\leq k \leq j \leq l}\alpha_{ij}^{\ot}\alpha_{kl}^{\ot} \big( -X_{ij} -X_{kl} + X_{ik}+X_{jk}+X_{il}+X_{jl} \big)~.
    \label{explicit loop integration 2}
\end{align}
The other two surface Symanzik polynomials are as follows:
\begin{align}
    \mathcal{U}^{\text{Mob}} = \sum_{i\leq k} \alpha^{\ot}_{ik}~, \hspace{2cm} 
    \mathcal{Z}^{\text{Mob}} = \sum_{i,k} \alpha_{ik}X_{ik}~.
\end{align}
The loop integrated curve integral for the Möbius strip, with the external momentum dependence coming through $X_{ik}$, is as follows:
\begin{align}
    A_n^{\text{Mob}}(\{p_i\}) = A_n^{\text{Mob}}(\{X_{ik}\})= \int_{t_0>0}\dd^n \vec{t} \,\frac{1}{\mathcal{U}^{D/2}}\,e^{-\mathcal{F}/\mathcal{U}+\mathcal{Z}}~. 
    \label{explicit loop integration 3}
\end{align}
Surface Symanzik polynomials obtained from explicit loop integration matches with the ones constructed from spanning subsurfaces.

\subsection*{Explicit $g$-vectors and headlight functions for $n=3$ with wheel as reference triangulation}
The following are the $g$-vectors for Möbius${}_3$ with the triangle graph $\big(C_{11}^{\ot},~C_{12}^{\ot},~C_{31}^{\ot}\big)$ as the reference triangulation:
\begin{equation}
\label{g-vectors for n=3 with wheel as reference}
    \begin{aligned}
            \vec{g}_{12} &= \big(0,1,-1\big) ~~, &\hspace{2cm} \vec{g}^{\ot}_{12} &= \big(0,1,0\big) ~~, \\
            \vec{g}_{23} &= \big(-1,0,-1\big) ~~, &\hspace{2cm} \vec{g}^{\ot}_{23} &= \big(-1,0,0\big) ~~, \\
            \vec{g}_{31} &= \big(1,-1,0\big) ~~, &\hspace{2cm} \vec{g}^{\ot}_{31} &= \big(0,0,1\big) ~~, \\
            \vec{g}_{11} &= \big(1,0,-1\big) ~~, &\hspace{2cm} \vec{g}^{\ot}_{11} &= \big(1,0,0\big) ~~, \\
            \vec{g}_{22} &= \big(-1,1,-2\big) ~~, &\hspace{2cm} \vec{g}^{\ot}_{22} &= \big(-1,1,-1\big) ~~, \\
            \vec{g}_{33} &= \big(0,-1,-1\big) ~~, &\hspace{2cm} \vec{g}^{\ot}_{33} &= \big(0,-1,0\big) ~~, \\
            & \hspace{2cm} & \vec{g}^{\ot} = \big(0,0,-1\big) ~. \qquad &
    \end{aligned}
\end{equation}
The following are the dual headlight functions:
\begin{equation}
\label{headlights for n=3 with wheel as reference}
    \begin{aligned}
        \alpha_{12} &= \mxx{\mxx{0,\,t_2,\,t_1+t_2}+\mxx{0,\,t_3,\,t_3+t_2}\ ,\ t_1+2t_2} - \mxx{0,\,t_2,\,t_1+t_2} \\
        &\qquad-\mxx{0,\,t_3,\,t_3+t_2} ~, \\
        \alpha_{23} &= \mxx{0\ ,\ \mxx{0,\,t_2,\,t_1+t_2}+\mxx{t_1,\,t_3,\,t_1+t_3}} - \mxx{0,\,t_2,\,t_1+t_2} \\
        &\qquad - \mxx{t_1,\,t_3,\,t_1+t_3} ~, \\
        \alpha_{31} &= \mxx{0,\,t_1,\,t_2,\,t_1+t_2} - \mxx{0,\,t_2,\,t_1+t_2}~, \\
        \alpha_{11} &= 2\,\mxx{0,\,t_2,\,t_1+t_2} + \mxx{0,\,t_3,\,t_2+t_3} - \mxx{0,t_2} \\
        &\qquad - \mxx{0,\,t_2,\,t_1+t_2,\,t_3+2\,\mxx{0,\,t_2,\,t_1+t_2}} ~, \\
        \alpha_{22} &= 2\,\mxx{0,\,t_2,\,t_1+t_2} + \mxx{0,\,t_3,\,t_2+t_3} + \mxx{t_1,\,t_3,\,t_1+t_3} \\
        &\quad - \mxx{t_1+\mxx{0,\,t_2,\,t_1+t_2} \ ,\ \mxx{0,t_1} + \mxx{0,t_2} + \mxx{0,\,t_2,\,t_1+t_2}+t_3 } \\
        &\qquad  - \mxx{0,\,t_3,\,t_2+t_3,\,t_1+t_2,\,t_1+t_2+t_3}~,  \\
        \alpha_{33} &= 2\,\mxx{0,\,t_2,\,t_1+t_2} + \mxx{t_1,\,t_3,\,t_1+t_3} - \mxx{0,t_1} \\
        &\qquad -\mxx{t_3\ ,\ t_2+\mxx{0,\,t_2,\,t_1+t_2} + \mxx{t_1,\,t_3,\,t_1+t_3} } ~, \\
        \alpha^{\ot}_{12} &= \mxx{0,\,t_3,\,t_2+t_3} - \mxx{0,t_3} ~, \\
        \alpha^{\ot}_{23} &= \mxx{t_1,\,t_3,\,t_1+t_3} - \mxx{0,t_3} - t_1~, \\
        \alpha^{\ot}_{31} &= \mxx{0,t_3} ~, \\ 
        \alpha^{\ot}_{11} &= \mxx{0,\,t_3,\,t_2+t_3,\, t_1+t_2+t_3} - \mxx{0,\,t_3,\,t_2+t_3} ~, \\
        \alpha^{\ot}_{22} &= \mxx{0,\,t_3} + \mxx{t_1\ ,\ t_3+\mxx{0,t_1} + \mxx{0,t_2}} - \mxx{0,\,t_3,\,t_2+t_3} \\
        &\qquad - \mxx{t_1,\,t_3,\,t_1+t_3} ~, \\
        \alpha^{\ot}_{33} &= \mxx{t_3+\mxx{0,t_2}\ , \ t_1+t_2+\mxx{0,t_3}} - t_2 - \mxx{t_3,\,t_1+\mxx{0,t_3}} ~, \\
        \alpha^{\ot} &= \mxx{t_3+2\,\mxx{0,\,t_2,\,t_1+t_2}+\mxx{t_1,t_3,t_1+t_3}\ ,\ 2t_1+t_2} - t_3  \\
        &\qquad -2\,\mxx{0,\,t_2,\,t_1+t_2} - \mxx{t_1,t_3,t_1+t_3} ~.
    \end{aligned}
\end{equation}
One can check that $\alpha_C\big(\vec{g}_C'\big)=\delta_{CC'}~.$

\newpage
\subsection*{Determining the momenta $P_{ik}^{\ot}$ for Mobius${}_3$ via mutations}
\begin{figure}[H]
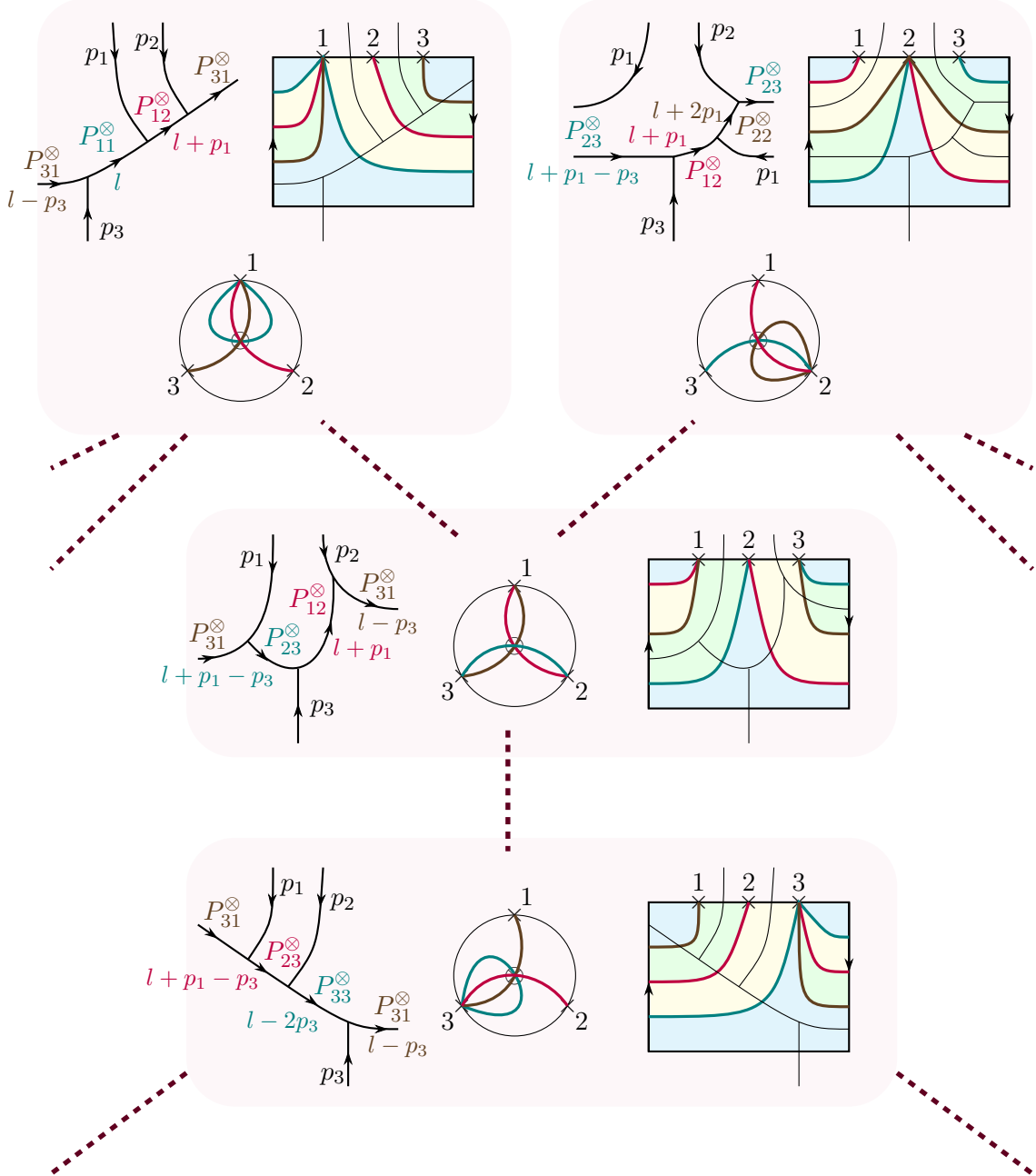

    \centering

\caption{A small region of the combinatorial polytope Möbius${}_{3}$ and the trivalent graphs associated to them. Choosing $P_{11}^{\ot}=l$ fixes the $P_{ik}^{\ot}$ through the mutations.}
    \label{fig:momenta-determination-Mobius3}
\end{figure}

Consider a Mobius strip with three marked points, and let us choose $P_{11} \equiv l$ as the loop momentum. In the trivalent graph $\{C_{11}^{\ot},~C_{12}^{\ot},~C_{13}^{\ot}\},$ it determines the momenta $P_{12}^{\ot}$ and $P_{13}^{\ot}$ by momentum conservation. By considering mutations, we obtain other diagrams with different curves, and we determine the associated momenta using momentum conservation. Figure \ref{fig:momenta-determination-Mobius3} depicts this. Using this, we find that $P_{ik}^{\ot} = l+P_{1i} +P_{1k}~.$

\newpage
\begin{figure}[H]
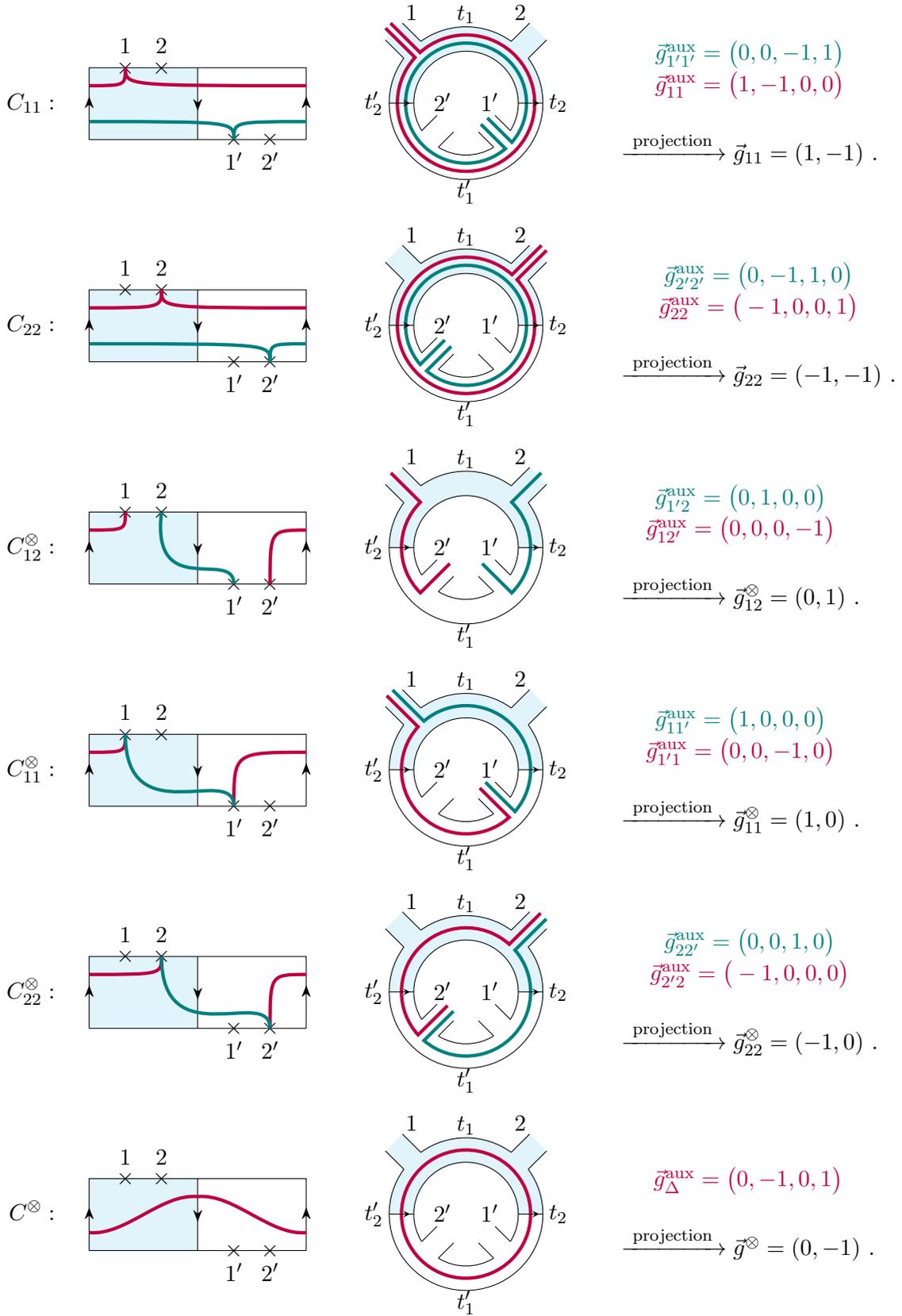

    \begin{align*}
C_{11}:\quad %
\end{align*}
\vspace{-0.7cm}
    \caption{Evaluation of $g$-vectors for Möbius${}_2$ by embedding it into an annulus.}
    \label{fig:g-vectors for Möbius2}
\end{figure}

\subsection*{Determining the $g$-vectors for Mobius${}_2$}
We embed the Möbius strip with two marked points into an annulus with two marked points on each boundary, and project out the $g$-vectors accordingly. Figure \ref{fig:g-vectors for Möbius2} depicts the calculation. The result is the $g$-vector fan in Figure \ref{fig:g-vector fan for Möbius1,2}.

\bibliographystyle{JHEP}
\bibliography{amplitudes}

\end{document}